\documentclass{eas}
\usepackage{graphicx}
\usepackage{fancybox}
\usepackage{amsmath,amssymb}
\usepackage{floatrow}

\begin{document}

\title{Influence of Rotation on Stellar Evolution} 
\author{A. Palacios}\address{LUPM - UMR 5299 - CNRS/Universit\'e
  Montpellier II - cc 072 - Place Eug\`ene Bataillon - F-34095
  Montpellier (France)}

\begin{abstract}
The Sun has been known to rotate for more than 4 centuries, and
evidence is also available through direct measurements, that almost
all stars rotate. In this lecture, I will propose a review of the
different physical processes associated to rotation that are expected
to impact the evolution of stars. I will describe in detail the way
these physical processes are introduced in 1D stellar evolution codes
and how their introduction in the modelling has impacted our
understanding of the internal structure, nucleosynthesis and global
evolution of stars.
\end{abstract}
\maketitle

\section{Foreword}

In his Ph.D. thesis, published in 2012, Dr A. Potter writes a humoristic
yet quite true sentence 
\begin{quotation}
 {\em Stars rotate because it is actually
  quite hard for them not to.}
\end{quotation} 
The molecular clouds that form stars are highly turbulent and part of their velocity dispersion is due to rotation. Although a substantial fraction of the total angular momentum should be lost during the collapse, the stars that are formed inherit angular momentum, and rotation is a natural feature among their characteristics (see P. Hennebelle and A. Belloche in this volume).\\
This fact being established conceptually and observationally (see \S~\ref{s-obs}), the question that remains is whether rotation significantly affects the structure, evolution and nucleosynthesis of stars, and if so, how and to what extent.\\
These are the questions within the scope of this lecture, where I will briefly summarize the observational evidences of stellar rotation (\S~\ref{s-obs}), review the expected effects on the stellar structure and how they can be taken into account in stellar evolution modelling (\S~\ref{s-struc}). I will also describe the main transport mechanisms that could be driven by rotation in stellar interiors (\S~\ref{s-transport}) and how angular momentum evolution can be followed in stellar evolution codes (\S~\ref{s-rotmod}). In the last part (\S~\ref{s-evolution}), I will present some key results obtained with rotating stellar evolution models.

\section{Observational evidences of stellar rotation}\label{s-obs}

\subsection{Measurements of stellar rotation}
The rotation of stars was first evidenced by Galileo Galilei (\cite{galilee}) in the case of the Sun, and in the late nineteenth - early twentieth century, the development of spectroscopy allowed astronomers to first envision to measure the rotation of stars by studying the broadening of the spectral lines that should result from it (\cite{abney,monck,souleyre,fowler}).\\
Direct measurements of the rotation rate of stars are accessible via 
\begin{itemize}
\item {\em photometry} light curve modulation by rotation allows to derive rotation periods P$_{\rm rot}$; 
\item {\em spectroscopy} spectral lines broadening is partly due to surface rotation and an estimation of the projected surface velocity on the line of sight, $\upsilon \sin i$ can be obtained from spectral analysis;
\item {\em spectropolarimetry and Doppler imaging} modulation of the polarized signal by rotation allowing to derive rotation periods P$_{\rm rot}$ and surface latitudinal differential rotation $\delta \Omega$;
\item {\em asteroseismology} splitting of oscillation modes due to rotation allows the
reconstruction of the rotation profile of some stars in addition to the Sun;
\item {\em interferometry} direct imaging of rapidly rotating non-spherical stellar surfaces allows to estimate the mean rotation rate as a fraction of the break-up velocity.
\end{itemize}

\begin{figure}[t]
\includegraphics[scale=0.5]{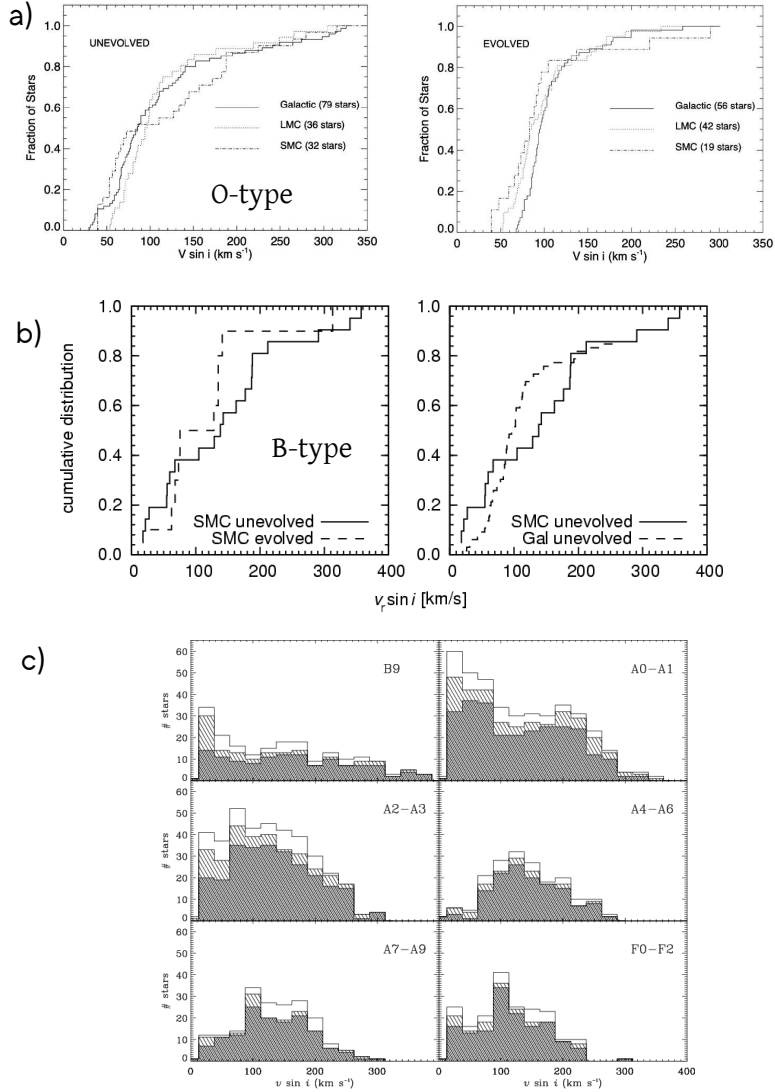}
\caption{Rotation of early-type stars. {\em a)} Projected velocity distribution of O-type stars ({\em From Mokiem et al. 2006}); {\em b)} Projected velocity distribution of B type stars ({\em From Penny \& Gies 2009}); {\em c)} Projected velocity distribution for A-type stars ({\em From \cite{royer07}})}
\label{f-obsrot1}
\end{figure}

\begin{figure}[t]
\includegraphics[scale=0.5]{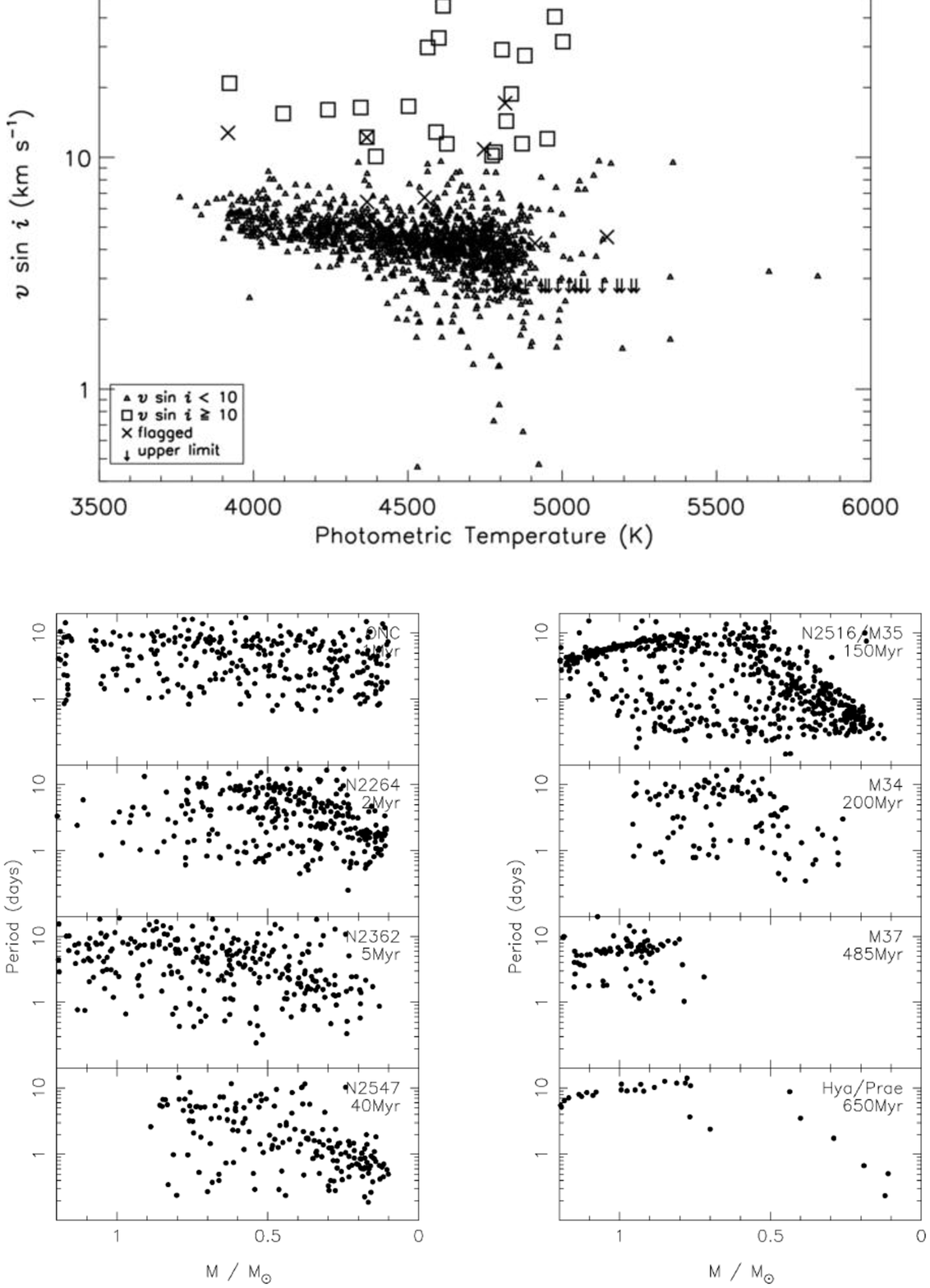}
\caption{Rotation of cool stars . {\sf Upper panel} Projected velocity as a function of surface temperature for a sample of K-type giants ({\em From Carlberg et al. 2011}); {\sf Lower panels} Rotation periods as a function of mass and age in a set of open clusters ({\em From Irwin \& Bouvier 2009}).}
\label{f-obsrot2}
\end{figure}

For some types of stars it is very difficult to actually derive a rotation rate, either because the spectral lines broadening is dominated by another process, as surface convection, which is the case of red giants and supergiants, or because the actual stellar surface is not accessible to observation, as it is the case for Wolf-Rayet stars, in which the dense winds yield the formation of spectral lines in the wind and not in the hydrostatic regions that could define an actual stellar surface, preventing any measurement of  stellar rotation.\\

The rotational study of stars based on these measurements allows us to draw some sort of rotational HR diagram :

\subsubsection*{\bf O and B-type stars (Fig.~\ref{f-obsrot1}, ab):}

 $\upsilon \sin i$ measurements in the Galaxy and in the Magellanic Clouds indicate that these are rapidly rotating stars, with a mean projected velocity of about 130 km.s$^{-1}$ for B-type stars (\cite{brag12,hunter08b,mokiem06,abt02}), and about 110 km.s$^{-1}$ for O-type stars  (\cite{penny96,penny09}). Some stars in the On subtype (with nitrogen enrichment) can reach projected velocities of about 420 km.s$^{-1}$ (\cite{walborn11}) but the number of these very rapid rotators is actually limited.
 
\subsubsection*{\bf A and F-type stars (Fig.~\ref{f-obsrot1}c)} 
A large survey of A and F type stars allowed Royer and collaborators (see \cite{royer07} and references therein) to establish a good statistic for their typical rotation. These are generally rapid rotators,  $\left\langle \upsilon \sin i \right\rangle \approx 150$ km.s$^{-1}$, although the velocity distribution is more complex when going into more details for each subtype. For instance, a significant fraction of late A-type stars rotate at more than 50\% of their break-up velocity. Actually part of the fast rotating stars imaged with interferometry and rotating near the break-up limit are A-type stars (see \cite{ZMC11}).

\subsubsection*{\bf G, K and M-type giants (Fig.~\ref{f-obsrot2})}
 Red giant stars are generally slow rotators, with surface velocities lower than about 10 km.s$^{-1}$ (\cite{medeiros04,carlberg11}). However, some of them present unusually higher projected velocities ($\gtrsim 10$ km.s$^{-1}$) as evidenced by \cite{rucinski90,barnbaum95,carlberg11,carlberg12}. Also evidenced from asteroseismic data, they have been interpreted as the result of some peculiar events that may occur during this phase.
 
\subsubsection*{\bf Low-mass pre-main sequence and main sequence stars (Fig.~\ref{f-obsrot2})}
A large effort has been dedicated to measure rotation periods via photometric surveys of open clusters sampling a wide range of ages (see MONITOR project for instance, and J. Bouvier this volume for more details). These studies show that from an essentially bi-modal distribution of P$_{\rm rot}$ at young ages, the rotation converges to a single low P$_{\rm rot}$ (for solar-type stars) or high P$_{\rm rot}$ (for very low-mass stars - M-type) at the age of the Hyades.\\

These direct measurements testify the stellar rotation and indicate in some cases as those shown in Fig.~\ref{f-gravdark}, that it can lead to significant departures from the spherical symmetry. They also show that the stellar angular velocity (and angular momentum) differs according to spectral type, and thus stellar mass but also time, and that it evolves with time along the stellar life.

\begin{figure}
\includegraphics[scale=0.48]{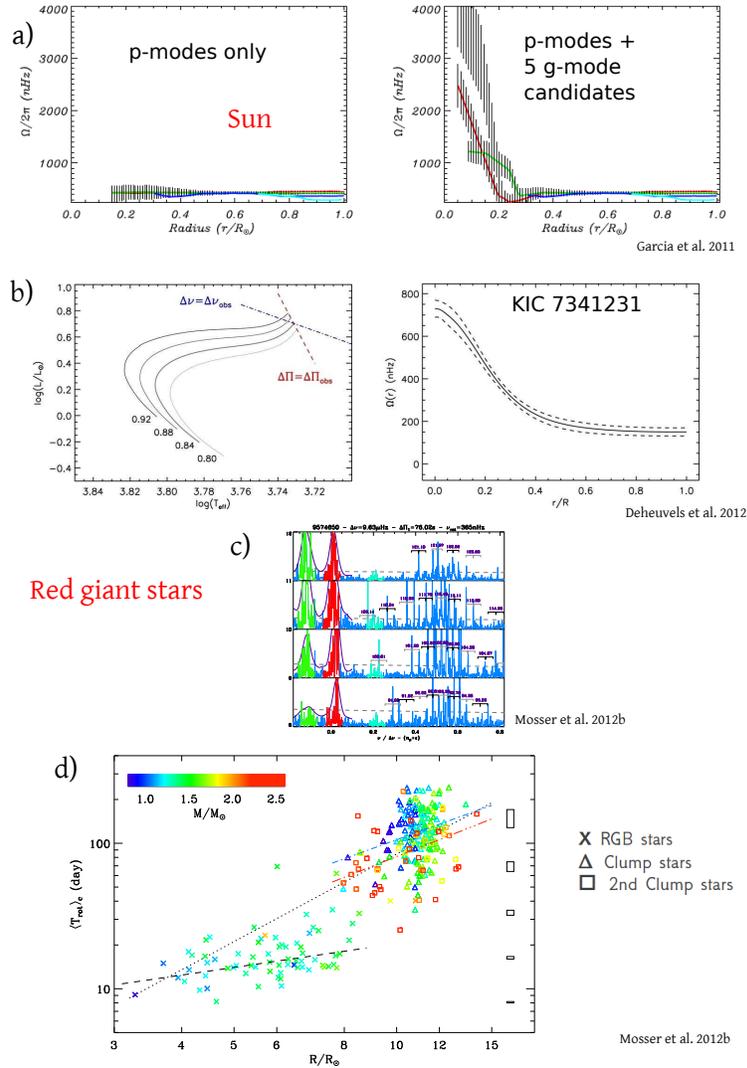}
\caption{Probing the internal rotation of stars through solar-like oscillations : {\em a)} Inversion of the internal rotation profile of the Sun using 4608 days of MDI and GOLF data; {\em b)} Evolutionary status and internal rotation profile as deduced from mixed-modes for the subgiant KIC 7341231; {\em c)} Rotational splittings of the mixed modes corresponding to radial orders 8 to 11 in the giant KIC 9574650; {\em d)}  Mean period of core rotation as a function of the asteroseismmic stellar radius for about 300 Kepler giants.}
\label{f-astero}
\end{figure}

\subsubsection*{\bf Internal rotation of the Sun and of red giants (Fig.~\ref{f-astero})}

Thanks to helioseismic data from the SoHO satellite (\cite{garcia11}) and much more recently to the instruments on-board the CoRoT and KEPLER satellites (\cite{deridder09,beck12,deheuvels12,mosser12a,mosser12b}), an insight to the {\em internal rotation profile} of the Sun and of a large number of red giant stars has become available. This opens a completely new road for our understanding of stellar structure and angular momentum distribution and constitutes extremely valuable constraints to stellar models.\\
Solar-like oscillations are due to the trapping of sound waves excited by convection. Assuming a spherical star, such oscillations appear as a discrete spectrum of degenerate modes characterized by their radial order $n$ and their spherical harmonic degree $l$.  The azimuthal number $m$ is degenerated, a degeneracy that is completely lifted by rotation as well as other aspherical perturbations. \\
In the Sun, the pressure modes ($p$-modes) are trapped in relatively superficial cavities and probe the convective envelope and part of the underlying radiative interior, the solar core remaining unprobed. Gravity modes on the other hand ($g$-modes) are generally confined to inner cavities, and scarcely appear at the surface. They could probe the solar core but do not appear at the solar surface and are extremely difficult to detect : after 14 years of SoHO data integration, only $g$-mode candidates have been tentatively identified (\cite{garcia11}). In red giant stars, the situation was discovered to be unexpectedly simpler, since mixed gravity/pressure modes have been detected. These modes correspond to the coupling of gravity waves in the interior to pressure waves in the envelope and give unprecedented insight on the internal properties of red giant stars.\\
Figure~\ref{f-astero} displays the inverted radial rotation profile of the Sun (\cite{garcia11}) and of a subgiant in the Kepler field (\cite{deheuvels12}), as well as a core rotation period - asteroseismic stellar radius plot for a sample of about 300 red giants from the Kepler field (\cite{mosser12b}). Red giants appear to experience differential rotation in their interior, and hints exist that a similar behaviour could exist in the solar central regions as can be seen from panels a) and b).\\ Another important point coming out of these data is a clear distinction between clump (e.g. core helium burning stars) and red giant stars (e.g. shell hydrogen burning stars) in terms of core rotation period (panel d)). This provides a new powerful tool to discriminate between stars belonging to these very different evolutionary phases that occupy overlapping regions in the Hertzsprung-Russell diagram and are thus very difficult to distinguish from classical surface parameters as temperature or magnitude.

\subsection{Indirect probes of stellar rotation}

\begin{figure}
\includegraphics[scale=0.45]{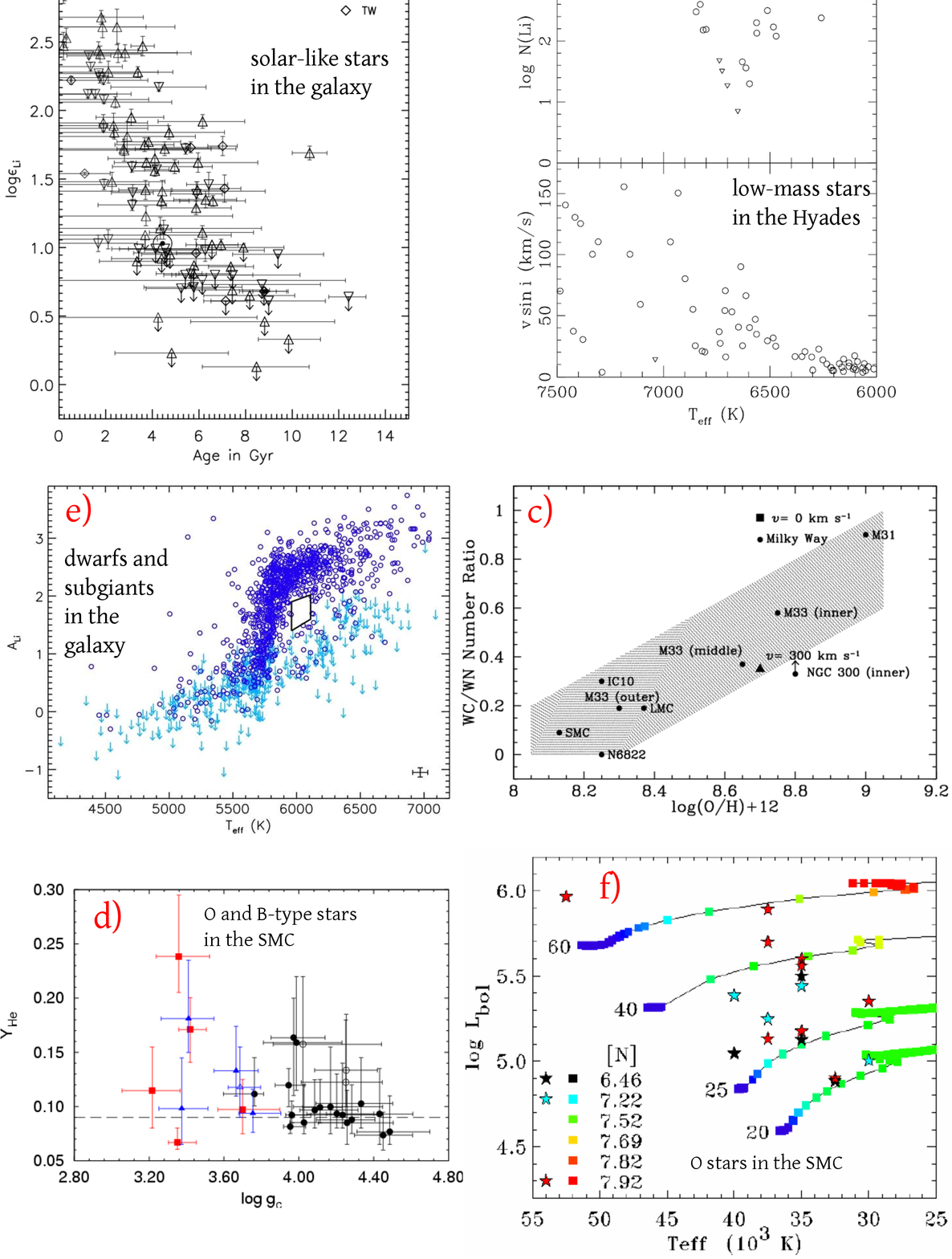}
\caption{{\bf a} Lithium abundances in solar-type stars as a function of age ({\em from} \cite{bauman10}); {\bf b} Lithium abundances and projected rotation velocities as a function of T$_{\rm eff}$ in stars of the Hyades ({\em from \cite{TC98}}); {\bf c} Number population of two types of Wolf-Rayet stars as a function of metallicity traced by the oxygen abundance ({\em from \cite{MM03}}); {\bf d} He overabundances in O-type stars of the Magellanic Clouds ({\em from \cite{mokiem06}}); {\bf e} Li abundances as a function of T$_{\rm eff}$ for a sample of dwarfs and subgiants in the Galaxy ({\em from \cite{mish12}}); {\bf f} Nitrogen abundances in O-type stars of the SMC ({\em from \cite{heap06}})}
\label{f-obsabund}
\end{figure}

In addition to the direct measurements, the analysis of the surface chemical properties of stars may constitute a set of evidences for transport processes occurring inside the stars, some of them being attributed to rotation-driven processes.\\
"{\em Abundance anomalies}", e.g. departures from the predictions of classical\footnote{Stellar evolution models including no transport of matter in the radiative regions are referred to as {\em classical} models. The only {\em mixed} regions are convectively unstable ones. {\em Standard} stellar evolution models account for microscopic diffusion processes that may occur on long time scales in radiative regions such as gravitational settling or radiative levitation.} stellar evolution models are powerful diagnostic tools to probe the efficiency of transport processes that pollute the stellar atmosphere. Such transport can be of internal origin and connect regions with efficient nucleosynthesis to the surface, or from an external source when material of different chemical composition is accreted onto the star.\\As all stars rotate, it is natural to first look for rotation-induced transport processes to explain the discrepancies between abundance observations and predictions of classical and standard stellar evolution models. One should nevertheless bear in mind that a degeneracy exists concerning the nature of the transport processes itself, and that if rotation-driven mechanisms can be found to be dominant in some cases, they are certainly not the only ones acting.\\

Those abundance variations observed at the surface of stars that reflect nucleosynthetic processes that are known to occur inside these very same stars can be considered as evidences for transport / mixing processes connecting the surface to the regions where the nucleosynthesis occurs. Indeed, the abundances of the products of hydrogen burning via the pp-chains and the {\em extended} CNO cycle (including NeNa- and MgAl- chains) can be used to trace internal mixing when variations are observed at evolutionary phases where they are not expected from standard stellar evolution models. In red giant stars, the presence of $s$-process\footnote{The $s$-process is a nucleosynthesis path that involves the production of heavy elements by slow neutron captures (\cite{Clayton84}).} at the surface may also be considered as evidence for internal mixing.
Figure~\ref{f-obsabund} presents a set of observed abundance patterns that are used as indirect probes for internal (rotational) mixing: \\

\begin{itemize}
\item Lithium depletion during phases where no convective dredge-up is expected $\longrightarrow$ {\sf in low- and intermediate-mass stars} (\cite{boes89,boes86,gratton00,bauman10,CM11}\newline \cite{mish12})
\item Lithium enrichment $\longrightarrow$ {\sf in low- and intermediate-mass stars} (\cite{CB00,agnes09})
\item Boron and Beryllium depletion $\longrightarrow$ {\sf in low-, intermediate-mass, and massive  stars} (\cite{boes05,boes05a,boes09})
\item Carbon depletion and carbon isotopic ratio decrease $\longrightarrow$ {\sf in low-mass red giant stars, in main sequence and post main sequence massive stars} (\cite{gratton00})
\item Nitrogen enrichment $\longrightarrow$ {\sf in low-mass red giant stars, in main sequence and post main sequence massive stars} (\cite{CEMP05,heap06,fab09})
\item Helium enrichment $\longrightarrow$ {\sf in massive O-type stars at low metallicity} (\cite{mokiem06})
\item s-process elements enrichment $\longrightarrow$ {\sf in intermediate-mass AGB stars}
\end{itemize}

\noindent The fact that abundance anomalies increase with decreasing metallicity may also be viewed as an evidence for rotation-driven transport inside the stars, as some of these mechanisms are expected to be more efficient as  stars become more compact and bluer at low metallicity.\\
Population counts, in particular for the subtypes of massive stars at different evolutionary phases (subtypes of supergiants, of Wolf-Rayet and O and B type stars)  can also be used as indirect probes of internal mixing (\cite{egg02,MM03,fab09}).\\

Figure~\ref{f-obsabund} presents a set of observed abundance patterns that are used as indirect probes for internal (rotational) mixing.\\

In \S~\ref{s-evolution} we will show how stellar evolution models of rotating stars confront to direct and indirect probes.


\section{Modelling the structural evolution of rotating stars}\label{s-struc}

It appears from observations that rotation may actually modify the
shape of stars, their lifetimes, their surface parameters and
abundances. These evidences constitute a strong incitement to actually
include rotation and associated physical processes in stellar
evolution modelling.\\
In this section, I present the different approaches,
prescriptions and techniques that are used in stellar evolution
codes to account for the direct effects of rotation on the stellar structure and evolution, leaving rotation-induced transport of angular momentum and nuclides for \S~\ref{s-transport}.

\subsection{Stellar Structure Equations : standard formulation}\label{s-stdeq}
A star is a continuous plasma that should be described by fully tri-dimensional
hydrodynamical equations :

\begin{eqnarray}
\textrm{Continuity equation} & & \frac{\partial \rho}{\partial t} = -
  \vec{\nabla} \cdot \left( \rho \vec{v}\right)\\
\textrm{Motion equation} & &\rho \frac{d \vec{v}}{dt} = -
  \vec{\nabla}P - \rho \vec{\nabla}\Phi\\
\textrm{Energy conservation equation} & & T \frac{d \rho s}{dt} = \rho
  \epsilon_{prod} - \vec{\nabla} \cdot \vec{F}\\
\textrm{Heat transfer equation} & & \vec{F} = - {\cal K} \vec{\nabla}
T\\
\textrm{Abundance conservation equation} & & \frac{\partial n_i}{\partial t} + \vec{\nabla} \cdot
  n_i \vec{v_{i}} = - \sum_j {\cal R}_{ij} n_i n_j + \sum_{k,l}
  {\cal R}_{kl}n_kn_l~~~ \forall i
\end{eqnarray}

where $\rho$ is the density; $\vec{v}$ is the macroscopic velocity; P is the total pressure (given by the equation of state); $\Phi$ is the gravitational potential derived from the Poisson
equation $\nabla^2 \Phi = 4 \pi G \rho$; $s$ is the specific entropy (defined by the second principle of
thermodynamics); T is the local temperature; $\epsilon_{prod}$ is the
energy production rate par unit mass and time; $\vec{F}$ is the heat
flux; $\cal{K}$ is the thermal conductivity tensor; $n_i$ is the
volumetric abundance of chemical species $i$; $\vec{v_i}$ is the microscopic velocity of particles of species
$i$; ${\cal R}_{kl}$ is the rate of destruction (if $k$ or $l$
are equal to $i$) or production of
species $i$ by unit volume and time for a particle of species $i$.

When studying a non-rotating, weakly or non-magnetic single star,
pressure and gravity are the only forces acting on a mass
element. The resulting spherically symmetric configuration allows to reduce the problem of
stellar evolution to the resolution of a set of equations that only
depend on time and one spatial coordinate.\\
Instead of the natural eulerian formulation that would derive from
expressing the above listed equations as a function of the radius $r$ only,
the Lagrangian formulation is normally preferred. The spatial
coordinate is now the mass $m$ associated to a radius $r$ and the resulting
set of stellar structure equations is the following:

\begin{eqnarray}
\frac{\partial r}{\partial m} & = & \frac{1}{4 \pi r^2 \rho} \label{eq-continuity}\\
\frac{\partial P}{\partial m} & = &   -\frac{1}{4 \pi r^2} \left( \frac{D
  v}{Dt} + \frac{\partial \Phi}{\partial m}\right)= -\frac{1}{4 \pi
  r^2} \left( \frac{D v}{Dt} + \frac{G m}{r^2}\right) \label{eq-motion}\\
\frac{\partial L}{\partial m} & = & \epsilon_n - \epsilon_\nu - C_p
\frac{\partial T}{\partial t} + \frac{\delta}{\rho} \frac{\partial
  P}{\partial t} =  \epsilon_n - \epsilon_\nu + \epsilon_g \label{eq-energy}\\
\frac{\partial T}{\partial m} & =& - \frac{G m T}{4\pi r^4 P} \left.\frac{d \ln
  T}{d \ln P}\right|_{medium} = - \frac{G m T}{4\pi r^4 P} \nabla  \label{eq-heat}\\
\frac{\partial X_i}{\partial t} & = & \frac{m_i}{\rho} \left(\sum_j
r_{ji} - \sum_{k} r_{ik} \right)~~ \forall i \label{eq-nucl}\\
L & = & 4 \pi r^2 \left(F_{conv} + F_{rad}\right) = \frac{16 \pi a c G
m T^4}{3 \kappa P} \nabla_{rad}
\end{eqnarray}

\noindent where  $\left.\frac{D v}{Dt}\right|_m$ is the Lagrangian derivative equal to
  $\frac{\partial v}{\partial t} + v \frac{\partial v}{\partial r}$; $\epsilon_n$, $\epsilon_\nu$ and $\epsilon_g$ are the nuclear,
  neutrinic, and gravitational energies resp.; $\delta = \left(\frac{d \ln \rho}{d \ln T} \right)_P$ entering
  the equation of state (\cite{KW90}); $X_i$ is the mass fraction of
  nuclide $i$; $m_i$ is the mass of nuclide $i$; $r_{ji}$ is the rate for the nuclear reaction between nuclide
  $j$ and $i$; $a = \frac{4 \sigma}{c} = 7.5657 \times 10^{-15}$ erg.cm$^{-3}$.K$^{-4}$ is the radiation constant and
  $\sigma = 5.6704 \times 10^{-8}$ erg.cm$^{-2}$.s$^{-1}$.K$^{-4}$ is the Stefan's constant.\\

These equations need to be complemented by (1) an equation of state
$\rho = \rho(P,T,\mu)$, (2)  an expression for the mean Rosseland opacity of the plasma,
$\kappa = \kappa(P,T,\mu)$ (in most stellar evolution codes, opacities
are tabulated), (3) a network of nuclear reactions rates $r_{ij}$,
(4) an equation for energy production/losses by neutrinos
$\epsilon_{\nu}$ and (5) a description of the
energy transport by convection to obtain $\vec{F}_{conv}$.\\
The convective flux, hence the temperature gradient $\nabla$, is generally derived from the
mixing length theory (\cite{BV58}). The formulation by Kippenhahn \& Weigert (1970) gives:
\begin{equation}
F_{conv} = \rho C_p T \sqrt{g \delta} \frac{l^2}{4 \sqrt{2}}
H_p^{-3/2} \left(\nabla - \nabla_e \right)^{1/2}
\end{equation}

\noindent where $H_p = - \frac{dr}{d \ln P}$ is the pressure scale height. $l =
\alpha_{\rm MLT} H_p$
is the so-called ``{\em mixing length}'' over which a fluid elements
dissolves completely into its surroundings, with $\alpha_{\rm MLT}$ a
parameter of order unity that is calibrated by having a 1 M$_\odot$
model reproducing the Sun's
radius, luminosity and temperature at the age of the Sun. $\nabla_e = \left.\frac{d
  \ln T}{d \ln P}\right|_e$ is the
temperature gradient of the fluid element.

\subsection{Stellar Structure Equations : the case of rotating stars}

The mechanical equilibrium of a star is modified when rotation is
taken into account, because the rotating star is not spherical
any more. This departure from spherical symmetry translates into 
\begin{itemize}
\item the deformation of the equipotential surfaces;
\item the reduction of the effective gravity away from the rotation
  axis by centrifugal forces;
\item the variation of the radiative flux on equipotential surfaces
  (Von Zeipel effect).
\end{itemize}

\noindent This list reveals immediately that rotation calls for a
multi-dimensional description of the stellar structure. However, this
is a real challenge for stellar evolution modelling, stellar evolution
being meant here as the {\em integration of all evolutionary phases
  from birth to final stages}. The difficulty arises from the huge
variety of spatial and temporal scales involved (about 10$^{11}$ and
10$^{17}$ orders of magnitude for spatial and temporal scales
respectively for solar-type star if microscopic processes and
convection are to be treated properly).\\Although great efforts are
devoted to move from the 1D picture to a 2D (see
  also Rieutord this volume, and \cite{REL13}) or 3D modelling
(\cite{Turcotte03,Bazan03}), the stellar evolution models presently in
use to interpret most of observations are the result of 1D modelling \newline

\begin{figure}[t]
\floatbox[{\capbeside\thisfloatsetup{capbesideposition={right,top},capbesidewidth=6cm}}]{figure}[\FBwidth]
{\caption{Distortion of a rotating star with the Roche model. Each
  color corresponds to the mentioned values of the ratio $\omega =
  \Omega/\Omega_{\rm crit}$ with $\Omega_{\rm crit} = \sqrt{\frac{8 G
      M}{27 R_P^3}}$ , M the stellar mass, $\Omega$ the angular
    velocity at the surface and $R_p$ the polar radius.{\em Courtesy
      A.T. Potter (\cite{potterPhD}).}}\label{f-roche}}{\includegraphics[scale=0.5,angle=-90]{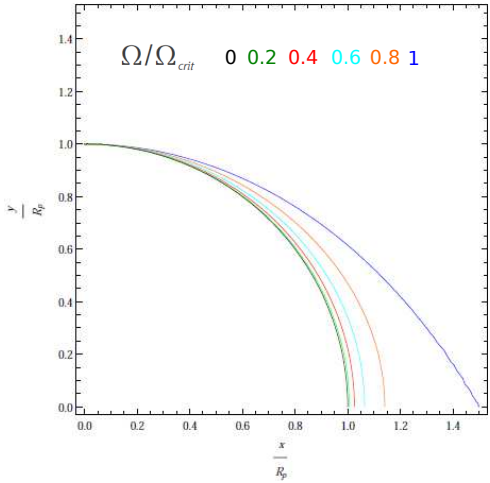}}
\end{figure}

\subsubsection{The Roche model for distorted rotating stars}

Two families of historical approaches exist to evaluate the shape of a
rotating fluid : The {\em MacLaurin's spheroids}, which assume that the
rotating fluid, e.g. the star, is a body of constant density and
uniform angular velocity, and the {\em Roche equipotentials} which
assume that the gravitational potential $\Phi$ is the same as if the
total mass of the star were concentrated at the centre.\\
In each of these families, the total potential writes as the sum of
the gravitational potential and of the potential from which the
centrifugal acceleration derives when the star is {\em
  barotropic}.

Although none of these visions are correct, the Roche model is much
more appropriated to model a star and is very commonly used.\\

For a barotropic star, the centrifugal acceleration derives from a
potential $V$
\begin{equation}
V = - \frac{1}{2} s^2 \Omega^2
\end{equation}
with $s$ the distance from the rotation axis, and $\Omega$ the angular
velocity.\\
On the other hand, the gravitational potential is defined by 
\begin{equation*}
-\vec{\nabla}\Phi = - \frac{Gm}{r^2} \frac{\vec{r}}{r}, 
\end{equation*}
and is spherically symmetric in the Roche approximation so that
\begin{equation}
\Phi = -\frac{Gm}{r}.
\end{equation}
Here $G$ is the gravitational constant and $m$ is the mass inside the
sphere of radius $r$.

In the Roche approximation, the total potential thus writes 
\begin{equation}
\Psi(r,\theta) = \Phi + V = - \frac{Gm}{r} - \frac{1}{2}  \Omega^2 r^2
\sin^2 \theta .
\label{eq-rochepot}
\end{equation}

The stellar surface is an equipotential, $\Psi =$ const. 
As the centrifugal force vanishes at the poles, the following
equation can thus be used to define the stellar surface :

\begin{equation}
\frac{GM}{R} + \frac{1}{2}  \Omega^2 R^2
\sin^2 \theta = const = \frac{GM}{R_p},
\end{equation}

where $M$ and $R$ are the total mass and radius of the star, and $R_p$
is the polar radius. Its critical value, obtained where the gravity
exactly compensates the centrifugal acceleration in the equatorial
plane is :
\begin{equation}
\Omega_{P,\rm crit} = \sqrt{\frac{8 G M}{27 R^3_{P,\rm crit}}}
\end{equation}

Figure~\ref{f-roche} represents the stellar surface in reduced radius
units for different rotation rates as indicated on the graph. We can
see that up to $\omega \approx 0.6$, the equatorial radius increase is
lower than 10\%, and only for the very rapid rotators should we expect
a strong distortion. This remark is important as it justifies a
perturbative approach to model the evolution of most rotating stars.

\subsubsection {Kippenhahn \& Thomas (1970): Case of conservative rotation law}\label{sub-KT70}
Noticing that pressure $P$ and density $\rho$ are constant on
equipotentials if the potential is conservative, Kippenhahn \& Thomas (1970) propose
to re-write the stellar structure equations on these equipotentials,
that can be distorted and thus account for the spherical symmetry
departure induced by rotation.\\
So let's $\Psi$ be a conservative potential, and equipotentials be
defined as surfaces where $\Psi =$ const. For the potential to be
conservative, it means that the centrifugal acceleration derives from
a potential.\\ In Kippenhahn \& Thomas (1970), $\Psi$ is the Roche potential from Eq.~(\ref{eq-rochepot}).

\noindent In Endal \& Sofia (1976), the potential is a Kopal-type potential (\cite{Kopal59}), divided in three
parts, 
\begin{equation}
\Psi = \Psi_s + \Psi_r + \Psi_d
\end{equation}
 with $\Psi_s$, the spherically symmetric part of the gravitational
potential, $\Psi_r$, the cylindrically symmetric potential
directly due to rotation and $\Psi_d$ the cylindrically symmetric
part of the gravitational potential due to the distortion of the star's
shape :
\begin{eqnarray}
\Psi_{s} &=& - \frac{G m}{r}\\
\Psi_r &=& \frac{1}{2} r^2 \sin^2 \theta \Omega^2\\
\Psi_d& =& \sum_{j=2}^\infty  \frac{4 \pi
  G}{(2j +1) r^{j+1}} \int_0^{r_0} \rho \frac{\partial}{\partial
  r_0'}(r_0'^{j+3}Y_j)dr_0'
\end{eqnarray}

\noindent where $r_0$ is the radius of the equipotential with angle $\theta_0$
so that $P_2(\cos \theta_0) = 0$, with $P_2$ the second order Legendre
polynomial and $Y_j$ is the axisymmetric tesseral harmonic relating
$r$ to $r_0$ on an equipotential (see Appendix A in \cite{ES76} for further details).

For a conservative potential, we can define
a mean radius for the equipotential, $r_\Psi$ as the radius of the
sphere containing the same volume $V_\psi$ as the equipotential of
surface $S_\Psi$
\begin{equation}
V_\Psi = \frac{4}{3} \pi r_\psi^3
\label{eq-vpsi}
\end{equation}

\noindent This expression preserves the form of the continuity equation,
where $r$ and $m$ are replaced by $r_\Psi$ and $m_\Psi$.\\

\noindent The mean of any physical quantity $u$ that is not constant on the equipotential
surfaces can be written as 
\begin{equation}
\langle u \rangle = \frac{1}{S_\Psi} \int_{\Psi = cst} u dS_\Psi.
\end{equation}

\noindent The gravity derives from the total potential, that is 
\begin{equation}
\vec{g} = - \vec{\nabla} \Psi \longrightarrow g = \frac{d \Psi}{d n}
\end{equation} 
with $dn$ the distance between equipotentials $\Psi$ = cst and $\Psi +
d\Psi$ = cst. It is not constant over the equipotential surfaces,
yielding :

\begin{eqnarray}
\langle g \rangle =  \frac{1}{S_\Psi} \int_{\Psi = cst} \frac{d
  \Psi}{dn} dS_\Psi & \textrm{and} & \langle g^{-1} \rangle =  \frac{1}{S_\Psi} \int_{\Psi = cst} \left(\frac{d  \Psi}{dn}\right)^{-1} dS_\Psi.
\end{eqnarray}

\noindent We can also define a volume element $d V_\Psi = d\Psi \int_{\Psi =
  cst}  \left( \frac{dn}{d\Psi}\right) dS_\Psi$, and thus relate the
potential derivative to the density and the mean of the inverse
gravity:

\begin{equation}
d\Psi = \frac{1}{S_\Psi \langle g^{-1} \rangle} \frac{dm_\Psi}{\rho}.
\label{eq-dpsi}
\end{equation}

\smallskip
\noindent The modification of the stellar structure equations is then
straightforward from Eqs.~(\ref{eq-vpsi}) and (\ref{eq-dpsi}). As an
example, the hydrostatic equilibrium (Eq.~\ref{eq-motion} with $v =
0$) becomes :
\begin{eqnarray}
\vec{\nabla}P &=& - \rho \vec{\nabla}\Psi \nonumber\\
\frac{\partial P}{\partial m_\Psi} & = & -\rho \frac{\partial
  \Psi}{\partial m_\Psi} = - \rho \frac{G m_\Psi}{r_\Psi^2}\frac{1}{4 \pi
  r_\Psi^2 \rho} \nonumber\\
 \frac{\partial P}{\partial m_\Psi}& = & - \frac{G m_\Psi}{4 \pi r_\psi^4} f_P
\end{eqnarray} 
with
\begin{equation} 
\boxed{f_P = \frac{4 \pi r_\Psi^4}{G m_\Psi S_\Psi} \frac{1}{\langle g^{-1} \rangle}}
\end{equation}

The complete new set of stellar structure equations \footnote{We
  omit here the abundance conservation equation that can be written as in Eq.~(\ref{eq-nucl}).} finally writes:
\begin{eqnarray}
\frac{\partial r_\Psi}{\partial m_\Psi} &=& \frac{1}{4 \pi r_\Psi^2
  \rho} \label{eq-controt}\\
\frac{\partial P}{\partial m_\Psi}& = & - \frac{G m_\Psi}{4 \pi r_\psi^4} f_P \\
\frac{\partial L_\Psi}{\partial m_\Psi} & = & \epsilon_n -\epsilon_\nu + \epsilon_g \\
\frac{\partial T}{\partial m_\Psi} & = & - \frac{G m_\Psi T}{ 4 \pi
  r_\Psi^{4} P} \nabla_\Psi \label{eq-heatrot}\\
L_\Psi & = & - \frac{4 a c T^3}{3 \kappa} \langle g^{-1} \rangle
S_\Psi^2 \langle g \rangle \frac{\partial T}{\partial m_\Psi}
 \end{eqnarray}

with 
\begin{equation}
\nabla_\Psi = - \frac{3 \kappa}{ 16 \pi a c G} \frac{P}{T^4}
\frac{L_\Psi}{m_\Psi} \frac{f_T}{f_P}
\end{equation}
and 
\begin{equation}
\boxed {f_T = \left( \frac{4 \pi r_\Psi^2}{S_\Psi}\right)^2 \frac{1}{\langle g\rangle \langle g^{-1} \rangle}.}\\
\end{equation}
\smallskip

This set of equations is very similar to the standard set of
stellar structure equations, with the addition of {\em
  deformation} factors $f_P$ and $f_T$ to the motion and heat transfer
(radiative equilibrium through the gradient $\nabla_\Psi$) equations.\\

\begin{figure}[t]
\includegraphics[width=6cm]{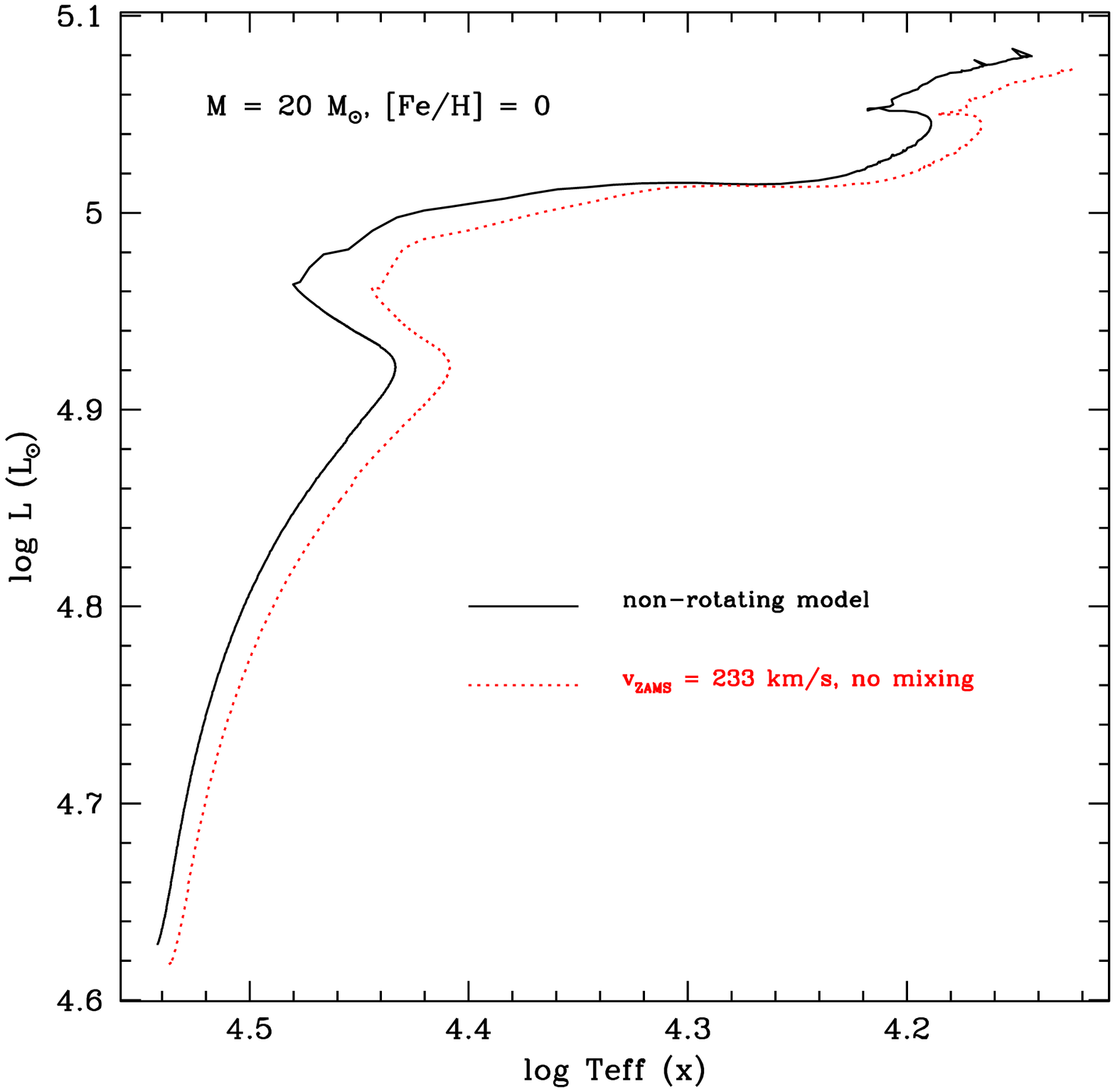}
\includegraphics[width=6cm]{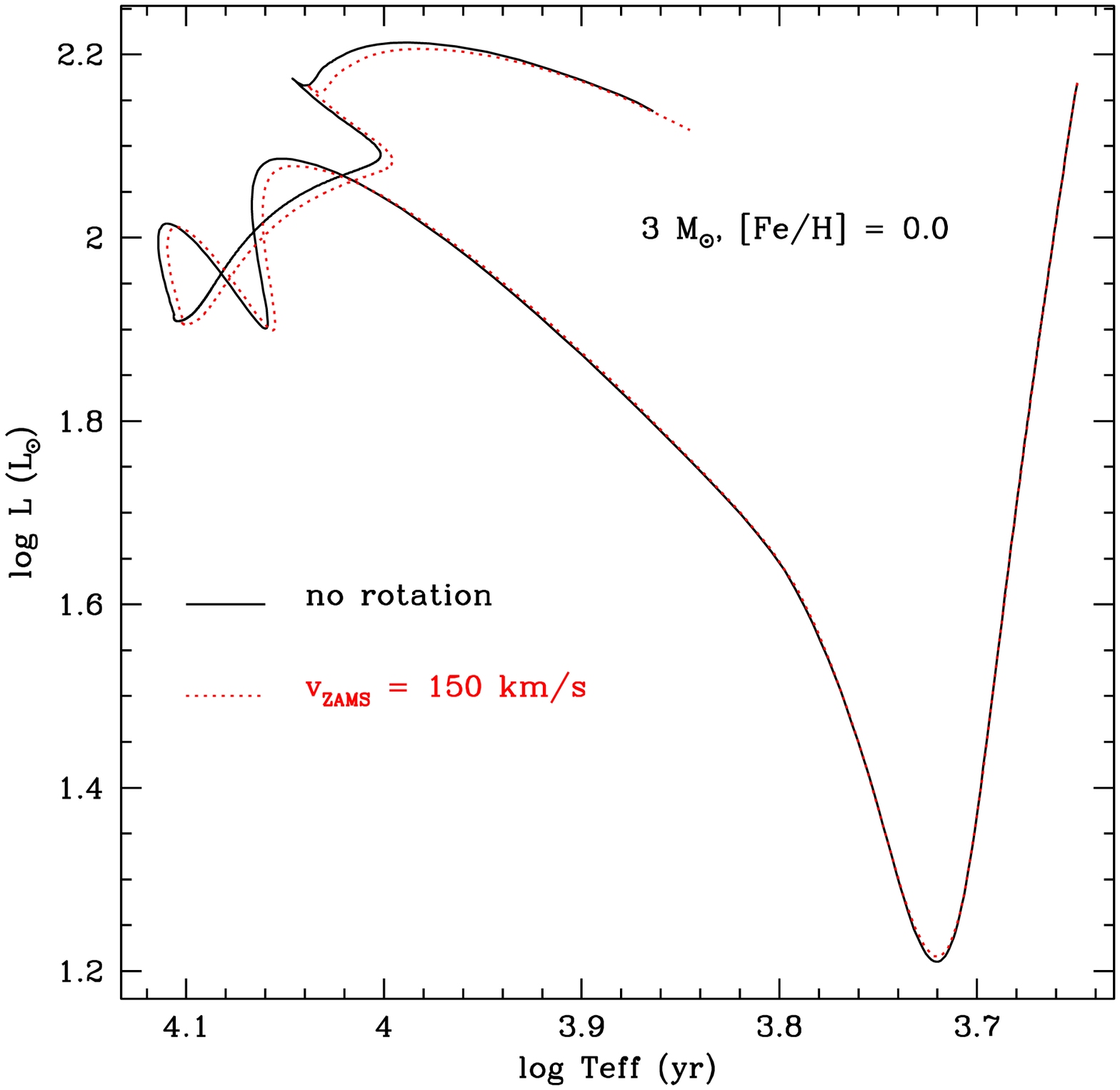}
\qquad
\floatbox[{\capbeside\thisfloatsetup{capbesideposition={right,top},capbesidewidth=6cm}}]{figure}[\FBwidth]
{\caption{Effect of the centrifugal acceleration on the main sequence evolutionary
  track of a 20 M$_\odot$ model at solar metallicity ({\em top left}),
  on the evolutionary track of a 3 M$_\odot$ model at solar metallicity,
  and on the pre-main sequence evolutionary track of a 1 M$_\odot$
  model at solar metallicity ({\em bottom left}). Black solid lines are for the
  non-rotating models and red dotted lines for the models including
  the modifications of the stellar structure equations and radiative
  equilibrium according to Endal \& Sofia (1976). No rotational mixing is
  included. Rotating models are assumed to experience solid-body
  rotation.}\label{f-evorotES}}{\includegraphics[width=6cm]{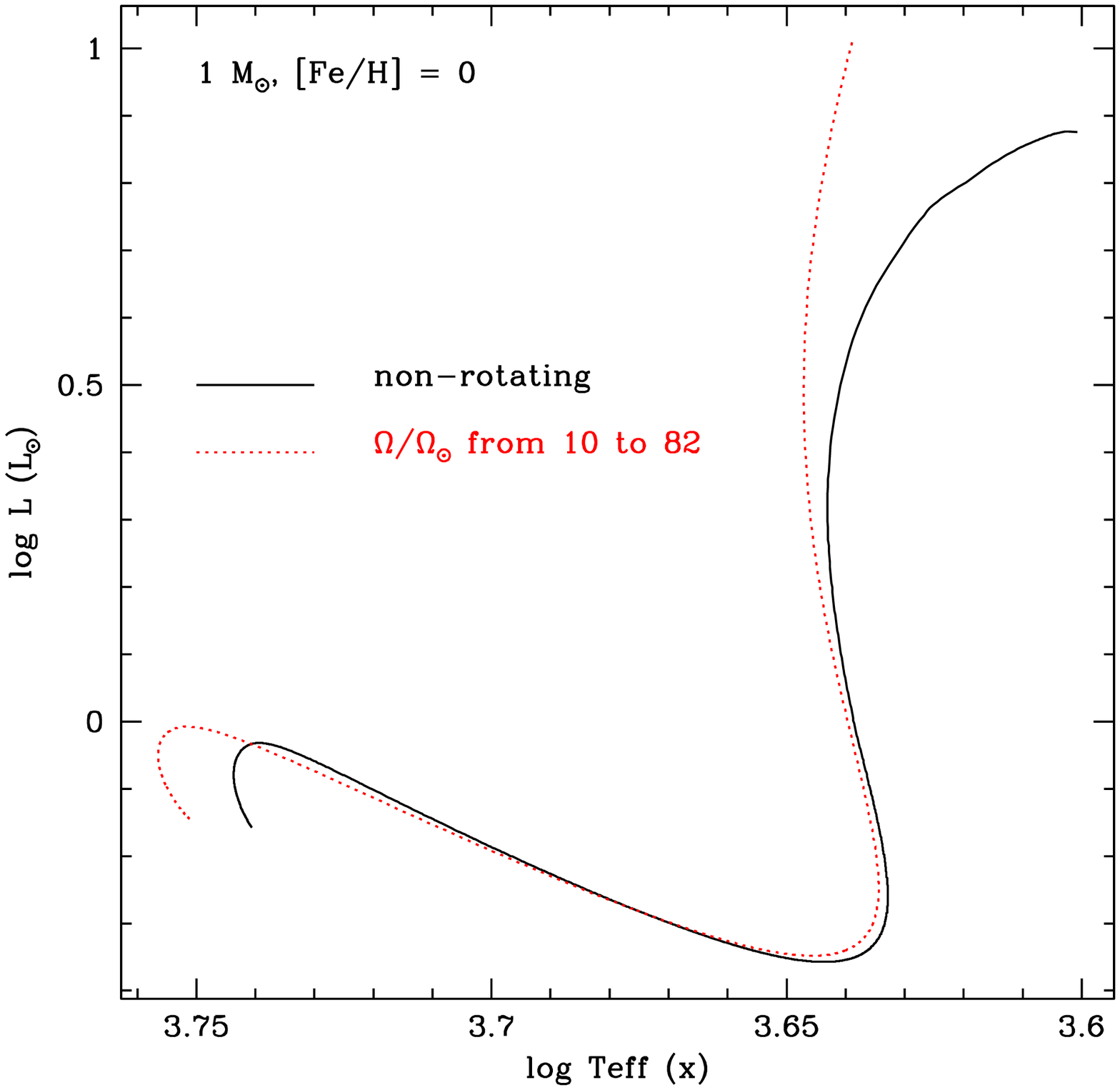}}
\end{figure}

\noindent Figure~\ref{f-evorotES} shows the impact of the centrifugal
acceleration on the evolutionary track of a main sequence massive (20
M$_\odot$) star, of an intermediate-mass star (3 M$_\odot$) from the pre-main sequence (hereafter PMS) to the subgiant phase, and of a PMS low-mass
star (1 M$_\odot$).\\ 

Centrifugal acceleration lowers
effective gravity, which in turn translates into a decrease of both
the luminosity and the effective temperature. Including the
centrifugal force in the total potential of the star makes it behave
as if it were a non rotating, slightly less massive star. This effect
is the same for the different masses and evolutionary stages presented on the
figure (see also \cite{pins89,mc96,mendes99}, \newline \cite{egg10,egg12}). As we will
show later, the hydrostatic effects constitute the main effect of
rotation on the evolutionary tracks of low- and intermediate-mass stars
beyond the turn-off.

\subsubsection{Meynet \& Maeder (1997): Case of non-conservative rotation law}

The formalism developed by Kippenhahn \& Thomas (1970) is only valid for conservative
rotation and needs to be revised for non-conservative rotation laws.
Among those, the {\em shellular rotation} introduced by
\cite{Zahn92} (see also \cite{CZ92}) provides the first
self-consistent formalism for the transport of angular momentum by
meridional circulation and shear turbulence (see \S~\ref{s-rotmod}).\\ Shellular rotation is
expected to occur in radiative stellar interiors, where turbulence develops (because the
plasma is highly inviscid) and where it is supposed to be highly
anisotropic.  Horizontal turbulent motions are presumably more
vigorous than vertical ones, and provided this anisotropy is large
($\nu_v \ll \nu_h$), the
angular velocity profile depends very weakly on colatitude , so that
$\Omega \equiv \Omega(r)$. \\
Such  rotation law is non-conservative and the centrifugal acceleration does not derive from a potential
any more. Meynet \& Maeder (1997) present an extension of Kippenhahn \& Thomas (1970)
formalism to integrate this fact.\\ 

The main difference between conservative and non-conservative rotation laws is the following :
\begin{itemize}
\item in the case of solid-body rotation , $\Omega = cst$, which is a
  conservative law, the isobars coincide with the equipotential
  surfaces, and the star is {\bf barotropic}. The equations of the
  stellar structure can then be projected on equipotentials;
\item in the case of shellular rotation, $\Omega = \Omega(r)$, which
  is a non-conservative law, the isobars do not coincide with the
  equipotentials nor with isotherms (they are inclined with respect to
  the horizontality defined by the equipotentials), and the star is
  {\bf baroclinic}. The equations of the stellar structure are then
  better projected on the {\em isobars}. On these surfaces $\Omega$ is
  constant.
\end{itemize}

The approach of Meynet \& Maeder (1997) is to use a formalism similar to that of Kippenhahn \& Thomas (1970), replacing the
equipotentials by the isobars. These surface  have the same shape as the equipotential in the conservative case. They are defined as surfaces where $\Psi_P$ is constant with
\begin{equation}
\Psi_P = -\Phi + \frac{1}{2}\Omega^2 r^2 \sin^2\theta
\end{equation}
and $\Phi$ the gravitational potential in its Roche form, $r$ the radius and $\theta$
the colatitude.\\
\\The subscripts $\Psi$ in
Eqs.~(\ref{eq-controt}) to (\ref{eq-heatrot}) are thus replaced by a
subscript $P$, and provided some reasonable assumptions\footnote{see
  Equations (A.23) to (A.26) in Meynet \& Maeder (1997)},
they finally obtain the following set of equations:

\begin{eqnarray}
\frac{\partial r_P}{\partial m_P} &=& \frac{1}{4 \pi r_P^2
  \bar{\rho}} \label{eq-controt2}\\
\frac{\partial P}{\partial m_P}& = & - \frac{G m_P}{4 \pi r_P^4} f_P \\
\frac{\partial L_P}{\partial m_P} & = & \epsilon_n -\epsilon_\nu + \epsilon_g \\
\frac{\partial T}{\partial m_P} & = & - \frac{G m_P T}{ 4 \pi
  r_P^{4} P} \nabla_P \label{eq-heatrot2}\\
L_P & = & - \frac{4 a c}{3} \left\langle g_{\rm eff}^{-1} \right\rangle
S_P^2 \left\langle \frac{T^3 g}{\kappa} \right\rangle \frac{\partial T}{\partial m_p}
 \end{eqnarray}
with 
\begin{equation}
\nabla_P = - \frac{3 \kappa}{ 16 \pi a c G} \frac{P}{T^4}
\frac{L_P}{m_P} \frac{f_T}{f_P}
\end{equation}
and 
\begin{equation}
\boxed {f_p = \frac{4 \pi r_P^4}{G m_P S_P} \frac{1}{\langle g^{-1} \rangle}.}
\end{equation}
\begin{equation}
\boxed {f_T = \left( \frac{4 \pi r_P^2}{S_P}\right)^2 \frac{1}{\langle
    g \rangle \langle g^{-1} \rangle}.}
\end{equation}\\

\noindent The effective gravity, noted $g$ for sake of simplicity, also has a different expression. It depends on the radius $r$ and on the colatitude
$\theta$ :
\[
 \vec{g} \equiv \vec{g}_{\rm eff}  = \vec{\nabla}\Psi_P - r^2 \sin^2 \theta \Omega \vec{\nabla}\Omega.
\]
 It cannot be expressed as the derivative of $\Psi_P$, since
this quantity is not a potential in the baroclinic case. From the
hydrostatic equilibrium equation 
\[
\vec{P} = - \rho \left( \vec{\nabla}\Psi_P -  r^2 \sin^2 \theta \Omega
\vec{\nabla}\Omega \right)
\] 
and noticing that $\vec{\nabla}\Omega$ is parallel to
$\vec{\nabla}\Psi_P$, an expression for $g_{\rm eff}$ can be given by
\begin{equation}
g \equiv g_{\rm eff} = \left( 1 - r^2 \sin^2 \theta \Omega \alpha
\right) \frac{d\Psi_P}{dn}
\label{eq-geff}
\end{equation}
where $\alpha = \frac{d\Omega}{d\Psi_P}$.\\

\noindent The set of equations of \S~\ref{sub-KT70} is actually
recovered if one assumes that all quantities do not depend on the local density
and temperature, but on their volume averaged means between two
isobars, $\bar{\rho}$ and $\bar{T}$, with 
\begin{equation}
\bar{\rho} = \frac{\rho \left( 1 - r^2 \sin^2 \theta \Omega \alpha
  \right)\left \langle g^{-1}\right \rangle}{\left \langle
  g^{-1} \right \rangle - \left \langle g^{-1} r^2
  \sin^2 \theta \right \rangle \Omega \alpha}.
\end{equation}

Zeng (2002) proposed a complete generalization of this formalism to the case of
unspecified rotation law. The projection of the general set of
equations for stellar structure is done on isobars again, but tunable
parameters (4) are added which somehow weakens the approach.\\

\subsection{Radiative transfer : the case of rotating stars}\label{s-mdot}

\subsubsection{Gravity darkening}

The radiative flux of a star writes as 
\begin{equation}
\vec{F} = - \frac{4 a c}{3 \kappa \rho} T^3 \vec{\nabla}T = \frac{4 a
  c}{3 \kappa \rho} T^3 \frac{dT}{d\Psi} \vec{\nabla}\Psi.
\end{equation}
When the star is rotating, the total potential $\Psi$ depends on the
angular velocity $\Omega$ and on the colatitude $\theta$. Its gradient
is not constant on an equipotential because the spacing between
equipotentials decreases when moving from the equator to the polar
regions. It follows that
$\vec{\nabla} \cdot \vec{F} = f((\nabla \Psi)^2) = 0$ cannot be verified at all colatitudes
and the radiative equilibrium is broken.\\

The Von Zeipel theorem, established in 1924 (\cite{vz24}) gives the
relation between the radiative flux and the local effective
gravity. For solid-body rotation, the hydrostatic equilibrium leads 
\begin{equation}
\vec{F} = \- A \vec{\nabla} T = A \frac{d T}{d P} \vec{\nabla}P
\underset{\rm hydro. equ.}{=} - \rho A \frac{d T}{d P} \vec{g}
\label{eq-radflux}
\end{equation}

\noindent where $A = \frac{4 a c t^3}{3 \kappa \rho}$ and $T, F, P$ and $g_{\rm
    eff}$ depend on $\Omega$ and on the colatitude $\theta$.\\
In parallel the total luminosity on an equipotential surface $S_P$ is
by definition the integral of the flux on that surface :
\begin{equation}
L = \iint_{S_\Psi} \vec{F} \cdot \vec{n} dS_\Psi
\underset{solid. rot.}{=}  \rho A \frac{d T}{d P} \iint_{S_\Psi}
\vec{\nabla}\Psi \cdot \vec{n} dS_\Psi
\end{equation}

Using the Poisson equation $\Delta \Psi = 4 \pi G \rho - 2 \Omega^2$ ,
we can express the luminosity as a function of the density and the
angular velocity:
\begin{equation}
L = \rho A \frac{d T}{d P} \iiint_{V_\Psi} \left( 4 \pi G \rho - 2
\Omega^2\right) dV_\Psi
\end{equation}

Putting back the result of the integral in Eq.~(\ref{eq-radflux}), one
finally finds

\begin{equation}\label{eq-fvz}
\vec{F} = \frac{L}{4 \pi G M_\star} \vec{g}
\end{equation}

\noindent where $M_\star = M \left(1 - \frac{\Omega^2}{2 \pi G
  \rho_m}\right)$.\\

The so-called {\em gravity-darkening effect} directly results from
this relation that implies a new relation between effective temperature and gravity

\begin{figure}[t]
\includegraphics[width=0.65\textwidth,angle=-90]{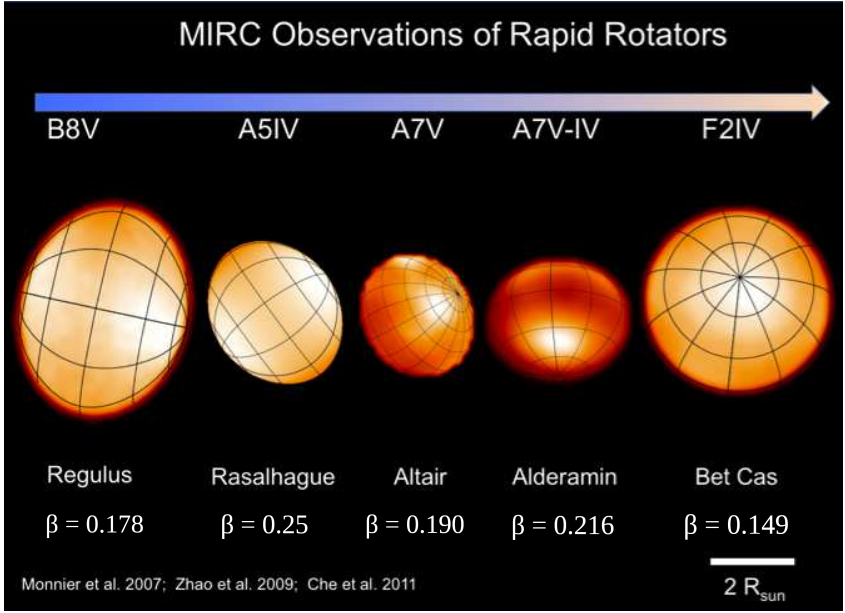}
\caption{CHARA/MIRC imaging of 6 rapid rotators, spanning a wide
  spectral range from B8 to F2 spectral types. The stars are scaled relatively to
  their linear sizes. The stars are oblate due to their rapid
  rotation. The polar areas of these stars are bright and their
  equatorial areas are dark because of the gravity darkening
  effect. The $\beta$ exponent for the best-fit models from Zhao et al. (2011) is also
  indicated. {\em Adapted from M. Zhao, public comm.}}\label{f-gravdark}

\end{figure}

\begin{equation}
T_{\rm eff} = \frac{L}{4 \pi G M_\star} g^{1/4}.
\label{eq-vzTeff}
\end{equation} 

As the effective gravity $g \equiv g_{\rm eff}$ is lower at the equator
due to the centrifugal force, the poles of a rotating star should be
hotter than the equatorial regions. This effect is called {\em gravity darkening}.\\

This derivation assumes solid-body rotation, and thus corresponds to a
{\em barotropic star}. For {\em a baroclinic star}, as is the case
when $\Omega = \Omega(r)$, the expression is slightly modified as
demonstrated by Maeder (1999); the radiative flux then depends on
the effective gravity and on the gradient of the angular
velocity $d \Omega / dr$.\\

The Von Zeipel theorem concerns the radiative flux, so that the
relation $T_{\rm eff} \propto g_{\rm eff}^\beta$, with $\beta = 0.25$ should be only valid
for stars with radiative envelopes (stars of type A and earlier). A
similar relation was established by Lucy (1967) for stars with a
convective envelope and yields 
\begin{equation}
T_{\rm eff} \propto g_{\rm eff}^{0.08}.
\label{eq-lucyTeff}
\end{equation}

\noindent The gravity-darkening has actually been directly observed by
NIR interferometric imaging as shown in Fig.~\ref{f-gravdark} (see also
\cite{ZMC11} and references
therein for a recent review on the
interferometric studies of rapid rotators). However, Eq.~(\ref{eq-vzTeff}) appears to give a bad fit to the data and overestimates the
difference between the polar and equatorial temperature. Based on the
presumed variation of the $\beta$ exponent from 0.25 for radiative
envelopes to 0.08 for convective envelopes, a general expression in
the form $$T_{\rm eff} \propto g_{\rm eff}^\beta$$ is generally used,
with $\beta$ an adjustable parameter varying between 0.1 and 0.25. \\

\noindent Very recently, Espinosa Lara \& Rieutord (2011) have proposed another expression for the
gravity-darkening effect which is more suited for fast rotators stars targeted by interferometric imaging\footnote{Von Zeipel's
  and Lucy's expressions are derived implicitly assuming a small
  deviation from the spherical symmetry, which should not be the case for very
rapid rotators.} :

\begin{equation}
\vec{F} = - \frac{L}{4 \pi G m} F_\omega(r,\theta) \vec{g}_{\rm eff},
\end{equation}  

\noindent where r and $\theta$ are the radial and latitudinal coordinates at the
stellar surface in spherical geometry, and $F_\omega(r,\theta)$ is a
non-dimensional function that is determined within the framework of
the Roche model for the mass distribution. $\omega = \sqrt{\Omega^2
  R_e^3/GM}$ is the only parameter controlling the latitudinal
variation of the effective temperature, and it is defined as the ratio
of the angular velocity to the keplerian velocity at the equator $\Omega_k = \sqrt{g_e/R_e}$.

\subsubsection{Mass loss of massive and luminous rapid rotators}

Let us summarize here the work by Maeder \& Meynet (2000a) concerning the effect of
rotation on radiative equilibrium of luminous massive stars and on mass loss.\\

In massive early-type stars, the radiative flux is important and needs
to be taken into account to define the total gravity, which is written:
\begin{equation}
\vec{g}_{\rm tot} = \vec{g}_{\rm eff} + \vec{g}_{\rm rad}
\end{equation}
where $g_{\rm eff}$ is the sum of the gravitational and centrifugal
accelerations and $g_{\rm rad}$ is the radiative acceleration. Both
the centrifugal and the radiative accelerations counteract the
gravity, and may yield to the break-up limit, when $\vec{g}_{\rm tot} =
\vec{0}$.\\ 
One can define the Eddington limit or $\Gamma$ limit as the
break-up limit when rotation is negligible, e.g. when $\vec{g}_{\rm  grav} + \vec{g}_{\rm rad} = \vec{0}$.\\
If rotation (e.g. centrifugal acceleration) is important, this
$\Gamma$ limit becomes what Maeder \& Meynet (2000a) call the $\Omega \Gamma$
limit defined by $\vec{g}_{\rm tot} = \vec{0}$.\\

As mentioned before, for a {\em baroclinic} rotating star ($\Omega =
\Omega(r)$), the von Zeipel theorem is slightly modified to account
for baroclinicity:

\begin{eqnarray}
\vec{F} & =& - \frac{L(P)}{4 \pi G M_\star} \vec{g}_{\rm eff}
  \underbrace{\left( 1 +
  \zeta(\theta) \right)}_{\textrm baroclinicity}\label{eq-vzmod}\\
\textrm{where}~~~ \zeta(\theta) & = & \left[\left(1 -
  \frac{\chi_T}{\delta}\right) \Theta + \frac{H_T}{\delta} \frac{d
    \Theta}{dr}\right] P_2(\cos \theta). \nonumber
\end{eqnarray}

$M_\star$ is the same as in Eq.~(\ref{eq-vzTeff}), $\delta$ is defined
by the equation of state (see \S~\ref{s-stdeq}), $H_T$ is the
temperature scale height, $\chi_T = \frac{\partial}{\partial T}\left(
\frac{4 a c T^3}{3 \kappa \rho}\right)$, $\theta$ is the colatitude,
$P_2(\cos \theta)$ is the Legendre polynomial of degree 2, and 
  $\Theta = \frac{\tilde{\rho}}{\rho}$ represents the density
    fluctuation on an isobar and is proportional to the angular
    velocity gradient $\frac{\partial \Omega}{\partial r}$ (see
    \S~\ref{s-rotmod}).\\

Combining Eqs.~(\ref{eq-vzmod}) and ~(\ref{eq-fvz}), we obtain the
following expression for the total gravity :

\begin{equation}\label{eq-loceddfac}
\vec{g}_{\rm tot} = \vec{g}_{\rm eff} \left[ 1 - \frac{\kappa(\theta) L(P)
    \left(1+\zeta(\theta)\right)}{4 \pi c G M \left(1 -
    \frac{\Omega^2}{2 \pi G \rho_m}\right)} \right] = \vec{g}_{\rm
  eff} \left( 1 - \Gamma_\Omega(\theta)\right).
\end{equation}

$\Gamma_\Omega(\theta)$ is called the local Eddington factor, defined as the
ratio of the actual flux to the limiting local flux, and is used to
define the $\Omega \Gamma$ limit.\\
From Eq.~(\ref{eq-loceddfac}), we see that $\Gamma_\Omega(\theta)$
reaches a maximum when the opacity is maximum, which occurs at the equator
where the temperature is lower. The $\Omega \Gamma$ limit is reached
where $\Gamma_\Omega(\theta) = 1$ and when $\vec{g}_{\rm tot} =
\vec{0}$. This limit defines the break-up limit for luminous
fast-rotating massive stars and should be the one considered when
studying early-type stars (OB), Luminous Blue Variable stars
(hereafter LBV), Red SuperGiants (hereafter RSG) and Wolf-Rayet stars
(hereafter WR stars).\\

The radiative wind theory applied in the context described in this
section leads the following expression for the total mass loss:

\begin{equation}
\frac{\dot{M}(\Omega)}{\dot{M}(0)} = \frac{(1 -
  \Gamma)^{\frac{1}{\alpha} - 1}}{\left[1 - \frac{\Omega^2}{2 \pi G \rho_m}\right]^{\frac{1}{\alpha}
    -\frac{7}{8}} \left[1 - \Gamma_\Omega\right]^{\frac{1}{\alpha} - 1}}
    \label{eq-Omlos}
\end{equation} 

where $\Gamma_\Omega = \frac{\Gamma}{1 - \frac{\Omega^2}{2 \pi G
    \rho_m}}$ with $\Gamma$ the Eddington ratio corresponding to
electron scattering opacity in a non-rotating star, and $\alpha$ is a
force multiplier parameter that can be determined empirically (see
\cite{mbook} for full details).\\

Using empirical force multipliers from the literature ($\alpha < 1$),
Eq.~(\ref{eq-Omlos}) leads to very large mass loss for massive stars near
classical break-up ($\Gamma \gtrsim 0.7$) , in the domain of LBV stars,
that should lead to the disruption of the star.\\ For moderate rotation,
the mass loss remains moderate.\\

\subsection{Mass loss of cool low-mass rotating stars}

\begin{figure}[t]
\includegraphics[width=6cm,angle=-90]{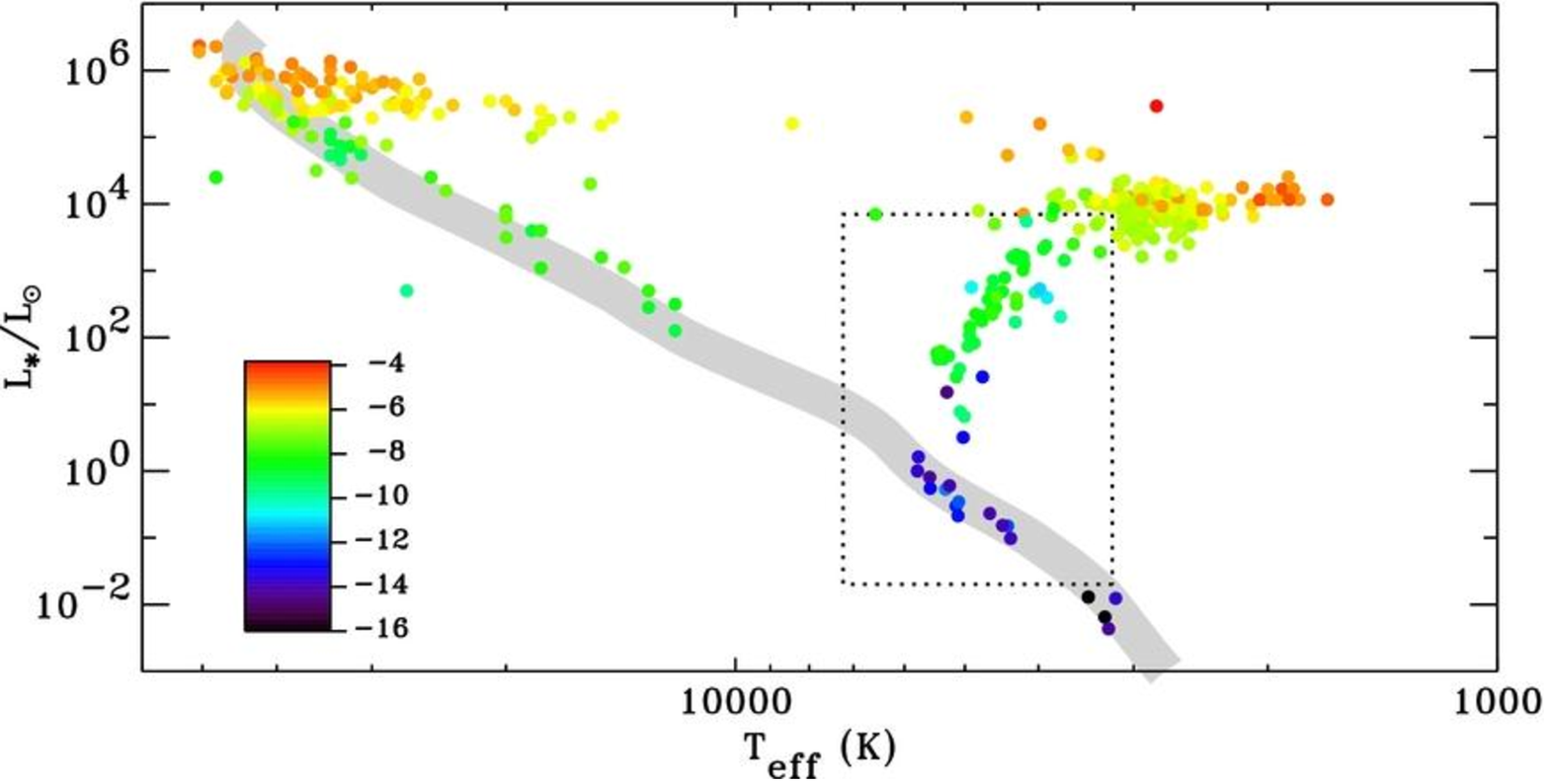}
\qquad
\qquad
\floatbox[{\capbeside\thisfloatsetup{capbesideposition={right,top},capbesidewidth=6cm}}]{figure}[\FBwidth]
{\caption{{\sf Upper panel} Definition domain for the \cite{CS11} mass-loss prescription
    in the HR diagram
  {\sf Lower left panel}. Theoretical predictions of the mass loss rate of
  main sequence cool stars as a function of temperature for
  different rotation periods. The
  Sun is indicated by the open circle. The mass loss can be strongly enhanced in the temperature range of G to F type stars in case of rapid rotation. {\em
  Reproduced from \cite{CS11}}}\label{f-CS}}{\includegraphics[width=6cm,angle=-90]{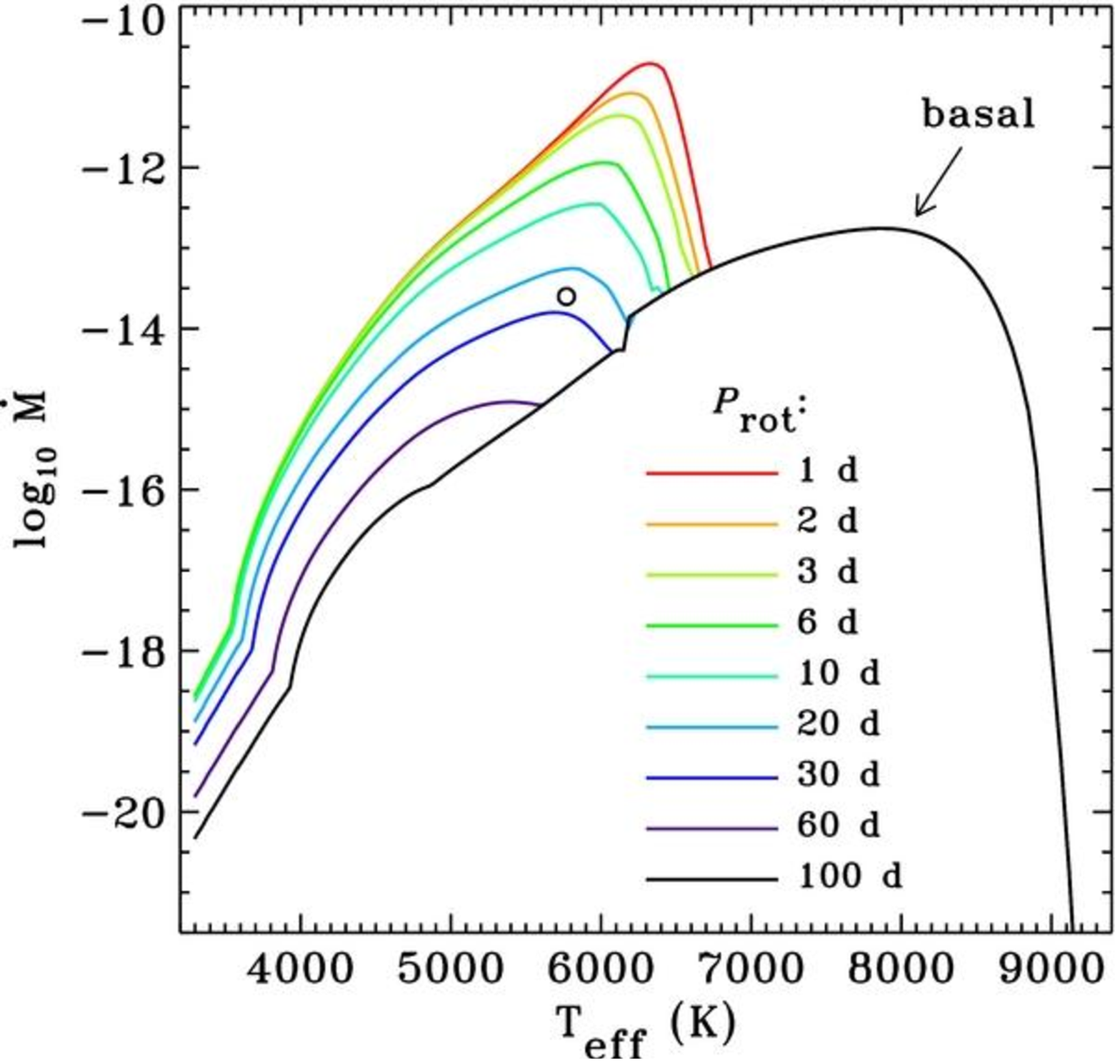}}
\end{figure}

Mass loss is an important process that can even drive the evolution
of stars at certain phases of their lives. For massive stars, mass loss
can be very strong (this is the case during the WR and LBV phases for
instance) and is driven by radiation processes. We have seen that it
can be modified, actually amplified by rotation and will thus probably
strongly affect the evolution of these stars (see
\S~\ref{s-evolution}).\\

Moving now towards cooler lower mass stars, mass loss can be driven by
(1) radiation pressure (with mass loss rates much lower than those
reached by OB stars due to the lower temperature and luminosity), (2)
acoustic waves (so-called {\em $\kappa$-mechanism}) for red giant
stars and (3) vibrational instability in pulsating stars. For low-mass
main sequence stars, the semi-empirical mass-loss prescription by
Reimers (1975) is usually used, and it does not include any rotational
effect :

\begin{equation}
\dot{M} = -3.98 \times 10^{-3} \eta \frac{L}{L_\odot}
\frac{R}{R_\odot} \frac{M_\odot}{M} ~~M_\odot.yr^{-1}
\label{eq-reimers}
\end{equation}
with a parameter $\eta$ calibrated in the range $1/3$ to $3$.\\

For long, this was one of the only prescription available to
use in stellar evolution models of main sequence dwarfs, and for these
objects it yields mass-loss rates in the range $10^{-15}$ to $10^{-13}$
M$_\odot$ yr$^{-1}$ depending on the mass.\\

Cranmer \& Saar (2011) have recently proposed a
physically motivated model of time-steady mass loss suited to the case of cool main
sequence and evolved giants (all late-type stars; see Fig.~\ref{f-CS}
upper panel), which can be used as an alternative to Eq.~(\ref{eq-reimers}).\\ This model considers MHD fluctuations to be the only
sources of energy and momentum to accelerate winds from cool
stars. From this postulate, mass loss is then associated to supersonic
winds emanating from (1) gas pressure in a hot corona, (2) wave pressure
in a cool extended chromosphere, so that 

\begin{equation}
\dot{M} \approx \dot{M}_{\rm cold} + \dot{M}_{\rm hot} \exp(-4 M_{A,TR})
\end{equation}

where $M_{A,TR} = u/V_A$ is the Alfv\'en Mach number in the transition
region between the cool chromosphere and the hot corona, $u$ is the
radial outflow speed of the wind, $V_A$ is the Alfv\'en
photospheric speed, directly proportional to the magnetic field
strength and configuration, $\dot{M}_{\rm cold}$ is the chromospheric
mass-loss rate and $\dot{M}_{\rm hot}$ is the hot coronal mass-loss
rate.\\
The mass-loss rate thus depends directly on the magnetic field
strength and configuration, which in turn depends on the rotation period
$P_{\rm rot}$. An approximate expression for this mass-loss rate would
be :

\begin{equation}
\frac{\dot{M}}{10^{-10} M_\odot yr^{-1}} \approx \left(
  \frac{R_*}{R_\odot}\right)^{16/7}\left(
  \frac{L_*}{L_\odot}\right)^{-2/7} \times \left( \frac{F_{A*}}{10^9
    erg cm^{-2} s^{-1}}\right)^{12/7} f_*^{(4+3\theta)/7}
\end{equation} 

where $f_*$ is the photospheric magnetic filling factor (associated to
flux tubes that emerge in the chromosphere) that directly depends on
the rotation period, with $f_* \propto P_{rot}^{-1.8}$ (according to
\cite{saar96}), $\theta$ is a dimensionless constant with values between 0 and 1, and $F_{A*}$ is the flux of energy in kink/Alfv\'en waves
     in stellar photospheres. \\

This mass-loss rate prescription directly depends on rotation, and yields very good
agreement with observations (for a sample of 29 stars with reliable
measurements). In Fig.~\ref{f-CS} we reproduce the predicted mass-loss rates as
a function of effective temperature and rotational period
$P_{\rm rot}$ . In the range $T_{\rm eff} \in [3800 K;6700 K]$, the
mass-loss rates are very sensitive to the rotation period and can be
greatly enhanced (up to 3 orders of magnitude) for rapid
rotators.\\ Let us mention here that this prescription is not valid
for (classical) T-Tauri stars, where the strong interaction of the
star with its near environment implies a very intricate interplay
between accretion and mass loss that remains to be fully understood
(see J. Ferreira, this volume). It may nonetheless be used to model
mass loss in weak T-Tauri stars (WTTS) for which the disk coupling and
associated mechanisms are not dominant anymore (see J. Bouvier, this
volume and \S~\ref{s-PMS}).


\section{Transport processes in rotating stars}\label{s-transport}

In the previous section we have seen how rotation modifies the basic
stellar structure equations via the modification of the total
potential by the centrifugal force, and the resulting modification of
the radiative equilibrium of the rotating stars.\\
In addition to these {\em direct} effects, the fact that the star
rotates also triggers hydrodynamical instabilities and large-scale
motions of the plasma in the radiative regions, resulting in the transport of angular
momentum and of chemical species. If the chemical stratification is
modified in the regions where nucleosynthesis occurs, these
transport processes are likely to affect the evolutionary path of the
stars.\\
In this section, I briefly present the different physical
transport processes that can exist in a rotating star (extended
descriptions have been published in the proceedings of EES 2006 :
see \cite{talon08}), and describe the formalisms developed to include them
in 1D stellar evolution codes.\\

\subsection{Hydrodynamical instabilities}\label{s-hydroinst}

According to the time-scales involved, we may separate the hydrodynamical instabilities that are
likely to redistribute angular momentum and modify the abundance
profiles in evolving stars in two classes :
\begin{itemize}
\item {\bf Dynamical instabilities --} They operate on typical time-scales
  of the order of the free-fall time-scale $\tau_{ff} =
  \sqrt{\frac{R_*^3}{G M_*}}$, defined as the necessary time for a star
  to have its radius divided by 2. These instabilities thus act on
  very short time-scales and the associated perturbations can be
  considered to be adiabatic.  
\item {\bf Secular instabilities --} They operate on typical time-scales
  of the order to the Kelvin-Helmholtz time-scale $\tau_{KH} =  \frac{G
    M_*^2}{2 R_* L_*}$, defined as the time for which a star can
  sustain its luminosity by the sole gravitational contraction. These
  instabilities act on time-scales of the order that of stellar
  evolution, and the associated perturbations are expected to allow
  energy exchange over nuclear time-scales. They should be explicitly
  accounted for.
\end{itemize}

\subsubsection{Solberg-H\o iland instability}

\begin{figure}
\includegraphics[width=6cm]{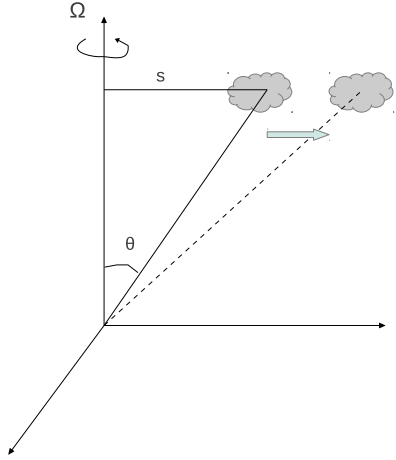}
\caption{Sketch of the configuration required for the Solberg-H\o iland
instability associated to the displacement of a fluid element in a
rotating fluid, the direction of which is indicated by the arrow. $s = r \sin \theta$ is the distance to the rotation
axis.}
\label{f-SHBV}
\end{figure}

This dynamical instability is related to Coriolis force and
applies to a fluid in cylindrical rotation. It is also known as the
axisymmetric baroclinic instability. It arises when the net
force (gravity + buoyancy + centrifugal force) applied to a fluid
parcel in an adiabatical displacement has components only in the
direction of the displacement.\\
The stability criterion against axisymmetric displacements is twofold
:
\begin{enumerate}
\item The Rayleigh stability criterion for a rotating fluid must be
  verified :
$$N^2 + N^2_\Omega \sin \theta \geq 0.$$ 
\begin{equation}
N^2 = N^2_T + N^2_\mu = \frac{g}{H_P} \left(\delta (\nabla_{\rm ad} -
\nabla) + \phi
\nabla_\mu \right)
\label{eq-BV}
\end{equation}
with $\delta = -(\partial \ln \rho / \partial \ln T)_{P,\mu}$ and $\phi = (\partial \ln \rho / \partial \ln \mu)_{P,T}$ the coefficients derived  from the
general equation of state (are unity for a perfect gas).

$N$ is the Brunt-V\"ais\"al\"a frequency at which a fluid element
oscillates when it is vertically adiabatically displaced (see
Fig.~\ref{f-SHBV}). $N^2 > 0$ ensures the local stability against
convection in non-rotating fluid.
\begin{equation}
N^2_\Omega = \frac{1}{s^3} \frac{d (s^4 \Omega^2)}{ds}
\end{equation}
defines the Rayleigh frequency associated to the Rayleigh criterion defining the stability of an
angular velocity distribution in a fluid.
\item The specific angular momentum $j = s^2 \Omega$ increases outward
  (e.g. increases from pole to equator along isentropic surfaces (S constant)).
\end{enumerate}

\noindent In the non-rotating case, the Ledoux stability criterion for
convection is recovered.\smallskip

 \centerline {\em Rotation has a stabilizing effect against
convection provided $d j / ds > 0$.}\smallskip

\noindent The Solberg-H\o iland instability is a baroclinic instability that
develops for axisymmetric and adiabatic perturbations only if the
Rayleigh criterion ($N_\Omega^2 > 0$) is violated. It occurs only in
regions with decreasing specific angular momentum gradient and is
efficiently quenched in stably stratified regions (where $\nabla <
\nabla_{ad} + \frac{\phi}{\delta} \nabla_\mu$). {\em It evolves on a dynamical characteristic time-scale.}

\subsubsection{Baroclinic instabilities}

These instabilities can develop in the radiative regions of stars experiencing non-cylindrical rotation
laws, that is in {\em baroclinic radiative stellar interiors}.\\
They are related to axisymmetric perturbations of the fluid and
act on secular time-scales. Only {\em linear} stability criteria exist
to describe them. These criteria can be viewed as
modifications of the Solberg-H\o iland criterion when the weakening
effect of thermal diffusivity on the stabilizing effect of the
temperature gradient (GSF instability) and the weakening effect of molecular diffusivity on the
stabilizing effects of the chemical stratification (ABCD instability)
are taken into account :

\begin{itemize}
\item {\sf GSF instability criterion} (\cite{gs67,f68,a78}) for axisymmetric
  perturbation of a rotating fluid with finite viscosity $\nu$ and
  thermal diffusivity $K_T$:
\begin{equation*}
\frac{\nu}{K_T} N^2_T + N^2_\mu < 0 ~~\textrm{or}~~ \left|s
\frac{\partial \Omega}{\partial z}\right|^2 > \frac{\nu}{K_T}N^2_{T,ad}
\end{equation*}
\item {\sf ABCD instability criterion} (\cite{ks83}) :
\begin{equation*}
\frac{\nu}{K_T} \left(N^2_T + N^2_\mu\right) + N^2_\Omega  < 0.
\end{equation*}
\end{itemize} 
These instabilities are hindered by large Prandtl numbers $\cal{P} =
\frac{\nu}{K_T}$, which is the case when horizontal turbulence
is strong.

\subsubsection{Shear instability}
 
\begin{figure}
\includegraphics[width=6cm,angle=-90]{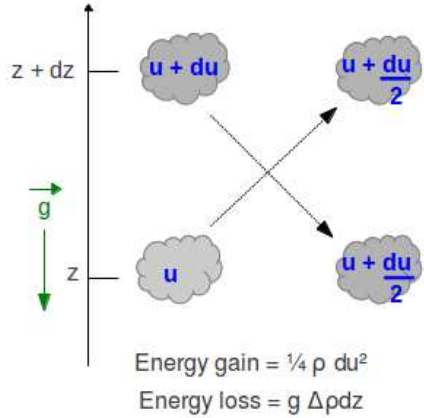}
\caption{Sketch illustrating the shear instability according to 
  Richardson criterion.}\label{f-Ri}
\end{figure}

The stability of a differentially rotating fluid against shear is determined by the Richardson
criterion :
\begin{equation}
Ri \equiv \frac{N^2}{(d u/d z )^2} > Ri_{\rm crit} = \frac{1}{4}
\end{equation}
where $u$ is the velocity of the fluid elements (``{\em eddies}'') and
$z$ designates the vertical direction as illustrated on Fig.~\ref{f-Ri}. \\ It defines the ability of
the entropy gradient to stabilize a radiative region and is hindered
by density stratification. This criterion was derived by Chandrasekhar (1961) 
from energy considerations, and is thus representative of {\em
  non-linear} regimes.\\

\noindent \underline{\sf Dynamical shear instability : horizontal turbulence}\\

Along isobars, no restoring force exist and the shear is free to
develop and becomes turbulent whenever horizontal differential rotation
is present. The horizontal shear is a dynamical shear instability,
acting on {\em dynamical time-scales} and leading to a large horizontal turbulent
viscosity that ensures shellular rotation. \\
Very little is known about the non-linear regime of the shear
instability, but from laboratory experiments, turbulence seems to have
a tendency to suppress its cause, that is differential rotation when
the turbulence results from shear (\cite{rz99}). Assuming that the
same occurs in stellar radiation zones Zahn (1992) proposed that the
main role of the horizontal turbulent viscosity $\nu_h$ is to smooth
out horizontal shear produced by the advection of angular momentum
through the meridional circulation.\\
Three prescriptions, derived from laboratory experiments (\cite{mpz04}), energetic
considerations (\cite{maeder03}), or relying on arbitrary postulates
(\cite{Zahn92}), are available to express the horizontal shear
turbulent viscosity as a function of the meridional circulation
(defined hereafter in \S~\ref{s-MC}), and we give their expressions
here :
\begin{figure}[t]
\includegraphics[width=6cm]{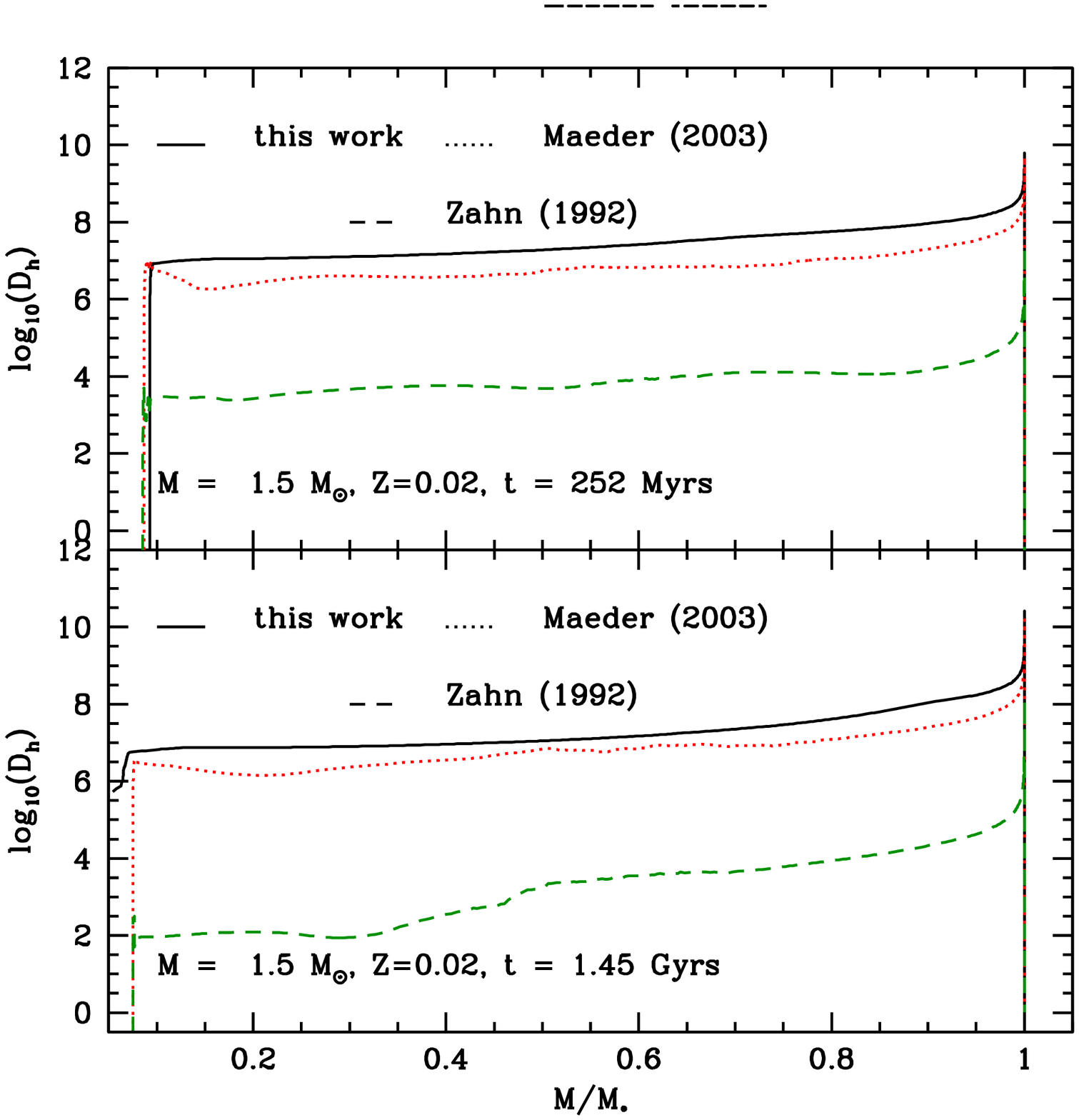}
\includegraphics[width=6cm]{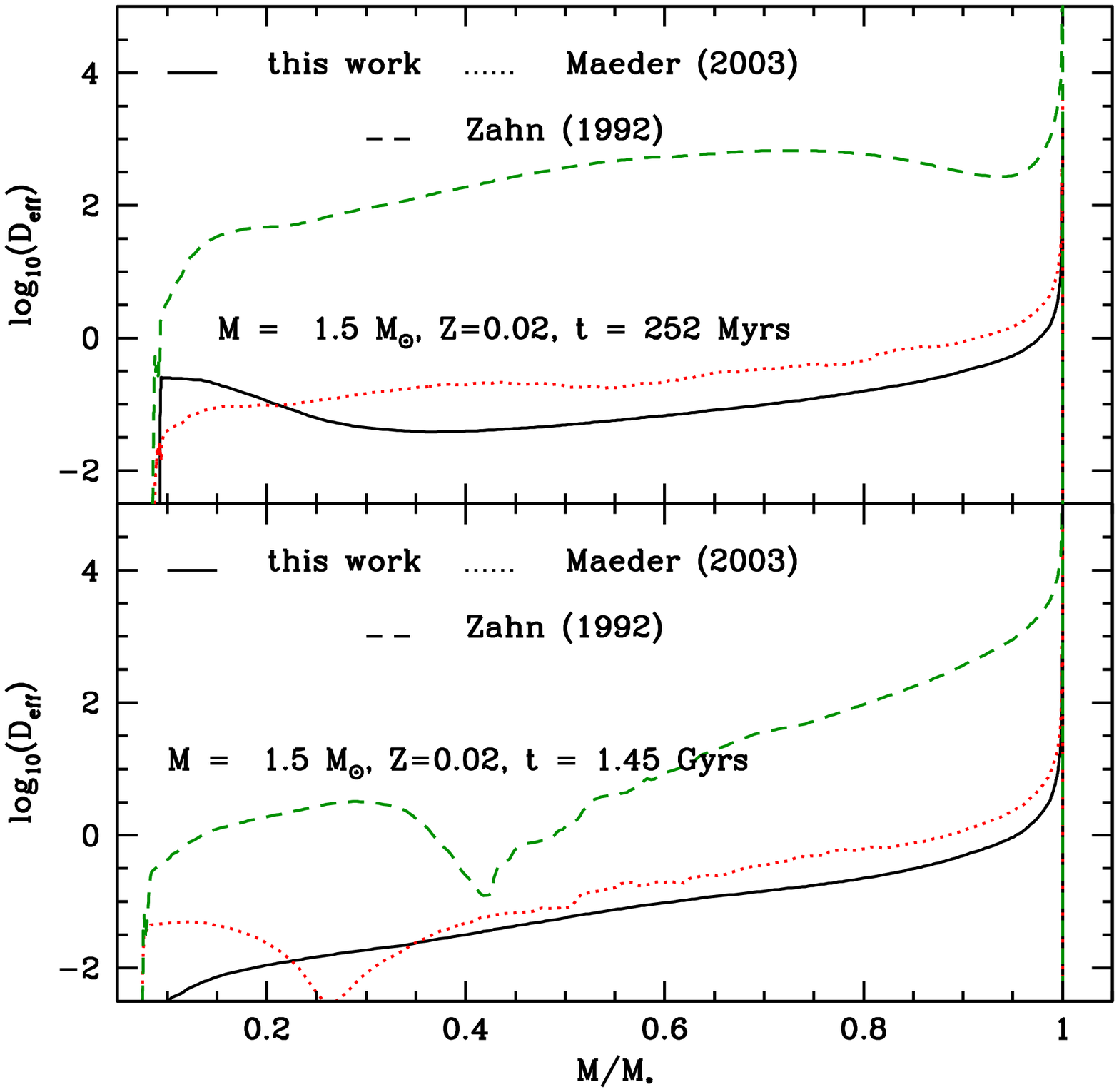}
\caption{Horizontal turbulent shear diffusivity ({\em left}) and
  effective diffusivity ({\em right}) profiles at two different ages in
  a 1.5 M$_\odot$ rotating model at solar metallicity. {\em Figure
    from Mathis et al. (2007)}}\label{f-Dh}
\end{figure}

\begin{figure}[t]
\includegraphics[scale=0.47,angle=-90]{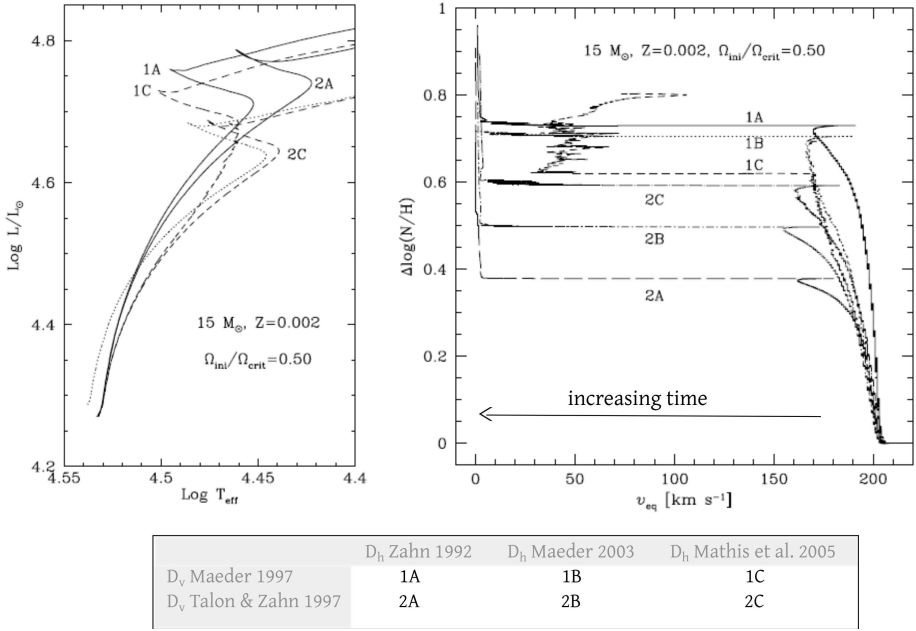}
\caption{Influence of the adopted prescriptions for the horizontal and vertical shear turbulent diffusion coefficient on the evolution and surface abundances of a 15 M$_\odot$ model at low metallicity and moderate rotation.{\em Adapted from Meynet et al. 2013.}}\label{f-prescr.}
\end{figure}

\begin{itemize}
\item {\sf Zahn (1992)}

\begin{equation}
\nu_h = \frac{r}{C_h}\left|\frac{1}{3 \rho r}\frac{{\rm d} (\rho r^2 U)}{{\rm d}
    r}-\frac{U}{2}\frac{{\rm d} \ln r^2\Omega}{{\rm d} \ln r}\right| \equiv
    \frac{r}{C_h}\left| 2 V - \alpha U \right| ,
\label{Dh}
\end{equation}
where $C_h$ is a free parameter of order 1, $V = \frac{1}{6 \rho r}\frac{\rm d}{{\rm d} r}\left( \rho r^2 U \right)$ is the
horizontal component of the meridional velocity, $U$ is the vertical
component of the meridional velocity (see hereafter); and $\alpha = \frac{1}{2}
\frac{{\rm d} \ln r^2 \bar{\Omega}}{{\rm d} \ln r}$.

\item {\sf Maeder( 2003) and Mathis et al. (2004)}
\begin{equation}
\nu_h = \chi^{\frac{1}{n}}~r~ \left( r~\bar{\Omega}(r)~V^k~[2V - \alpha U] \right) ^{\frac{1}{n}},
\label{Dhnew}
\end{equation}
where  ($\chi$~;~$n$~;~$k$) = ($\frac{3}{400 m
  \pi}$~;~3~;~1) in Maeder's expression (with $m$ = 1,3 or 5), and ($1.6
10^{-6}$~;~2~;~0) in Mathis et al. (2004).
\end{itemize}

In Fig.~\ref{f-Dh}, we reproduce the profiles of the horizontal shear
diffusivity assimilated to the horizontal shear viscosity ($\nu_h
\equiv D_h$) for the three prescriptions above, for the case of a low-mass main
sequence star with a surface velocity of 110 km.s$^{-1}$ on the Zero
Age Main Sequence (hereafter ZAMS). The more recent prescriptions by Maeder (2003) and Mathis et al. (2004)
predict a similar and much stronger horizontal shear (by 4 orders of magnitude!) in the differentially
rotating radiative interior of this star. This strongly impacts the
transport of nuclides as can be seen on the right panel of the same
figure. There the effective diffusion coefficient profiles associated to the
transport of nuclides by the meridional circulation (see below and
\cite{CZ92}) are presented and are up to 2 orders of magnitude lower
for these prescriptions of $D_h$.\\

Let us finally underline that the dynamical shear instability as defined by the sole Richardson
criterion does not account for the thermal effects and will thus only
develop in fluids that have a large Peclet number ${\cal P}e = \frac{v
  \ell}{K_T}$  that is the ratio of the thermal to the dynamical
time-scales for $v$ a typical velocity and $l$ the typical length scale
of fluid elements. As such it is thus expected to develop only in the
late evolutionary phases of rotating massive stars, being hindered by
the buoyancy in most of the cases.\\ 

\noindent \underline{\sf Secular shear instability : vertical turbulence}\\

\noindent Shear instability can also develop in stellar interiors over {\em
  secular time-scales}.\\
  
While on isobars the shear instability is free to grow, in the
direction of entropy stratification, the thermal and chemical
stratifications prevent the instability to develop. However, when the
horizontal shear is fully developed so that $\nu_h \gg \nu_v$, it
contributes to weaken the stabilizing effect of stratification, as
does thermal diffusivity. A modified instability Richardson criterion
was first proposed by Maeder (1997) to incorporate the destabilizing
effects of thermal diffusivity and it was further extended by
Talon \& Zahn (1997) to yield :

\begin{equation}
\left( \frac{\Gamma}{\Gamma +1} \right) N_T^2 + 
\left( \frac{\Gamma_\mu}{\Gamma_\mu +1} \right) N_\mu^2 < Ri_{\rm crit} 
\left( \frac{d u}{d z} \right) ^2
\label{eq-shear}
\end{equation}
where $\Gamma = v\ell/K_T$ and $\Gamma_\mu = v\ell/K_\mu$. $\Gamma$
may be understood as a turbulent Peclet number, where the turbulent
viscosity replaces the usual molecular viscosity. $Ri_{\rm crit}$ is
usually considered to be 1/4 as in the classical Richardson criterion,
but values of 1/6 or values closer to the unity can be found in the
literature (\cite{canut98}).\\
Let us emphasize that the criterion enunciated in Eq.~(\ref{eq-shear})
derives from energy considerations and as such also {\em describes the
non-linear (turbulent) growth of the instability}.\\

\noindent In order for the secular shear instability to grow, another
criterion must be complied, ensuring that the associated turbulent
viscosity $\nu_v$ is larger than the microscopic viscosity of
the medium. This condition is known as the Reynolds instability
criterion

\begin{equation}
\nu_v \geq \nu {\cal R}e_c
\label{eq-Re}
\end{equation}
where ${\cal R}e_c$ is the critical Reynolds number beyond which 
turbulence can develop freely and is not damped any more by the fluid's
viscosity. In stellar evolution codes, ${\cal R}e_c = 10$ is generally
adopted.\\
Finally an expression for the vertical shear turbulent viscosity
$\nu_v$ can be obtained from Eq.~(\ref{eq-shear}) (\cite{TZ97}):

\begin{equation}
\nu_v = \frac{1}{4} \frac{Ri_c \left( s \frac{d\Omega}{d r}
  \right)^2}{\frac{N^2_T}{K_T+\nu_h} + \frac{N^2_\mu}{\nu_h}}.
\label{eq-Dv}
\end{equation}
with $s = r \sin \theta$.\\

\noindent Figure~\ref{f-prescr.} shows the effect of the different possible combinations of prescriptions for the vertical and horizontal shear turbulent diffusion coefficients (\cite{meynet13}). The evolutionary tracks and the surface enrichments predicted for a given initial rotation rate are significantly affected by the choice of these prescriptions.\\ 
Comparison between models using different descriptions  for the rotation-driven instabilities should be done with caution. Actually, this kind of parameter-space exploration gives a hint on theoretical errorbars one should keep in mind when using stellar evolution models that have been computed with specific prescriptions and physical inputs. \\

\noindent As seen from the list above, rotation-driven instabilities are
numerous. Unfortunately most of them lack a reliable description in
their non-linear regime. This is a crucial point since mixing and
transport by hydrodynamical instabilities occur via the turbulence
they generate, and turbulence occurs in the non-linear regime.\\ For
those hydrodynamical instabilities that lack a non-linear criterion
(the baroclinic instabilities including the Solberg-H\o iland
instability for instance), caution is advised when evaluating the amount of
mixing they may generate from linear instability criteria.\\
The instability criterion derived for the shear instability on the
other hand comes from energy considerations and can be applied to the
{\em non-linear} regime. This ensures a better grounded evaluation of
the mixing generated by shear in stellar radiative interiors.\\

\subsection{Meridional circulation}\label{s-MC}
Meridional circulation occurs in convective and radiative regions in
stars, and we focus here on the radiative regions.
It was first introduced to solve the so-called {\em Von Zeipel paradox} by
Eddington (1925) and Vogt (1925). 

When considering the equilibrium of a rotating radiative region, von
Zeipel (1924) noticed that ``{\em a rotating star cannot achieve
  simultaneously hydrostatic equilibrium and solid-body
  rotation}''. This can be easily seen combining the energy
conservation equation and the hydrostatic equilibrium in spherical
coordinates for the axisymmetric case, with solid-body rotation :
\begin{equation}
\vec{\nabla} \cdot \vec{F} = \frac{d}{d\Psi} \left(\chi(\Psi) \frac{d
  T}{d\Psi}\right) \left(\vec{\nabla}\Psi\right)^2 + \chi(\Psi)
\frac{d T}{d \Psi} \Delta\Psi + \rho\epsilon = 0
\label{eq-vzpar}
\end{equation}
In this equation $\Psi$ is the total potential (gravitational +
centrifugal) and in the barotropic case considered, $\vec{\nabla}\Psi$
is the effective gravity which is not constant on equipotentials,
contrary to the other terms in $\frac{dT}{d\Psi}$. This leads to the
necessity for the energy generation to depend on the angular velocity, 
which did not suit Eddington and Vogt. They simultaneously proposed
in 1925 a solution to the Von Zeipel paradox that would not require
any specific form for the energy generation, by introducing the advection of
entropy in the heat equation

\begin{equation}
\rho T \vec{U} \cdot \vec{\nabla}S = 
\vec{\nabla} \cdot \left( \chi \vec{\nabla} T \right) + \rho \varepsilon 
\label{eq-EV25}
\end{equation}
where $\vec{U}$ is the meridional circulation velocity. Meridional
circulation carries away the excess energy from warm regions and brings energy to the otherwise cooling regions.
Although very largely adopted as the correct solution to von Zeipel paradox, this formulation is not correct as first shown
by Zahn (1992) and as clearly explained by Rieutord (2008) (see also
S. Mathis in this volume). It is nonetheless still in use to model
meridional circulation in some stellar evolution codes (see \cite{ES78,pins89,Heger00}).\\

\noindent Assuming strong anisotropic shear turbulence that ensures
that the horizontal shear viscosity (along the isobars) is much larger
than the vertical shear viscosity , $\nu_h \gg \nu_v$, the heat
equation must be complemented by an horizontal heat flux $\vec{F}_h$
that creates a thermal imbalance. This imbalance is compensated by the
advection of entropy, introducing here the meridional circulation (see 
\cite{Zahn92,MZ98} for complete description) :

\begin{equation}
\rho T \vec{U} \cdot \vec{\nabla}S = 
\vec{\nabla} \cdot \left( \chi \vec{\nabla} T \right) + \rho
\varepsilon - \vec{\nabla} \cdot \vec{F}_h.
\label{eq-heatMZ}
\end{equation}

$\vec{F}_h$ can be approximated by $$\vec{F}_h = - \nu_h \rho T
\vec{\nabla}S \approx - \nu_h \rho C_p \vec{\nabla}_h T.$$

The meridional circulation velocity is a vector field that can be
expanded over spherical functions for axisymmetric fields:

\begin{equation}
\vec{U}= \underbrace{\sum_{l>0} U_l(r) P_l(\cos \theta)}_{U_r} \vec{e}_r + \underbrace{\sum_{l>0}
V_l(r) \frac{dP_l(\cos \theta)}{d \theta}}_{U_\theta} \vec{e}_\theta,
\end{equation}
where $P_l$ is the Legendre polynomial of order $l$.
Stopping the expansion at the second order, we get
\begin{equation}
\vec{U} = U_2(r) P_2(\cos \theta) \vec{e}_r + V_2(r) \frac{dP_2(\cos\theta)}{d \theta} \vec{e}_\theta.
\end{equation}
$U_2$ and $V_2$ depend on the radius of the isobar only, which is very
convenient for the introduction of this formalism into 1D stellar
evolution code, and their expressions are as follows (see
\cite{Zahn92,MZ98} for full derivation) :

\begin{eqnarray}
V_2 & =& \frac{1}{6\rho r} \frac{d (r^2 \rho U_2(r))}{d r}~~~
\textrm{from~continuity~equation}\\
U_2 & = &\frac{P}{\rho g C_P T}\frac{1}{\left( \nabla _{\rm ad} -
  \nabla + \nabla _\mu \right)} \left[ \frac{L}{M} \left( E_\Omega+ E_\Theta + E_\mu \right) +  \frac {TC_P}{\delta} 
\frac{\partial \Theta}{\partial t}\right] P_2 \left( \cos \theta
\right)\nonumber\\
& & \textrm{From Eq.~(\ref{eq-heatMZ}).}
\end{eqnarray} 
Here $\Theta = \frac{\tilde{\rho}}{\bar{\rho}} = \frac{r^2}{3 g}
\frac{\partial \Omega^2}{\partial r}$ describes the density 
fluctuation on the isobar of radius $r \equiv r_P$, and $g \equiv
g_{\rm eff}$ is the effective gravity as defined by Eq.~(\ref{eq-geff}). $E_\Omega$, $E_\Theta$ and $E_\mu$ are terms depending explicitly on the angular velocity profile, density, and mean molecular weight fluctuations on an isobar respectively.\\
Meridional circulation transports both angular
momentum and nuclides, as the hydrodynamical instabilities
described earlier. Within the framework of shellular rotation, its
action on nuclides can be assimilated to a diffusion process. The
associated diffusion coefficient was derived by Chaboyer \& Zahn (1992):

\begin{equation}
D_{\rm eff} = \frac{| r U_2 |^2}{30 D_h}
\label{eq-Deff}
\end{equation}
 
\subsection{Internal Gravity Waves}\label{s-IGW}

Contrary to hydrodynamical instabilities or to meridional
circulation which are {\em rotation-driven} transport processes,
internal gravity waves (hereafter IGW) exist and propagate in stellar interiors even in
the absence of rotation. Rotation however affects their damping and
propagation, making IGW important vectors for the transport of angular
momentum.\\
The reader is refered to Talon (1998) and to Maeder (2009) for an in depth introduction to IGW. A brief phenomenological description of IGW is given in this section together with a more detailed and updated description of the available formalism to introduce these waves as an additional angular momentum transport process in 1-D stellar evolution codes.\\
\begin{figure}[t]
\includegraphics[scale=0.5,angle=-90]{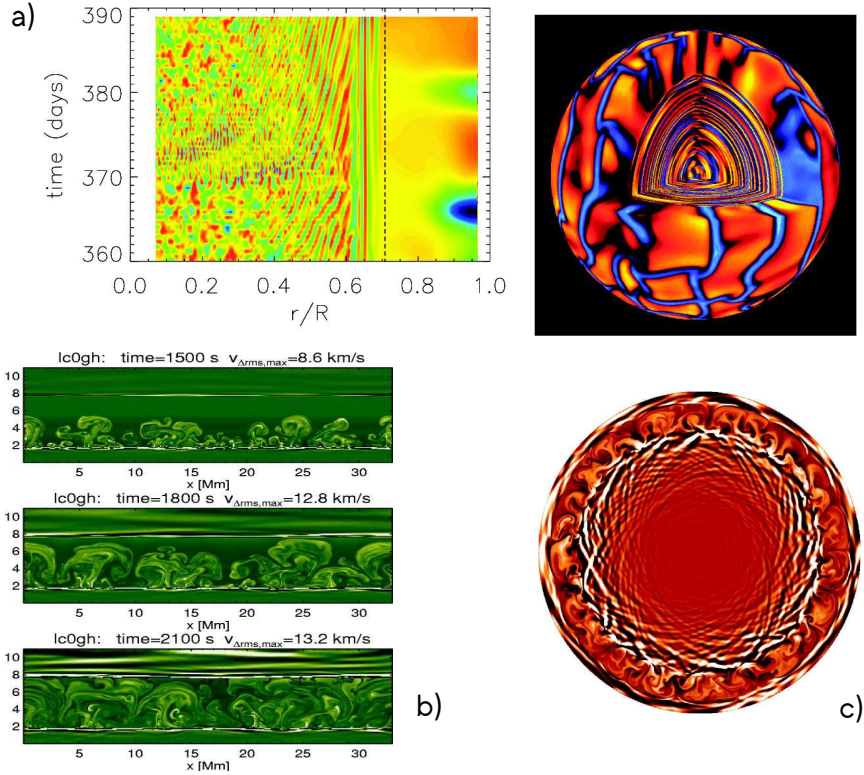}
\caption{Internal gravity waves in 2-D and 3-D MHD simulations. a) In the Sun ({\em From \cite{lucie2012}}). b) Propagation of IGW in the radiative layer above the convectively unstable helium burning region developing during the He flash in a low-mass red giant star ({\em From \cite{herwig06}}). c) IGW propagating below the solar convection zone as obtained in a 2-D hydrodynamical simulation ({\em From \cite{rogers06}}).}\label{f-IGW3D}
\end{figure} 

Fluid displacements generated by convective motions in the underlying stable and stratified radiative regions lead to the oscillation of the fluid elements displaced, thus yielding the propagation of IGW. 
Gravity and buoyancy are the restoring forces for these waves, which means that a favoured direction of propagation exists. When trapped in cavities, waves are referred to as {\em g-modes}. The IGW denomination is usually reserved to evanescent gravity waves.

Fig.~\ref{f-IGW3D} shows the results of hydrodynamical simulations for the Sun (sub-panels a and b) and in a red giant star during the helium flash (sub-panel c), where the generation of IGW is clearly seen in the radiative stable layers surrounding the convective regions where convection is fully developed.

Propagating vertically and horizontally in the fluid {\em IGW transport energy and angular momentum in stratified media}. As the natural oscillation frequency of a displaced fluid element in a stratified medium is the Brunt-V\"ais\"ail\"a frequency $N$ (see Eq.~\ref{eq-BV}), their frequency $\sigma$ is necessarily lower than $N$ with $$ 0 < \sigma < N.$$

\noindent The excitation processes of IGW are twofold (1) convective overshooting in a stable region and (2) excitation by the Reynolds stresses in the convection zone itself. The mechanism preferentially chosen within the framework of IGW propagation and damping modelling in stellar evolution codes is the later one because it has the advantage of being fully described by a formalism developed by Goldreich et al. (1994). \\ Using this formalism and considering that IGW are purely gravity waves, the kinetic energy flux per unit frequency for IGW writes (see \cite{TC05}):

\begin{eqnarray}
{\cal F}_E \left( \ell, \omega \right) &=& \frac{\omega^2}{4\pi} \int dr \frac{\rho^2}{r^2}
   \left[\left(\frac{\partial \xi_r}{\partial r}\right)^2 +
   \ell(\ell+1)\left(\frac{\partial \xi_h}{\partial r}\right)^2 \right]  \nonumber \\
 && \times  \exp\left[ -h_\omega^2 \ell(\ell+1)/2r^2\right] \frac{v_c^3 L^4 }{1
  + (\omega \tau_L)^{15/2}}
\label{eq-IGWflux}
\end{eqnarray}

with $\xi_r$ and $[\ell(\ell+1)]^{1/2}\xi_h$ the radial and horizontal displacement wave functions normalized to unit energy flux at the edge of the considered convection zone, $v_c$ the convective velocity, $L=\alpha_{\rm MLT} H_P$ the radial size of an energy bearing turbulent eddy,
$\tau_L \approx L/v_c$ the characteristic convective time, $H_P$ the pressure scale height $P/\rho g$, and $h_\omega$ the radial size of the largest eddy at depth $r$ with characteristic frequency of $\omega$ or higher ($h_\omega = L \min\{1, (2\omega\tau_L)^{-3/2}\}$). 
The radial and horizontal wave numbers  (respectively $k_r$ and $k_h$) are related by
\begin{equation}
k_r^2 = \left( \frac{N^2}{\omega^2} -1 \right) k_h^2 = 
\left( \frac{N^2}{\omega^2} -1 \right) \frac{\ell \left( \ell +1 \right)}{r^2} \label{kradial}
\end{equation}
where $N^2$ is the Brunt-V\"ais\"al\"a frequency.\\
According to Unno et al. (1989), for the waves in the low-frequency range ($\sigma \ll N$), the dispersion relation is given by 
\begin{equation}
k^2 \sigma^2 + N^2 k^2_h = 0.
\label{eq-IGWdopp}
\end{equation}
where $\sigma = \omega - m[\Omega(r) - \Omega_{CZ}]$ is the local Doppler-shifted frequency of the wave, and $\omega$ is the wave frequency in the reference frame of the convective zone where the angular velocity $\Omega_{CZ}$ is assumed to be constant.  \\

Using the energy flux defined in Eq.~(\ref{eq-IGWflux}), it is possible to establish the so-called " {\em angular momentum luminosity}" at the edge of the convective region generating the IGW 

\begin{equation}
{\cal L}_J(\ell, m) = 4 \pi r^2 {\cal F}_J (\ell, m, \omega) = 4 \pi r^2 \frac{2m}{\omega} {\cal F}_E (\ell, \omega)  
\label{eq-monochromLJ}
\end{equation}
where ${\cal F}_J (\ell, m, \omega) $ is the mean flux of angular momentum carried by a monochromatic wave of spherical order $\ell$ and local (i.e., emission) frequency $\omega$ ($m$ being the azimuthal order).\\
\begin{figure}[t]
\includegraphics[scale=0.5]{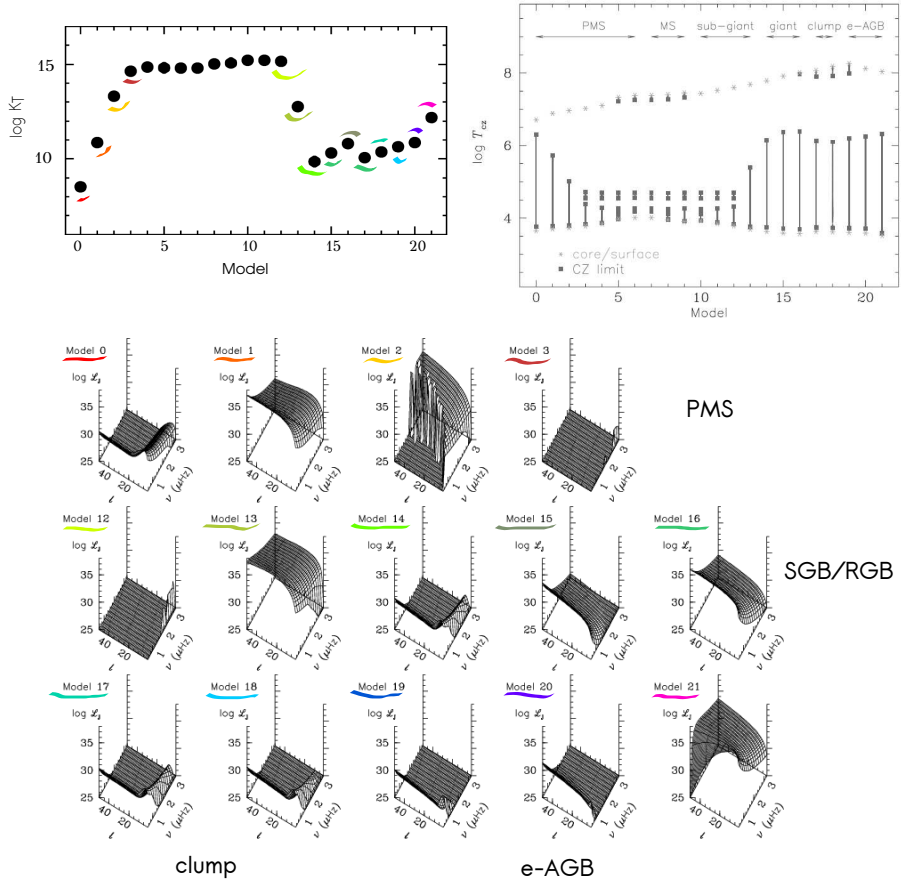}
\caption{Thermal diffusivity, temperature at the surface and the convective edges, and spectrum of angular momentum luminosity integrated over 0.2 $\mu Hz$ in a 3M$_\odot$ model at solar metallicity at different evolutionary points from the PMS to the early-AGB. ({\em Adapted from \cite{TC08}.})}
\label{f-IGWTC08}
\end{figure}

IGW are prone to radiative damping by thermal diffusivity and viscosity in corotation resonance (see \cite{Zahn97}), and deposit the angular momentum they carry at the location where they are damped. Angular momentum deposition can be positive or negative depending on the nature in terms of prograde or retrogade (azimuthal number $m >0$ or $<0$ respectively). In the absence of differential damping for prograde and retrogade waves, there will be no net local excess or deficit of angular momentum.\\ The local momentum luminosity at a given radius $r$ within a {\em differentially rotating} radiative region where the waves propagate, accounts for prograde and retrograde waves generated by the convective zone and is given by

\begin{equation}
{\cal L}_J(r) = \sum_{\sigma, \ell, m} {{\cal L}_J}_{\ell, m} \left( r_{\rm cz}\right)
\exp \left[ -\tau(r, \sigma, \ell) \right].
\label{eq-lumIGW}
\end{equation}
Here the '$cz$' index corresponds to the radiation/convection interface, and $\tau(r, \sigma, \ell) $ is the local damping rate given by

\begin{equation}
\tau(r, \sigma, \ell) = = \left[ \ell(\ell+1) \right] ^{3/2} \int_r^{r_c} 
\left( K_T + \nu_v \right) \; {N {N_T}^2 \over
\sigma^4}  \left({N^2 \over N^2 - \sigma^2}\right)^{1/2} \frac{r}{r^3} 
\label{eq-tauIGW}
\end{equation}
where $\sigma$ is the Doppler-shifted frequency defined earlier, $K_T$ is the thermal diffusivity, $\nu_v$ is the vertical shear turbulent viscosity (see \S~\ref{s-hydroinst}) and $N^2_T$ is the thermal part of the Brunt-V\"ais\"al\"a frequency (see Eq.~\ref{eq-BV}).\\

When Coriolis forces are neglected (no gravito-inertial waves), Eq.~(\ref{eq-lumIGW}) and (\ref{eq-tauIGW}) indicate that low-frequency ($\sigma \ll N$), low-degree (small $\ell$) IGW are the ones efficient to redistribute angular momentum in the radiative interiors. They penetrate in the  deepest radiative layers and their prograde and retrograde components are subject to strong differential damping, which is required for a net angular momentum deposition (or extraction). {\em This differential damping is due to the differential rotation present in the radiative interiors} (see \cite{Zahn97,TC03,TC05,TC08,CC13}). Equations also show that the actual efficiency of IGW to angular momentum redistribution strongly depends on the physical conditions at the edge of the convective regions, which in turn depend on the surface temperature of the star. For instance, the IGW angular momentum luminosity is reduced for shallow convective envelopes at the base of which the thermal diffusivity is large, leading to an efficient damping very close to the convective edge (see \cite{TC08} and Fig.~\ref{f-IGWTC08} adapted from the same paper).\\

Pantillon et al. (2007) provide modified expressions of $\cal{L}_J$ when accounting for the Coriolis acceleration  and the fact that not only pure gravity-waves but also gravito-inertial waves are generated and propagate in the radiative regions. They use the ``{\em traditional approximation}" where the radial component of the Coriolis acceleration is neglected in the equation of motion of the fluid.\\
In that case, the dispersion relation (Eq.~\ref{eq-IGWdopp}) and the total angular momentum luminosity (Eq.~\ref{eq-lumIGW}) associated to the waves are modified. Further modifications also appear when Lorentz forces are also accounted for in presence of a stellar magnetic field (see \cite{MdB12}).\\

\section{Modelling rotation-induced mixing in stellar evolution codes}\label{s-rotmod}

In the previous sections we have seen how rotation itself,
rotation-driven and rotation-influenced processes can modify the
stellar structure, the stellar radiative equilibrium, and generate
transport of angular momentum inside radiative regions essentially. We
have also described different formalisms that have been developed
across the years to model rotation in 1-D stellar evolution
codes.\\ 
Here we put together the two main approaches that have been
implemented in these very same 1-D stellar evolution codes to account
for the transport of angular momentum and nuclides by rotation-driven
and rotation-influenced physical processes.

\subsection{Diffusive approach}

This approach was first proposed by Endal \& Sofia (1978) in their
seminal paper. It is used in the stellar evolution codes KEPLER (\cite{Heger00}), STERN (\cite{yl05}) and MESA (\cite{mesa13}). The angular momentum and nuclides profiles evolution
due to secular rotation-driven instabilities are described by a set of
two diffusion equations, which have the advantage to be computationally
easy to implement :
\begin{eqnarray}
\frac{\partial \Omega}{\partial t} & = & \frac{1}{\rho r^4}
  \frac{\partial}{\partial r} \left( \rho r^4 \nu \frac{\partial
    \Omega}{\partial r}\right)\\
\frac{\partial X_i}{\partial t} & = & \frac{1}{\rho r^2}
  \frac{\partial}{\partial r} \left( \rho r^2 D \frac{\partial
    X_i}{\partial r}\right)
\end{eqnarray}
where $X_i$ is the mass fraction of nuclide $i$, $\nu$  and $D$ are respectively the total viscosity and the total diffusion coefficient defined as a sum of all diffusion coefficients associated to
the transport processes taken into account. Each of these diffusion
coefficients is built as the product of the velocity $v$ and the path
length $l$ of the redistribution currents with
\begin{equation}
l = min\left( r, \left| \frac{\partial r}{\partial \ln v} \right| \right).
\end{equation}
$\left| \frac{\partial r}{\partial \ln v} \right| $ is the velocity
scale height. The time-scale associated to the {\em redistribution}
over path scale is then simply $l^2/D = l/v$.\\
With this description, the viscosity $\nu$ is given by 
\begin{equation}
\nu =  D_{\rm conv} + D_{\rm sem} +  D_{\rm SSI} + D_{\rm DSI} + D_{\rm SHI} + D_{\rm ES} + D_{\rm GSF}
\end{equation}
 and the diffusion coefficient is usually written as the following sum
(\cite{Heger00})
\begin{equation}
D =  D_{\rm conv} + D_{\rm sem} + f_c \times\left( D_{\rm SSI} + D_{\rm DSI} + D_{\rm SHI} + D_{\rm ES} + D_{\rm GSF}
\right)
\end{equation}
$f_c$ is one of the efficiency parameters entering the diffusive
formalism. It is defined as the ratio of the
diffusion coefficient  to the turbulent viscosity $f_c \equiv D/\nu$
and it is calibrated on observations (solar lithium abundance in
\cite{pins89}, nitrogen abundances at the surface of B-type in the LMC
in \cite{Brott11a}). Another parameter, $f_\mu$ ($\in[0;1]$) is introduced to mimic
the sensitivity of the rotationally induced mixing to $\mu$-gradients, i.e.,
$\nabla_\mu$ is replaced by $f_\mu \nabla_\mu$.\\
These parameters were introduced by Pinsonneault et al. (1989) to
account for the ``{\em considerable uncertainties}'' (as explicitly
written in Heger et al. 2000) associated to
diffusion coefficient used as they derive from order-of-magnitude considerations.\\

The equation for the angular momentum evolution describes the angular velocity at all radii inside the star, with no distinction between radiative and convective regions. In the latter, solid-body rotation is assumed and actually results from the very large convective diffusion coefficient $D_{\rm conv}$ associated to the very short time-scales characterizing convection. \\
This equation can be modified to include the presence of a surface torque, for instance as that exerted by a disc coupling or a magnetized wind. We address this specific point in \S~\ref{s-wind} (see also J. Ferreira and J. Bouvier, this volume).

\subsection{Combined advective/diffusive approach}

The other family of stellar evolution models including rotation adopts 
the formalism developed by Chaboyer \& Zahn (1992), Zahn (1992) and
Maeder \& Zahn (1998), in which only the transport of nuclides is
modelled with a diffusion equation, but the evolution of the angular
momentum obeys an advection/diffusion equation. This formalism is used in the STAREVOL (\cite{Ana06,decressin09}), CESTAM (\cite{marques13}) and Geneva stellar evolution codes \newline(\cite{MM97,genevacode}). It is also included in a simplified hybrid form in the FRANEC stellar evolution code (\cite{cl13}).\\

The transport equations governing the evolution of angular momentum and nuclides {\em in the radiative interior} read :

\begin{eqnarray}
\rho \frac{d}{d t} \left[ r^2 \Omega \right] &=& \underbrace{\frac{1}{5r^2} 
\frac{\partial}{\partial r} \left[ \rho r^4 \Omega U \right]}_{ADVECTION} + \underbrace{ \frac{1}{r^4}
\frac{\partial}{\partial r} \left[ \rho \nu_v r^4 
\frac{\partial \Omega}{\partial r} \right]}_{DIFFUSION} + \underbrace{\frac{3}{8\pi r^2} \frac{\partial {\cal L}_J}{\partial r}}_{IGW} \label{eq-jtransp}
\end{eqnarray}
\begin{eqnarray}
\rho \frac{d c}{d t} &=& \underbrace{\rho \dot{c}}_{nuclear} + \underbrace{
\frac{1}{r^2} \frac{\partial}{\partial r} \left[ r^2\rho V_{ip}c \right]}_{microscopic proc.} + \underbrace{
\frac{1}{r^2} \frac{\partial}{\partial r} 
\left[ r^2 \rho \left( D_{\rm eff} + D_v \right) \frac{\partial c}{\partial r} \right]}_{shear + merid. circ.}.\label{eq-cdiff}
\end{eqnarray}
where $c_i$ is the concentration of nuclide $i$, and $V_{ip}$ is the microscopic diffusion velocity of nuclides of species $i$ with respect to protons.\\

These equations need to be complemented by an equation describing the evolution of the mean molecular weight relative fluctuations $\Lambda = \frac{\tilde{\mu}}{\bar{\mu}}$ on an isobar. 
As for the nuclides abundances/concentrations/mass fractions, we can write  a diffusion equation for the evolution of the mean molecular $\mu$ on an isobar provided that we write it using the perturbative approach $\mu = \bar{\mu} + \tilde{\mu}P_2(\cos \theta)$ :

\begin{equation}
\frac{\partial \tilde{\mu}}{\partial t} + U_2 \frac{\partial \bar{\mu}}{\partial r} = \frac{6}{r^2} D_h \tilde{\mu}
\end{equation}
which can be re-written into an equation for $\Lambda$ :
\begin{equation}
\frac{\partial \Lambda}{\partial t} =  \frac{U_2}{H_P} \nabla_\mu -\frac{6}{r^2} D_h \Lambda
\label{eq-lambda}
\end{equation}

With this advective/diffusive description of the angular momentum transport in the {\em radiative stellar interiors}, at each evolutionary time-step, it is necessary to solve a coupled system of five non-linear differential equations resulting from the splitting of the 4th order in $\Omega$ of the angular momentum transport equation (Eq.~\ref{eq-jtransp}) + the equation for mean molecular weight fluctuations evolution (Eq.~\ref{eq-lambda}). This set of equations is complemented by boundary equations that  include in particular the expression of the torque exerted by a wind for young low- and intermediate-mass stars.\\
The convective regions are assumed to have a solid-body rotation, they are treated as a bulk by integrating their inertia momentum at each time-step, and serve as boundary conditions to the radiative interior equations.
The diffusion equation for nuclides (Eq.~\ref{eq-cdiff}) is also solved at each time-step.

\subsection{Interaction with the environment during early evolution}\label{s-wind}

As indicated in the previous subsections, the transport of angular momentum, whether it is modeled by a diffusion or an advection/diffusion formalism, needs to be complemented by other mechanisms than the hydrostatic effects and the rotation-driven/influenced instabilities, that should also play a role in redistributing angular momentum.\\

This paper appears in the proceedings of the Evry Schatzman School 2012 on ``{\em Angular momentum transport mechanisms and their role on the formation and early evolution of stars}", where several contributions (J. Ferreira, J. Bouvier, S. Fromang) discuss the interaction between the low-and intermediate-mass stars with their environment during the T-Tauri phase.\\
The star is then surrounded by a disk with which it strongly interacts. In addition young stars appear to be highly magnetized, which reinforces the star-disk interaction and deeply influences the stellar winds. Facing the profound complexity of these processes, which undoubtedly play a major role in the angular momentum evolution during the early evolution phases, we lack from a consistent formalism to incorporate in stellar evolution codes.\\
Accretion of matter (and of angular momentum) from the disk onto the star is an important process that can alter the evolutionary paths, the ages and the angular momentum evolution of low-mass stars on the pre-main sequence. Yet this process is generally not accounted for when computing stellar evolution models for stars with M$_{\rm ini} \geq 0.6$ M$_\odot$ with some exceptions (\cite{ps92,ps93,siess96,siess97,siess99,tout99,baraffe09}). Note that for accretion rates of about $10^{-7} M_\odot yr^{-1}$, the influence of this process is negligible on the stellar tracks and ages. Its influence on the global angular momentum evolution is on the other end yet to be explored.\\

{\em To summarize it in one word, the T-Tauri (Classical T-Tauri phase) is not properly taken into account in stellar evolution codes as far as the angular momentum evolution is concerned. During this phase, and based on empirical (observational) evidence (see J. Bouvier in this volume), stellar evolution models just enforce the surface angular velocity to remain constant for a variable duration corresponding to what is abusively called ``disc-coupling"}.\\

When reaching the more advanced weak T-Tauri phase, the star-disk interaction is much weaker and there, several prescriptions exist for the stellar wind torque that is exerted at the stellar surface. The first popular prescription is based on extrapolation of a fit to a mean rotation velocity of stars in the Pleiades, Ursa Major, the Hyades and of the Sun as a function of age (\cite{sku72}). This law is of the form:
\begin{equation}
\Omega_{\rm eq} \propto t^{-1/2}
\label{eq-sku}
\end{equation}
where $\Omega_{\rm eq}$ is the equatorial angular velocity. \\
Assuming conservation of angular momentum and a certain amount of angular momentum at the solar age for instance one can thus calibrate the law to derive the evolution of the surface velocity from the ZAMS to the present Sun.\\

This law is now replaced by more realistic prescriptions for the torque due to a magnetized stellar wind. These prescriptions are based on the simple assumption that the stellar surface and the disc are in co-rotation at a specific radius, the Alv\'en radius $r_A$ beyond which the wind velocity becomes Alfv\'enic and the wind is free to flow. Three prescriptions are particularly used, for which the torque depends on the mass and radius of the star, but also on the angular velocity and on the magnetic configuration at the stellar surface. The effect of these prescription on the evolution of surface angular velocity is shown in Fig.~\ref{f-brake}.

\begin{itemize}
\item{\sf Kawaler (1988) }
This analytical prescription is the most popular one.
\begin{eqnarray}
\dot{J} & = & - K \Omega \Omega^2 \left( \frac{R}{R_\odot} \right)^{1/2} \left( \frac{M}{M_\odot} \right)^{-1/2} ~~~ \textrm{for} ~~~ \Omega < \Omega_{\rm sat} \nonumber\\
\end{eqnarray}
\begin{eqnarray}
\dot{J} & = & - K \Omega \Omega^2_{\rm sat} \left( \frac{R}{R_\odot} \right)^{1/2} \left( \frac{M}{M_\odot} \right)^{-1/2} ~~~ \textrm{for} ~~~ \Omega \geq \Omega_{\rm sat}
\label{eq-kawaler}
\end{eqnarray}

It depends only on the saturation value for rotation, $\Omega_{\rm sat}$, which is associated with a saturation of the stellar magnetic field and usually taken in the range [8 $\Omega_\odot$; 20 $\Omega_\odot$]. The parameter $K$ is a free parameter that needs to be calibrated on observations. It is typically of the order of $10^{47}$. Let us note that Kawaler (1988) actually provides a more general expression in which the mass-loss and the form of the magnetic field configuration are accounted for. The formula of Eq.~(\ref{eq-kawaler}) corresponds to the case of a field with a geometry close to the radial geometry (but slightly more complex) and of a total surface field strength that is directly proportional to the rotation rate.

\item{\sf Reiners \& Mohanty (2012)}

Very recently, Reiners \& Mohanty (2012) proposed a different analytical model for the magnetic torque exerted on late-PMS early main sequence low-mass stars considering that rotation is related to magnetic field strength instead of magnetic flux.
\begin{eqnarray}
\dot{J} & = & -C \left[\left(\frac{\Omega}{\Omega_{\rm crit}}\right)^4\Omega \left(\frac{R^{16}}{M^2}\right)^{1/3}\right]~~~\textrm{for}~~~ \Omega < \Omega_{\rm crit} \nonumber\\
& = & -C \left[\Omega \left(\frac{R^{16}}{M^2}\right)^{1/3}\right]~~~\textrm{for}~~~ \Omega \geq \Omega_{\rm crit}\\
\textrm{with}~~ C & \equiv & \frac{2}{3} \left( \frac{B^8_{\rm crit}}{G^2 K^4_V \dot{M}}\right)^{1/3}
\label{eq-rm}
\end{eqnarray}
with $K_V$ a dimensionless constant scaling factor relating the wind velocity $v_A$ to the Alfv\'enic velocity : $v_A = K_v \sqrt{GM/r_A}$.
This prescription considers a radial configuration for the magnetic field. It proposes a strong  dependence on the stellar radius, making any of its variations having a strong impact on the angular momentum evolution. There also is a strong dependency on $\Omega_{\rm crit}$, which corresponds to the angular velocity for which the magnetic field saturates ($B = B_{\rm crit}$). This free parameter, together with $C$ , is fitted on observations and the best combination found is $(\Omega_{\rm crit}, C) = (3 \Omega_\odot, 2660~g^{5/3} cm^{−10/3} s)$ 
\begin{figure}[t]
\includegraphics[width=4cm,angle=-90]{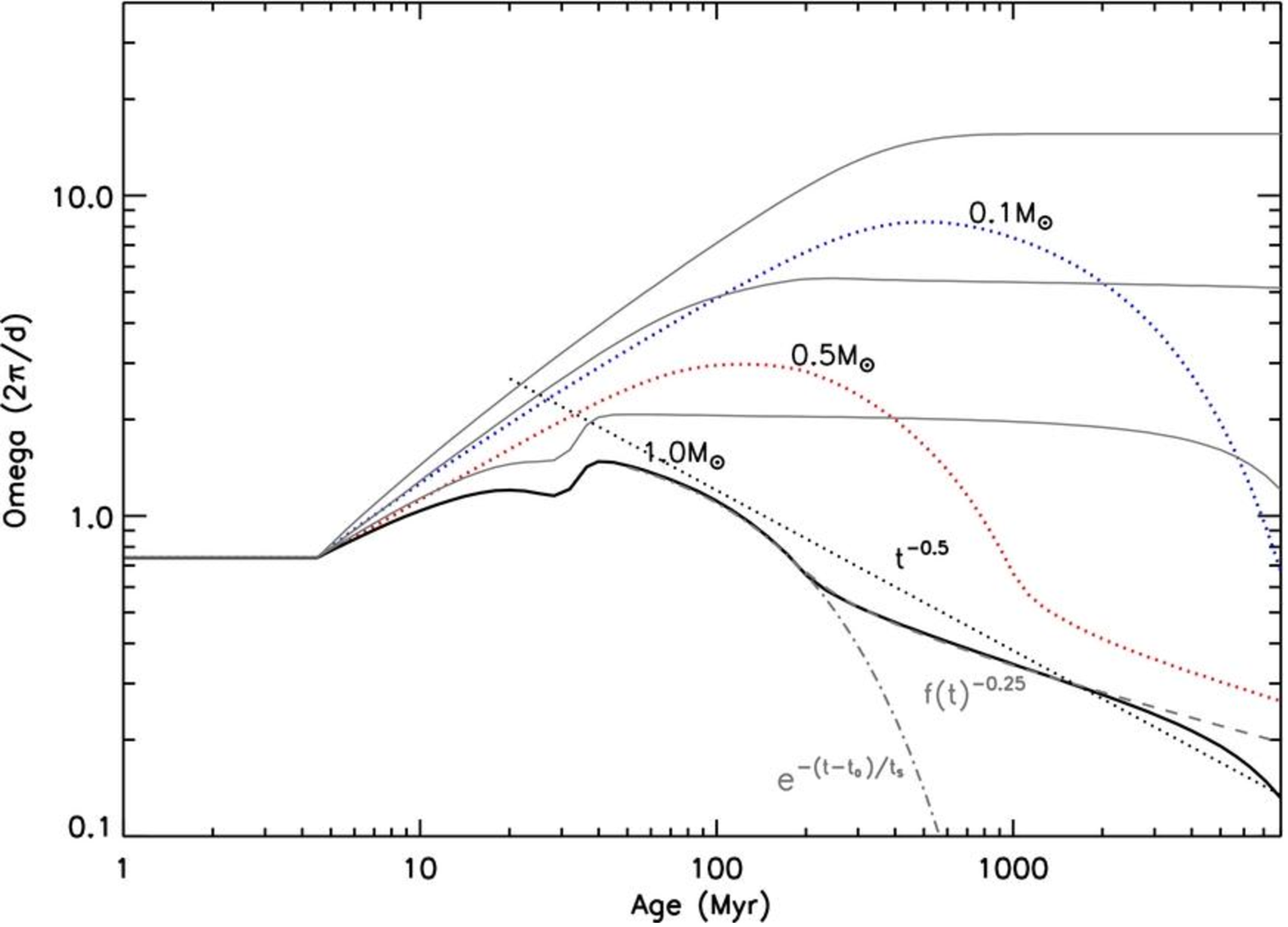}%
\includegraphics[width=4cm,angle=-90]{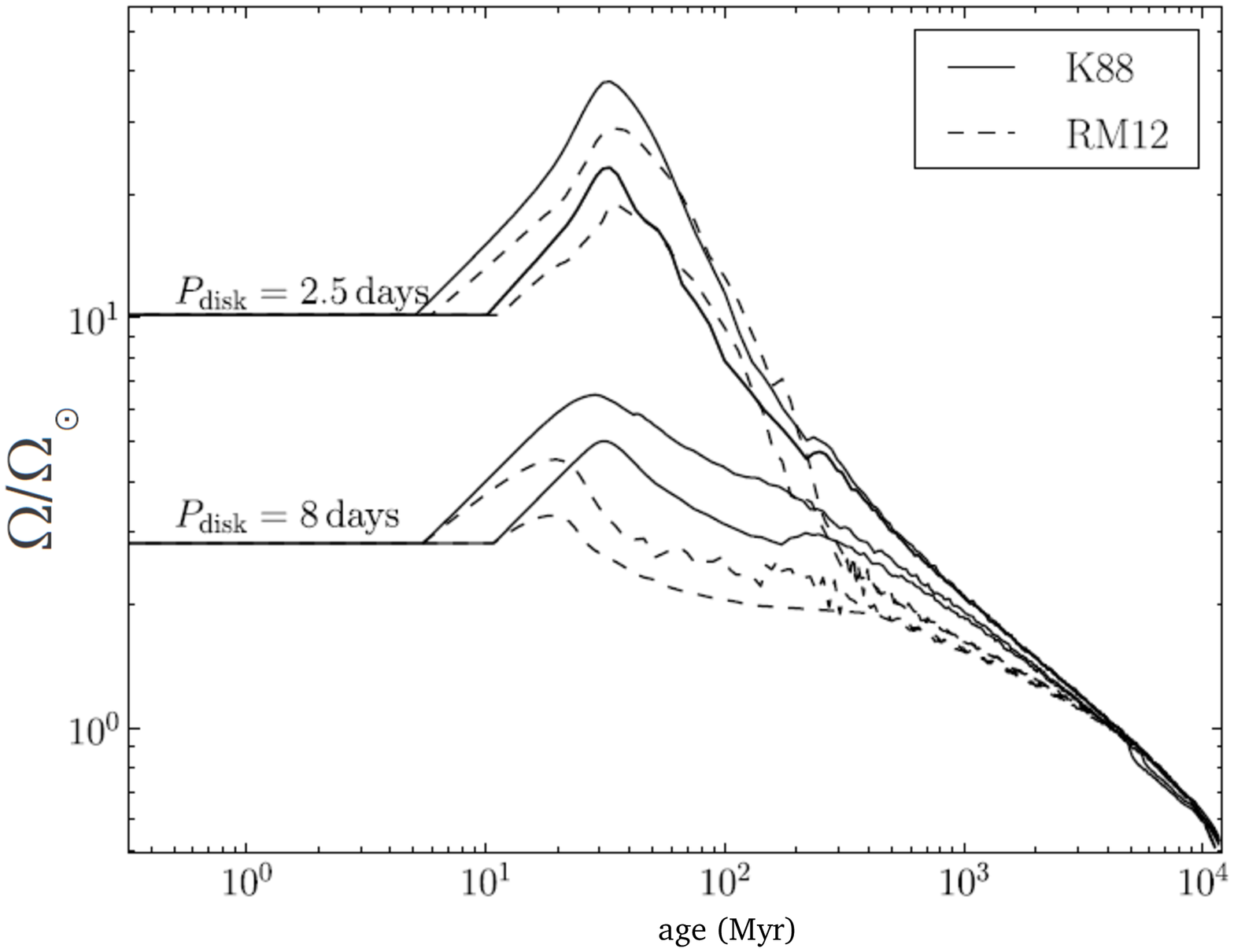}
\includegraphics[width=3cm,angle=-90]{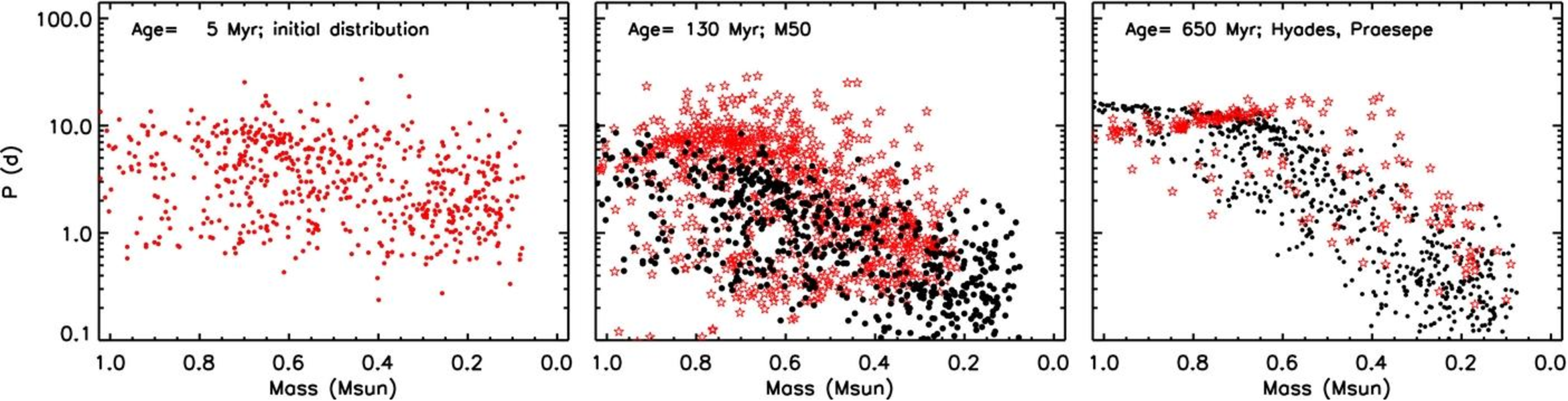}
\caption{{\sf Upper row - left panel} Figure from \cite{rm12} showing the evolution of the surface rotation period from the PMS to several Gyrs for different masses, with (bold black line, dotted lines) and without (solid grey lines) magnetized winds according to Eq.~(\ref{eq-rm}). {\sf Upper row - right panel} Figure from \cite{mg13} showing the impact of using Eq.~(\ref{eq-kawaler}) or Eq.~(\ref{eq-rm}) to describe the torque exerted by the magnetized stellar wind for a 1 M$_\odot$ model.{\sf Lower row} Figure from \cite{rm12} showing a comparison between models (in black) and observations (in red) of the rotational period evolution of open clusters stars. }\label{f-brake}
\end{figure}

\item{\sf Matt et al. (2012)}

Also in 2012, Matt et al. propose a semi-analytic derivation for the torque resulting from two-dimensional axisymmetric magnetohydrodynamical simulations that compute steady-state solutions for solar-like stellar winds from rotating stars with dipolar magnetic fields. This prescription fits all the simulations to a few percent.

\begin{eqnarray}
\Delta J & = &  K_1^2 B^{4m}_* \dot{M}^{1-2m}_W R^{4m+2}_* \frac{\Omega_*}{\left(K_2^2 v^2_{\rm esc} + \Omega^2_* R^2_*\right)^m}
\label{eq-matt}
\end{eqnarray}
where $K_2$ determines at what spin rate the stellar rotation becomes dynamically important for the wind; $K_1$ and $m$ are dimensionless fit constants with values of about 1.3 and 0.22 respectively. $B_*$ is the magnetic field strength at the stellar equator, $\dot{M}_W$ is the total mass loss rate,  $\Omega_*$ the (solid body) angular rotation rate of the stellar surface, $M_*$ and $R_*$ the stellar mass and radius, and $v_{\rm esc}$ is the gravitational escape speed.
\end{itemize}

\subsection{Angular momentum transport in convective regions}

As seen in the previous sections, provided the assumption of shellular rotation, it is possible to obtain an elaborate description of the angular momentum transport in the radiative stellar interiors via meridional circulation and shear turbulence. No similar formalism exists for convective regions that are in most cases roughly described by the phenomenological approach of the mixing length theory in 1-D stellar evolution codes.\\
The problem of how to describe the angular momentum distribution and angular velocity profile in convective regions arose already in the 70s. Tayler (1973) studied the effects of rotation in stellar convective zones and came to the
conclusion that meridional currents could develop and alter the rotation law in stellar convective regions. In that case the rotation regime could approach a uniform specific angular momentum in the convective regions ($j \propto r^2 \Omega(r)$). Another limiting case would be that of uniform angular velocity profile ($\Omega(r) = const$), and for long no real constrain was available in favour of one or another of these approaches so that the simplest was often adopted, that is uniform angular velocity , e.g. solid-body rotation. The inversion of the solar angular velocity profile from helioseismology later comforted this hypothesis : the Sun presents a fairly uniform {\em radial} angular velocity within its convective envelope (\cite{koso97}).\\
\begin{figure}[t]
\floatbox[{\capbeside\thisfloatsetup{capbesideposition={right},capbesidewidth=5cm}}]{figure}[\FBwidth]
{\caption{Mean angular velocity profiles as a function to the reduced radius from 3D hydrodynamical simulations of the inner 50\% of the convective envelope of a rotating RGB star with $R \approx 40 R_\odot$  and $L \approx 400 L_\odot$, with bulk rotation rates equal to 1/10$^{th}$ (cases A and B) and 1/50$^{th}$ (cases C and D) of the solar rotation rate 414 nHz. The profiles are obtained by averaging the latitudinal component of the velocity field over latitude, longitude, and time. {\em From models published in Brun \& Palacios (2009)}.}\label{f-rgbrot}}{\includegraphics[width=6cm,angle=90]{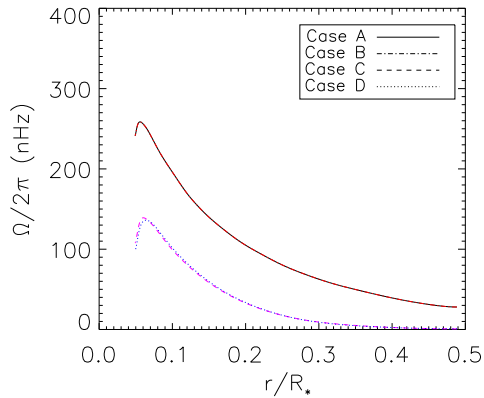}}
\end{figure}

Recent 3-D hydrodynamical simulations of rotating convective envelopes of stars at different evolutionary stages seem to indicate that not all convective regions behave as the solar one (see for instance Ballot et al. 2007 for young suns). 
In particular Brun \& Palacios (2009) have shown that the extended deep convective envelopes of red giant stars are likely to undergo strong radial differential rotation, with an angular velocity profile of the form $\Omega(r) \propto r^{-0.5}$ (see Fig.~\ref{f-rgbrot}). \\
Such a change in the adopted angular velocity profile in convective regions in stellar evolution models may impact the angular momentum and the nuclides transport in the underlying radiative region, with possible implications on the surface abundance patterns.

\subsection{The thermal relaxation and the transport loop in Zahn (1992) formalism }

\begin{figure}[t]
\includegraphics[width=7.5cm,angle=-90]{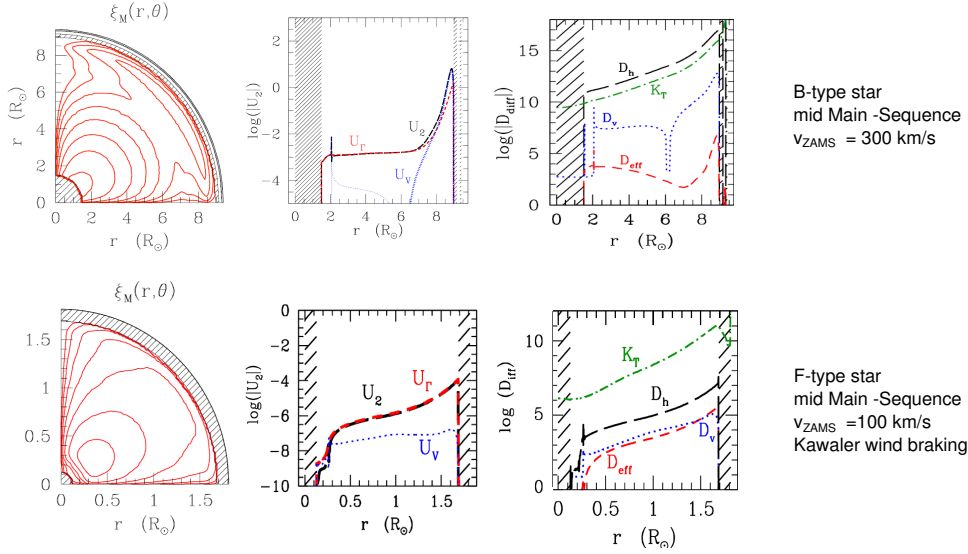}
\caption{Transport loop according to Maeder \& Zahn (1998) formalism in a typical B-type  star ({\sf upper row}) and in a typical F-type star ({\sf lower row}). 2-D meridional circulation velocity rendering, meridional circulation velocity $U_2$ and its two components $ U_\Gamma = \frac{5}{4\pi\bar{\rho}r^4\bar{\Omega}} \frac{d}{dt} \left[ \int_0^{m(r)} r'^2 \bar{\Omega} dm' \right]$ and $ U_V = -5 \nu_v \partial_r \ln \bar{\Omega}$  and diffusion coefficients (see \S~\ref{s-transport}) are represented on each row when the star is at mid-length on the main sequence ($X_c \approx 0.32$). {\em Adapted from Decressin et al. 2009.}}\label{f-AMtransp}
\end{figure}

The analysis of the transport loop as generated by meridional circulation and shear turbulence in the framework of Zahn's (1992) formalism allows to disentangle the relative importance of each of the processes at play for the angular momentum transport as well as for the transport of nuclides. This analysis is made easier by the use of the diagnostic tools developed by Decressin et al. (2009).\\
Here is an overview of the transport loop in typical B-type and F-type main sequence stars, that are summarised in Fig.~\ref{f-AMtransp}.\\

\noindent {\sf B-type star}\\

In B-type stars (and also in O-type stars), no torque is exerted at the surface\footnote{No strong / efficient braking mechanism operates during the main sequence.} although angular momentum losses occur at the surface through the important mass loss. The angular velocity profile is dominated by the angular momentum losses by winds at the surface and by the structural changes of the star during its main sequence evolution (in particular the radius inflation). Meridional circulation develops in the radiative interior that is driven by the local angular momentum transport in the high-density regions, and by shear turbulence in low-density sub-surface regions where a stationary solution is reached and the so-called Gratton-\"Opik term ($\propto \bar{\Omega}^2 / 2 \pi G \rho_m $) dominates, forcing the inversion of the angular velocity gradient. {\em Meridional circulation extracts angular momentum from the core} to compensate for the core contraction and envelope inflation. This extraction is accompanied by the advection of specific entropy, which results in a perturbation of the thermal equilibrium that generate temperature fluctuations on the isobars.\\ {\em The entropy advection by meridional circulation is compensated by the thermal readjustment.}\\
The transport of nuclides is on the other hand dominated by the shear turbulence.\\

{\em Angular momentum and nuclides transport are not dominated by the same processes.}\\

\noindent {\sf F-type star } \\

F- (and G-)type stars are prone to wind braking during the wTT phase and early main sequence, and this wind is one of the leading mechanisms that shape the internal velocity of the star on the main sequence. Meridional circulation is driven by the local transport of angular momentum everywhere in the radiative interior and the Gratton-\"Opik term, associated with the differential rotation rate, is negligible. {\em Meridional circulation is thus the dominant process for the transport of angular momentum} although it is weakened in the sub-surface regions  by the torque exerted by wind braking. This configuration is known as a {\em wind-driven meridional circulation}, after Zahn (1992). Meridional circulation also extracts angular momentum from the central regions.\\
The transport of nuclides is in this case due to  both meridional circulation and shear turbulence as the diffusion coefficients associated to both mechanisms are of the same order of magnitude.


\section{The evolution of rotating stars}\label{s-evolution}

In this section I review the main results that have been obtained in the past two decades, following the introduction of rotation and rotation-driven transport processes in stellar evolution codes, with a special emphasis on the models of rotating low-mass stars. 

\subsection{General trends of rotating stellar evolution models}\label{s-trends}

Stellar evolution models incorporating the transport of angular momentum and chemicals by rotation-driven processes all present the same general trends compared to standard stellar evolution models that are summarized here (see also \cite{MM00b,Ana06,egg10},\newline \cite{egg12} and references therein) :

\begin{itemize}
\item Hydrostatic effects described in \S~\ref{s-struc} define the evolutionary path of low- and intermediate-mass stars on the PMS and beyond the main sequence: rotating models are more compact and hotter than non-rotating ones;
\item Rotation-induced mixing efficiently modifies the hydrogen profile inside main sequence stars so that evolution tracks of moderate to rapid rotators on the Zero Age Main Sequence (ZAMS) are modified. At a given age, rotating models are more luminous and have larger radii, and the main sequence phase duration is longer;
\item Rotation-induced transport of nuclides leads to larger cores and longer lifetimes in the different evolutionary stages. For low- and intermediate-mass stars, the red giant branch is bluer and the RGB tip is more luminous in rotating models, which also have larger He core mass;
\item Mixing efficiency in terms of nuclides increases with increasing mass and initial rotation, and decreasing metallicity.
\end{itemize}

Rotation thus affects the stellar parameters not only via the hydrostatic effects that are essentially important during evolutionary phases when the departure from hydrostatic equilibrium is important (PMS, red giant phases), but also through rotation-induced mixing that modifies the elements stratification inside stars and thus impacts lifetimes and core extensions during the main core burning phases.
\begin{quotation}
{\em Rotation modifies stellar evolution and nucleosynthesis from the very early evolutionary phases. Late (post-main sequence) evolutionary phases should always be viewed as the result of the integrated evolution and treated consistently and in continuity with earlier ones.}
\end{quotation}

\noindent On top of these general trends, the impact of rotation on the mass and angular momentum losses will be different according to the type of stars and the initial rotation rates. We give hereafter more specific results of rotating stellar evolution models for different stellar types. 

\subsection{Massive stars}

With the work of the Geneva group and of Pr. N. Langer's group, stellar evolution with rotation has been shown to be successful in efficiently improve the comparison between models' predictions and observational data.\\

\noindent {\sf Surface abundance pattern}\\

\begin{figure}
\includegraphics[scale=0.35,angle=-90]{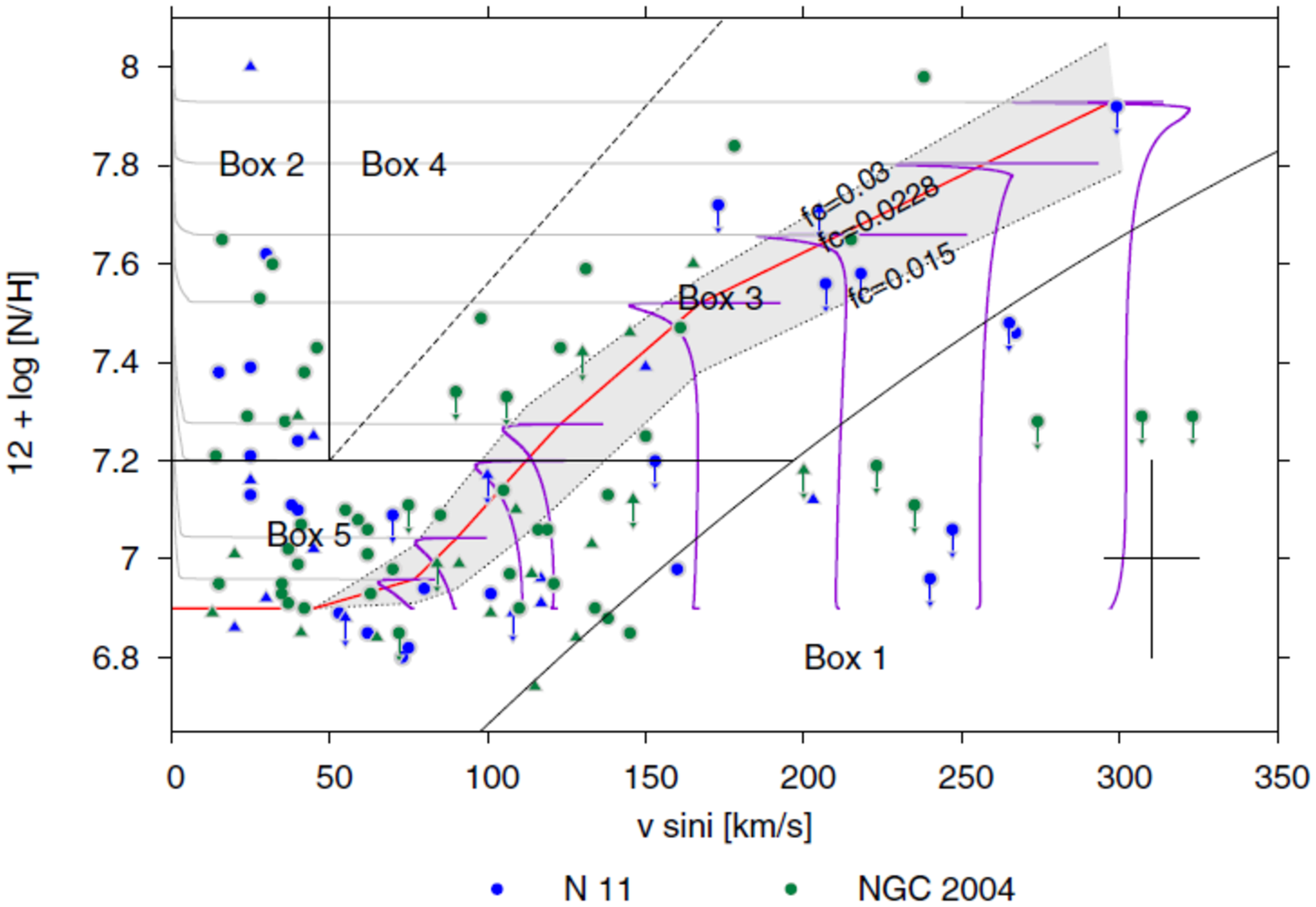}%
\includegraphics[scale=0.35,angle=-90]{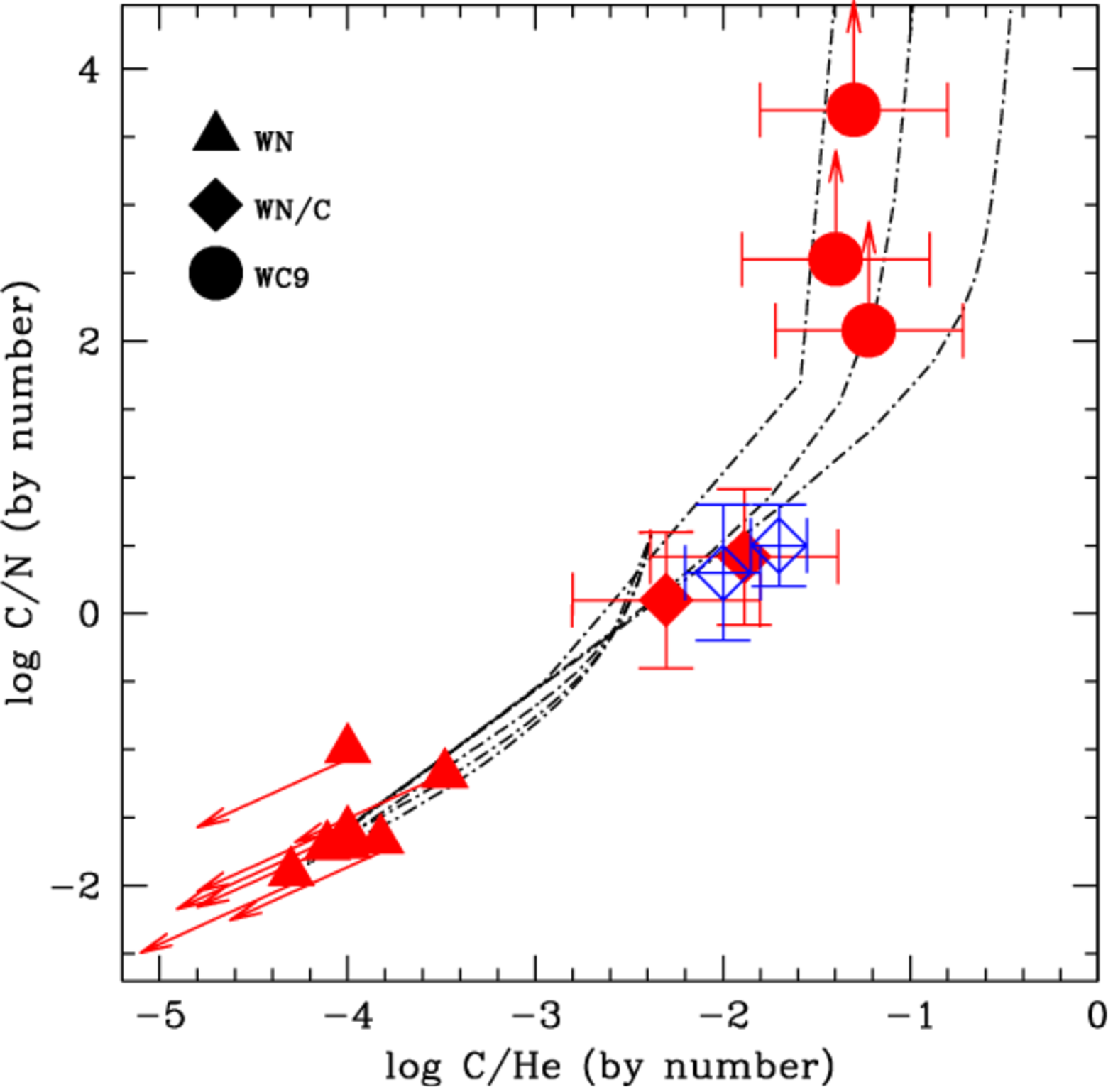}
\caption{{\sf Left panel --} Hunter diagram with N/H as a function of
  initial rotation velocity; data points are from Hunter et al. (2008a) and evolutionary tracks from Brott et al. (2011a). The grey area corresponds to the predicted values of N/H by models incorporating different mixing efficiency. These models use the diffusive approach for the transport of angular momentum and nuclides ({\em from \cite{Brott11a}}). {\sf Right panel --} C/N versus C/He for a set of Wolf-Rayet stars in WN/C Wolf-Rayet stars in the Galaxy (points) as compared to rotating models by Meynet \& Maeder (2005) (\em from \cite{fab07}).}
\label{f-Nobs}
\end{figure}

In massive stars, the high luminosity leads to strong mass-loss and this process is thus an important driver of evolution. We have seen in \S~\ref{s-mdot} that the modification of the total potential of the stars and the departure from spherical symmetry induced by rotation modify the radiative transfer and the mass loss of rotating stars, with a larger effect on the most luminous and fastest rotating ones.\\
Actually, it appears that mass loss is strongly enhanced by rotation for stars with initial masses larger than about 30 M$_\odot$ and dominates the evolution of the surface abundance pattern during their main sequence evolution. Indeed, the strong mass-loss leads to the exhibition at the surface of regions that have been nuclearly processed through the CNO cycle already on the main sequence. As a consequence, He and N enrichments are predicted from rotating models of O-type stars, in better agreement with observations than the predictions of non-rotating models, as can be seen in Fig.~\ref{f-Nobs}. For non-rotating models, the mass-loss is expected to be efficient enough to reveal internal composition only for stars beyond 60 M$_\odot$.

\begin{figure}
\includegraphics[scale=0.4]{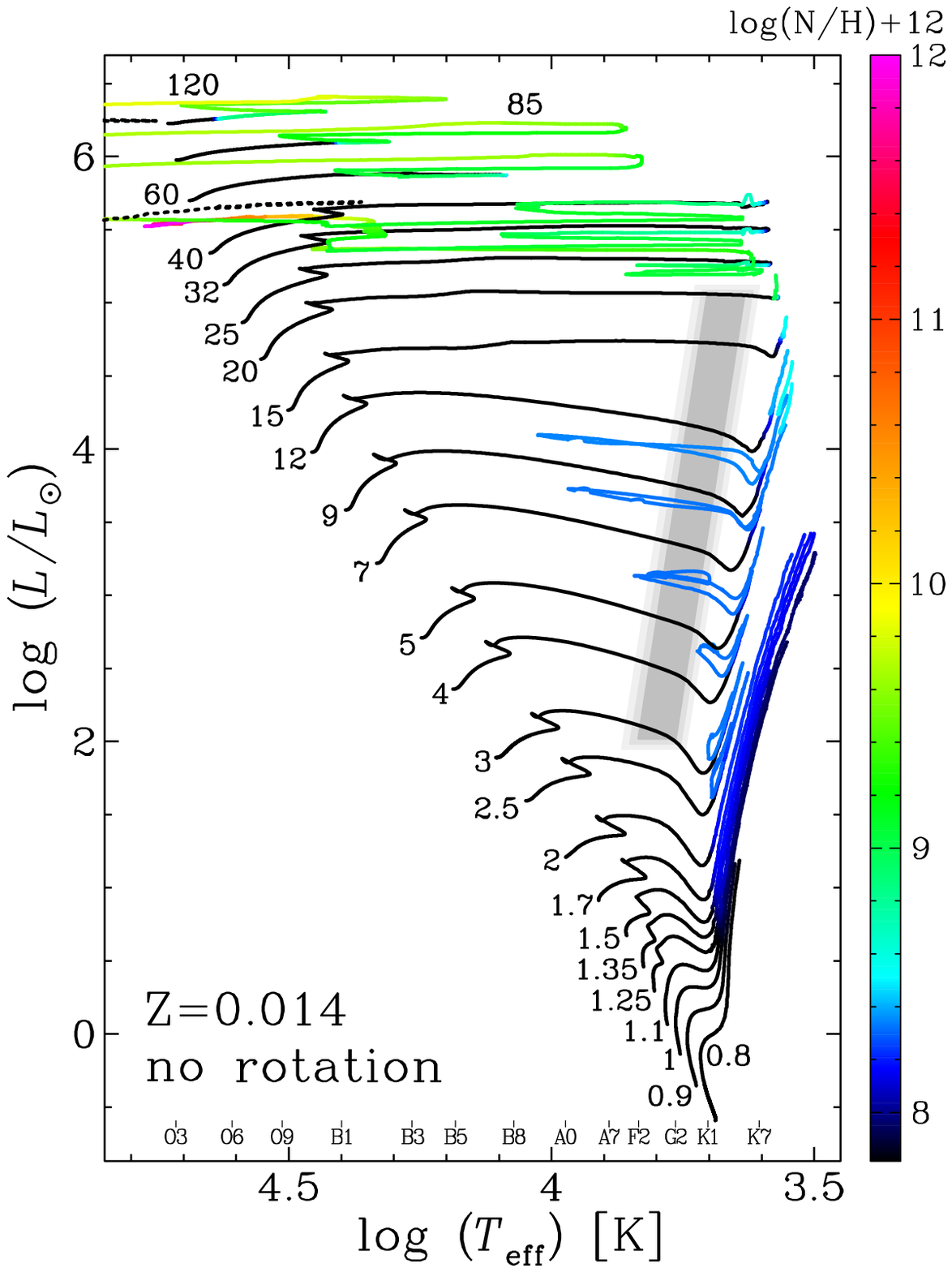}\qquad
\includegraphics[scale=0.4]{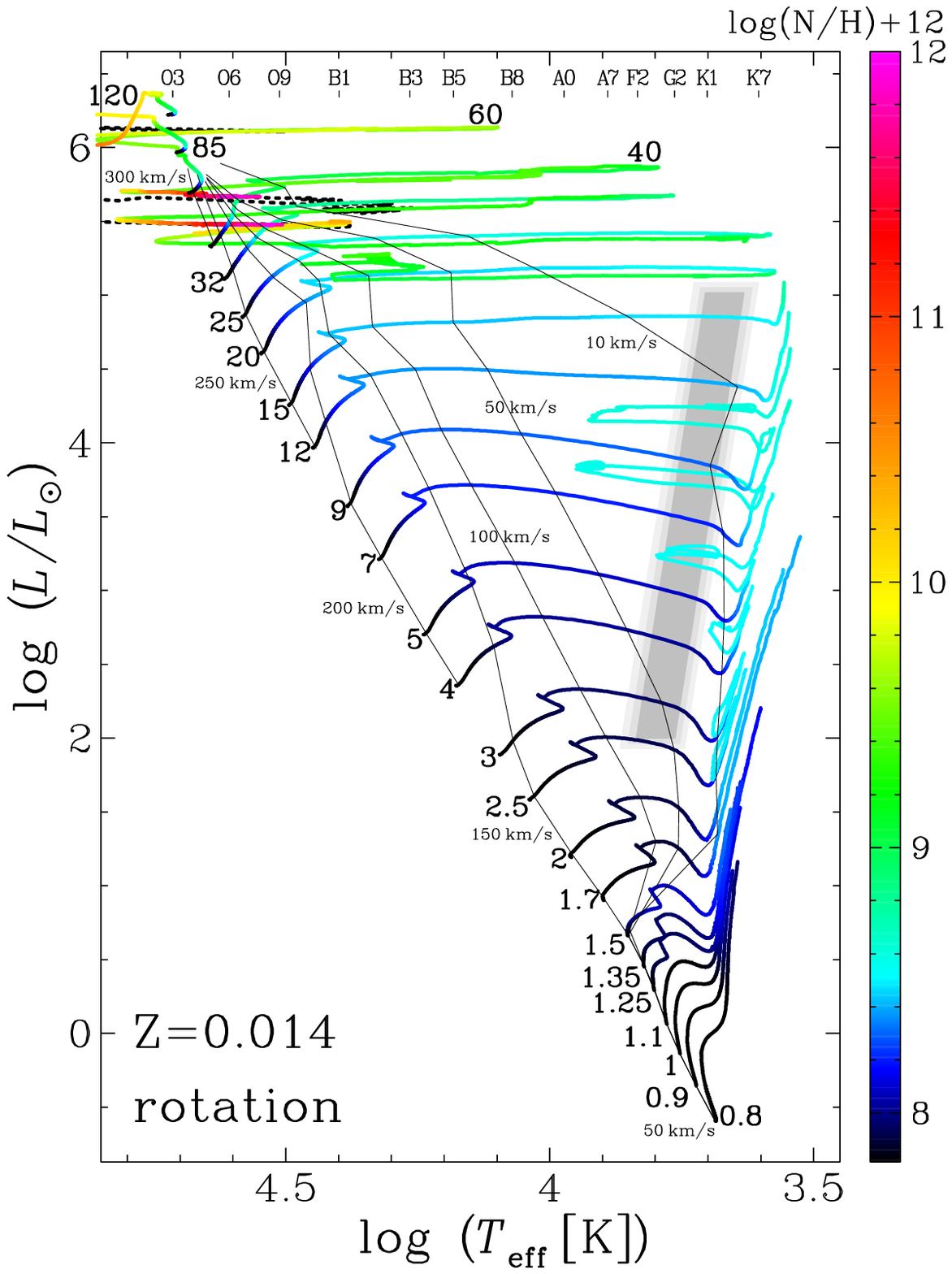}
\caption{Evolution of the surface N/H abundance in a grid of rotating ({\em right}) and non-rotating ({\em left}) massive stars at solar metallicity (Z = 0.014). The models are computed using the advective/diffusive approach.{\em From Ekstr{\"o}m  et al. (2012)}.}
\label{f-Ngeneva}
\end{figure}

\begin{figure}
\includegraphics[scale=0.4,angle=-90]{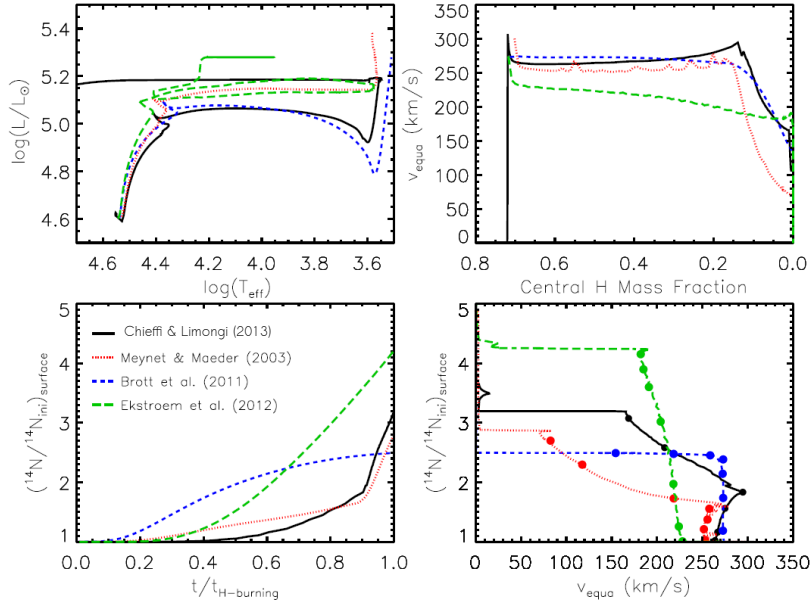}
\caption{HR diagram, surface velocity evolution as a function of central hydrogen mass fraction, surface nitrogen relative enhancement as a function of time on the main sequence and of surface velocity for 20 M$_\odot$ models from 4 different sources as listed on the diagram. The initial velocities and chemical compositions slightly differ from one to another as well as the treatment of rotational transport and  mass loss. {\em From \cite{cl13} (see paper for more details).}}
\label{f-compcodes}
\end{figure}

For stars with masses lower than about 30 M$_\odot$, mass loss on the main sequence is weaker and does not lead the evolution of the surface abundance pattern that is largely influenced by the rotation-driven transport processes, and in particular by the shear turbulence that dominates the transport of nuclides on the main sequence.\\
The shear allows to connect the surface to regions nuclearly processed through the CNO cycle :  nitrogen and to a lesser extent helium are enhanced at the surface while carbon is depleted. These abundance variations are also expected in non-rotating stars but will only appear during the later red supergiant phase when mass loss has uncovered inner processed regions. In this mass range, the earlier nitrogen enhancement predicted by rotating models also improves the comparison with observations (Fig.~\ref{f-Nobs}).\\

As can be seen from Fig.~\ref{f-Ngeneva} and Fig.~\ref{f-compcodes}, these features are qualitatively found using different approaches both for the transport of angular momentum and nuclides and for the mass loss, but substantial differences arise from one grid of models to another as can be seen in Fig.~\ref{f-compcodes} extracted from Chieffi \& Limongi (2013). These differences are partly due to different input standard physics (different initial chemical composition, different treatment of mass loss) but are also the result of different treatment of rotation and rotation-induced transport processes (see \cite{Brott11a,Ekstrom12,cl13}).\\

\noindent {\sf Number populations}\\

\begin{figure}
\includegraphics[scale=0.45,angle=-90]{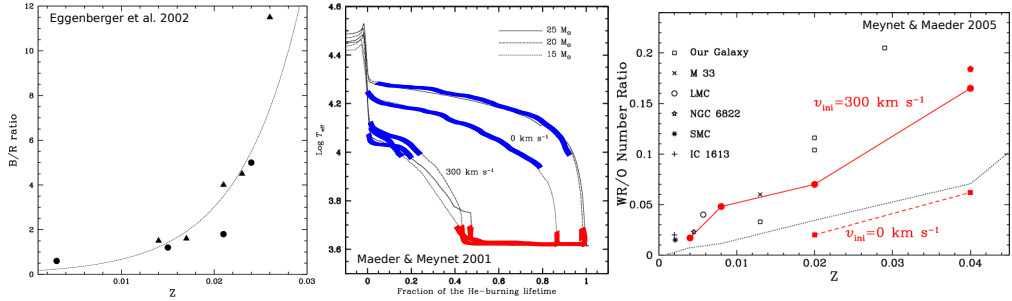}
\caption{Blue to red supergiants number counts in young open clusters of the Galaxy and in the SMC; model predictions with and without rotation for different masses at the metallicity of the SMC, with the blue and the red supergiant phases highlighted; proportion of Wolf-Rayet to O-type stars at different metallicities and predictions from rotating and non-rotating models.}
\label{f-RSG-WR}
\end{figure}

The number counts of different Wolf-Rayet stars (different subtypes) as well as that of blue and red supergiants as revealed by the observational census has been a long standing problem because the picture drawn by standard stellar evolution models is far from reproducing it (\cite{LM95,MM03}).\\
Including rotation in single stellar evolution models has greatly improved the comparison as direct consequence of the mass-loss enhancement and mixing described just above as shown in Fig.~\ref{f-RSG-WR}.
Rotation modifies the evolutionary sequence of massive stars in terms of spectral types and leads to a better yet not completely good agreement with observations. In particular the number of Wolf-Rayet stars predicted by models including rotation is larger, with a significant lowering of the minimum initial mass that a star should have to enter this phase.\\

\subsection{Low and intermediate mass stars beyond the ZAMS}
The introduction of rotation-driven transport in stellar evolution codes, and more notably using the exact same parameters for the shear turbulence modelling as those successfully employed for massive stars\footnote{This is only valid when the advective/diffusive formalism is used. In the diffusive formalism, the parameters tuning depends on the type of stars studied.}, also yields significant improvement of the comparison between models and observations.\\

\begin{figure}[t]
\includegraphics[scale=0.43,angle=-90]{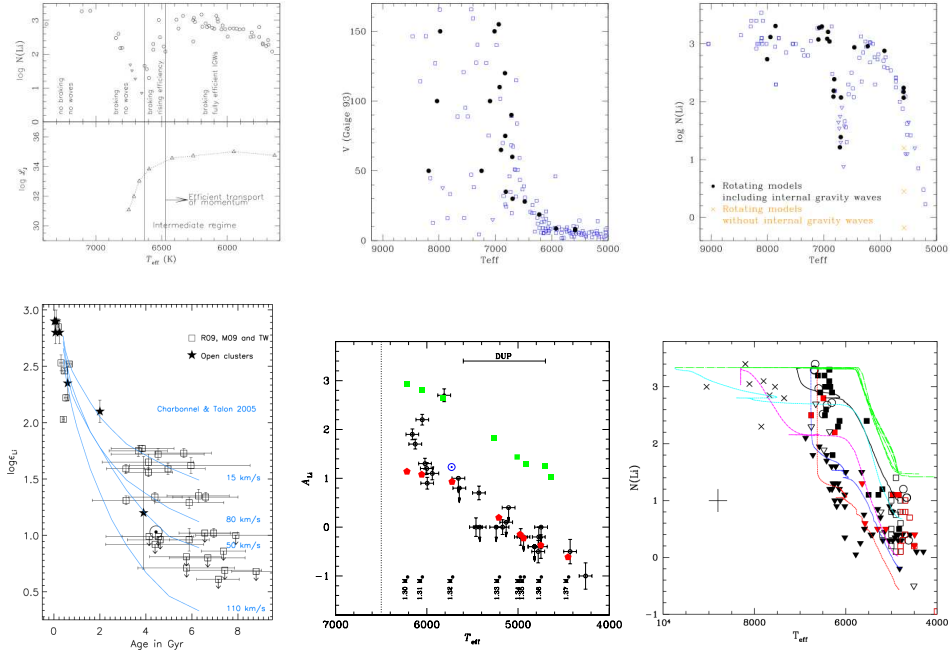}
\caption{Successes of stellar evolution models including rotation-driven transport and IGW to reproduce Li abundances in various environments.{\sf Upper row:} The Lithium dip revisited with models including rotation-driven instabilities and IGW. The waves luminosity, the calibrated surface velocities based on Gaig\'e (1993), and the resulting surface lithium abundances are displayed together with observational data for models in the mass range. ({\em From \cite{CT08}}). {\sf Lower row} {\sf Left} : Lithium abundance as a function of age for solar twins with over-plotted predicted evolution from \cite{CT05} ({\em From Baumann et al. 2012}); {\sf Middle} : Lithium abundance as a function of effective temperature for subgiant stars in the M67 open cluster compared to models with rotational mixing and no IGW for two different initial rotation velocities ({\em From \cite{CM11}}). {\sf Right} : Comparison of stellar evolution predictions for models without rotation (green) and with rotation (other colors) belonging to the hot side of the Li-dip on the main sequence and that evolve on the subgiant branch. Data are for single stars from open clusters and the galactic field ({\em From \cite{Ana03}}). }
\label{f-Limods}
\end{figure}

\noindent {\sf The rotation profile of the Sun}\\

Helioseismology has revealed that the bulk of the solar radiative interior rotates almost as a solid body (\cite{koso97}, see also Fig.~\ref{f-astero}), a profile that cannot be reproduced when shear turbulence and meridional circulation are the only efficient mechanisms to transport angular momentum.\\ Adding the transport by the IGW generated at the base of the solar convective envelope,Charbonnel \& Talon (2005) have shown that they efficiently extract angular momentum from the solar interior, flattening out the rotation profile built by meridional circulation and shear turbulence. IGW associated to rotation-driven instabilities can thus be viewed as efficient transport processes of angular momentum in solar type stars and propose a possible solution to this long-standing puzzle. Concomitantly, Eggenberger et al. (2005) have shown that magnetic fields could also enforce near solid-body rotation in the solar radiative interior, and more recently, the addition of g-mode candidates to the inversion of the solar angular velocity profile reveal a possible acceleration of the solar core (\cite{garcia11}, see also Fig.~\ref{f-astero}a), showing that this question is still to be fully understood.\\

\noindent {\sf Rotation of red giant stars}\\

The rotation of red giants is constrained by $\upsilon \sin i$ spectroscopic determinations and by the inversion of mixed gravity/pressure modes from asteroseismology data as pointed out in \S~\ref{s-obs}. These stars present slow projected surface velocities and internal differential rotation, both characteristics qualitatively  predicted by models including rotation (\cite{Ana06}), but quantitatively far from the observed values.
Indeed, the models including the transport processes described in the previous section predict a steep decrease of the surface velocity driven by the expansion of the stellar radius that basically dominates the transport of angular momentum in the very limited radiative interiors of red giants. This is valid under seemingly abusive assumption of solid-body rotation in the very extended (up to 90\% of the total stellar radius) and should be emphasized if differential rotation of the convective envelope is considered, as suggested by 3D hydrodynamical simulations (\cite{bp09}, see also Fig.~\ref{f-rgbrot}). In addition to that, Palacios et al. (2006) and Eggenberger et al. (2012b) have shown that the models including self-consistent transport of angular momentum in radiation interiors predict much steeper angular velocity profiles than that derived from asteroseismic data by Deheuvels et al. (2012) (Fig.~\ref{f-astero}b).\\
This discrepancy between models and observations for red giant stars points towards additional angular momentum transport processes that could efficiently redistribute angular momentum in these stars, accelerating the surface and decelerating the core. Planets accretion has been suggested recently by Carlberg et al. (2012) as one of the possible causes for rapid rotation of low-mass red giants, but no exploration of such processes have been made with stellar evolution models so far.\\

\noindent {\sf Lithium abundance}\\

For low- and intermediate-mass stars, introducing rotation-driven
transport processes and IGW allows to build a self-consistent picture
for the Li surface abundance evolution in various environments, as can
be seen in Fig.~\ref{f-Limods}.\\ In stars with shallow convective
envelopes on the main sequence, IGW are inefficient in the transport
of angular momentum and chemicals, which is dominated by meridional
circulation and turbulent shear. The degree of mixing affecting the Li
surface abundance depends on the degree of radial differential
rotation which feeds the shear. This is directly linked to the surface
rotation rate and to the torque applied at the surface by the
magnetized winds that these stars experience during their early
evolution on the main sequence. By calibrating it on the observed
rotation rates of low-mass stars in open clusters such as the Hyades
(the calibration concerns the constants that are introduced in the
wind braking prescriptions described in \S~\ref{s-wind}), Palacios et
al. (2003) and Canto Martins et al. (2011) have shown that the shear mixing efficiently operates on the main sequence, leading to lithium abundance reduction at the stellar surface as the star evolves on the main sequence and beyond, in excellent agreement with observations.\\
In cooler and lower-mass stars, the convective envelope on the main sequence is deep enough for the IGW to play a significant part in the transport of angular momentum and (indirectly) that of nuclides. The lithium dip showed in upper left panel of Fig.~\ref{f-Limods} is a characteristic feature of lithium abundances which is seen both in the field and in open clusters (e.g., Wallerstein et al. 1965; Boesgaard \& Tripicco 1986; Balachandran 1995). It corresponds to a narrow region in effective temperature (around $T_{\rm eff} \sim~$6650 K) where the surface Li abundances are reduced by up to $\sim $2.5 dex. While models including rotation-driven transport for stars on the hot side of the dip (more massive than about 1.35 M$_\odot$) lead to important lithium (and beryllium) destruction at the correct $T_{\rm eff}$ (\cite{TC98,Ana03}), they predict even larger Li depletion for stars on the red cold side of the dip, contrary to what is observed. When IGW are also taken into account for the transport of angular momentum in these models, as they should be, they appear to significantly reduce the shear mixing, thus partially preventing Li depletion, in agreement with observations (Charbonnel \& Talon 2005). Moreover, these models also reproduce correctly further evolution of Li surface abundances of low-mass main sequence stars in open clusters (lower left panel of Fig.~\ref{f-Limods}).\\

\noindent{\sf CNO abundances} \\

The successes in reproducing abundance patterns of CNO with rotating models is limited to A- and F- main sequence stars. For these stars contrary to models solely including microscopic diffusion of nuclides, rotation-driven mixing inhibits CNO atomic diffusion in the temperature range of the Li dip, in agreement with observations (\cite{Ana03,vm99,takeda98}).\\
On the other hand, the carbon depletion and carbon isotopic ratio decrease associated with the increase of nitrogen abundances in red giant stars that was tentatively considered a result of rotation-driven transport of nuclides connecting regions where the CN-cycle operates to the stellar surfaces, cannot be reproduced by models of the diffusive/advective type (\cite{Ana06}). These abundances variations are attributed to thermohaline mixing occurring in these hydrogen shell burning stars (\cite{CL10}) .

\subsection{Solar-type stars on the PMS}\label{s-PMS}

To close this overview, let us turn to the angular momentum evolution of young (pre-main sequence and early-main sequence) solar-type stars for which a large number of rotation periods have been measured (see Figs~\ref{f-obsrot2} and ~\ref{f-brake}) in clusters with ages ranging from 1 Myr to 1 Gyr and can serve to elucidate the dominant processes controlling the angular velocity evolution from the ``{\em disc-locking}" phase to the early-main sequence.\\

\begin{figure}[t]
\includegraphics[scale=0.4,angle=-90]{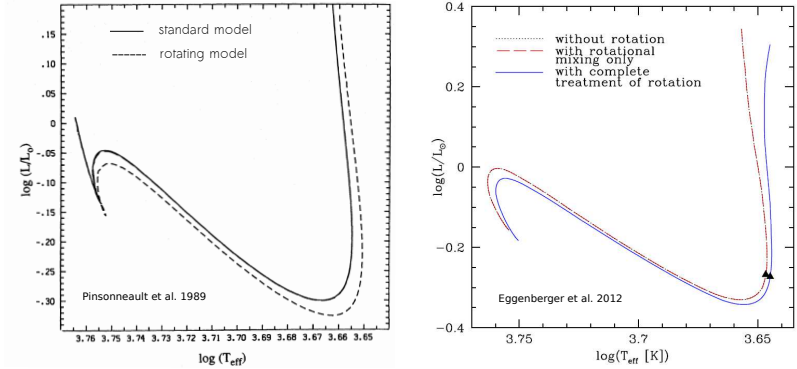}
\caption{Hertzsprung diagram for rotating and non-rotating models of a 1 M$_\odot$ star including both the hydrostatic effects described in \S~\ref{s-struc} and rotational transport using the diffusive approach ({\sf left panel}) and the advective/diffusive approach ({\sf right panel}).}
\label{f-HRD1M}
\end{figure}
\noindent{\sf HR diagram}\\

As explained earlier (see \S~\ref{s-trends}), the hydrostatic effects (e.g. the mechanical and thermal distortion associated with rotation) dominate the impact of rotation on the evolutionary path of low-mass stars on the pre-main sequence. Rotation globally lowers gravity and makes the star appear as if it was a less massive non-rotating one as can be seen in Fig.~\ref{f-HRD1M}. The transport of angular momentum and chemicals do not modify the evolutionary path, yet they obviously impact the surface rotation and chemical pattern. The more rapid the rotation at disc decoupling, the larger the shift between non-rotating and rotating tracks.\\
 On the main sequence, apart from the slight shift of the ZAMS inherited from the PMS different evolution, no noticeable difference on the tracks is expected (see \cite{pins89,egg12}).\\
 
\noindent {\sf Lithium abundance}\\

Lithium nuclear burning occurs during the PMS as soon as temperature above 2.5 10$^6$ K can be reached inside the star. These temperatures are reached when the star is still fully convective, so that the destruction of lithium in the core regions immediately shows up at the surface. When the radiative core appears, Li burning may still be going on but in the absence of any transport mechanisms in the radiative interior to connect the now receding convective envelope, Li surface abundance is expected to remain constant. When rotational mixing is taken into account, Mendes (1999), Eggenberger et al. (2012), and Marques \& Goupil (2013) obtain an additional surface Li depletion (as expected if the rotational mixing is efficient) that remains marginal (does not exceed 0.25 dex according to \cite{egg12}).\\
An important point made by Marques \& Goupil (2013)  is that despite the small effect on the PMS, rotation may significantly affect the evolution of the lithium abundance further at the ZAMS and beyond, depending on the initial conditions and particularly on the assumed rotation period during the CTT (disc-locking) phase. This effect is seemingly due to the gradients of angular velocity hence the amount of shear that can develop in the radiative regions where Li is being processed.
It is thus crucial to properly account for rotation-induced transport during the early phases.\\

\noindent {\sf Surface angular velocity}\\

\begin{figure}[t]
\includegraphics[scale=0.3]{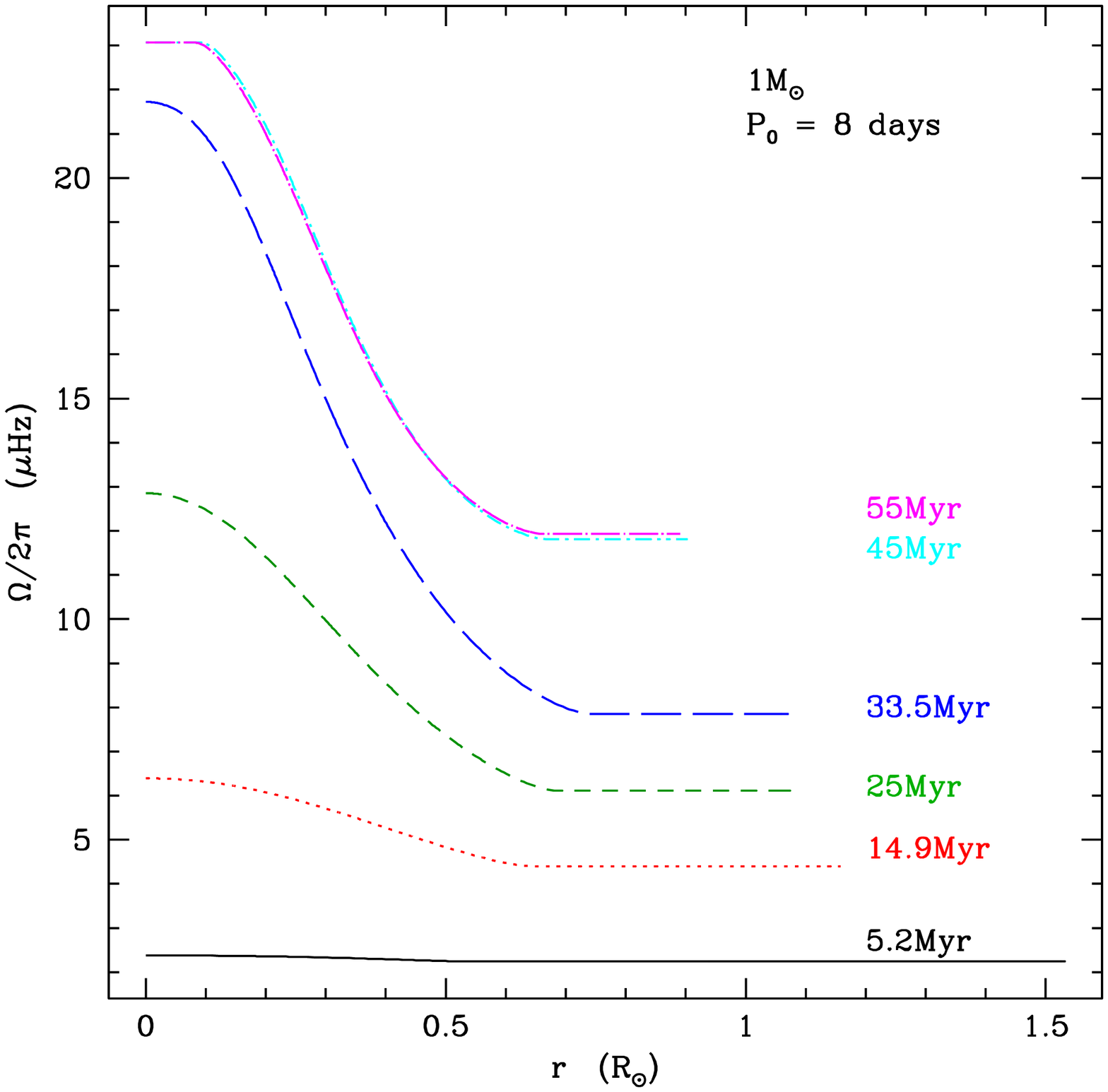}%
\includegraphics[scale=0.3]{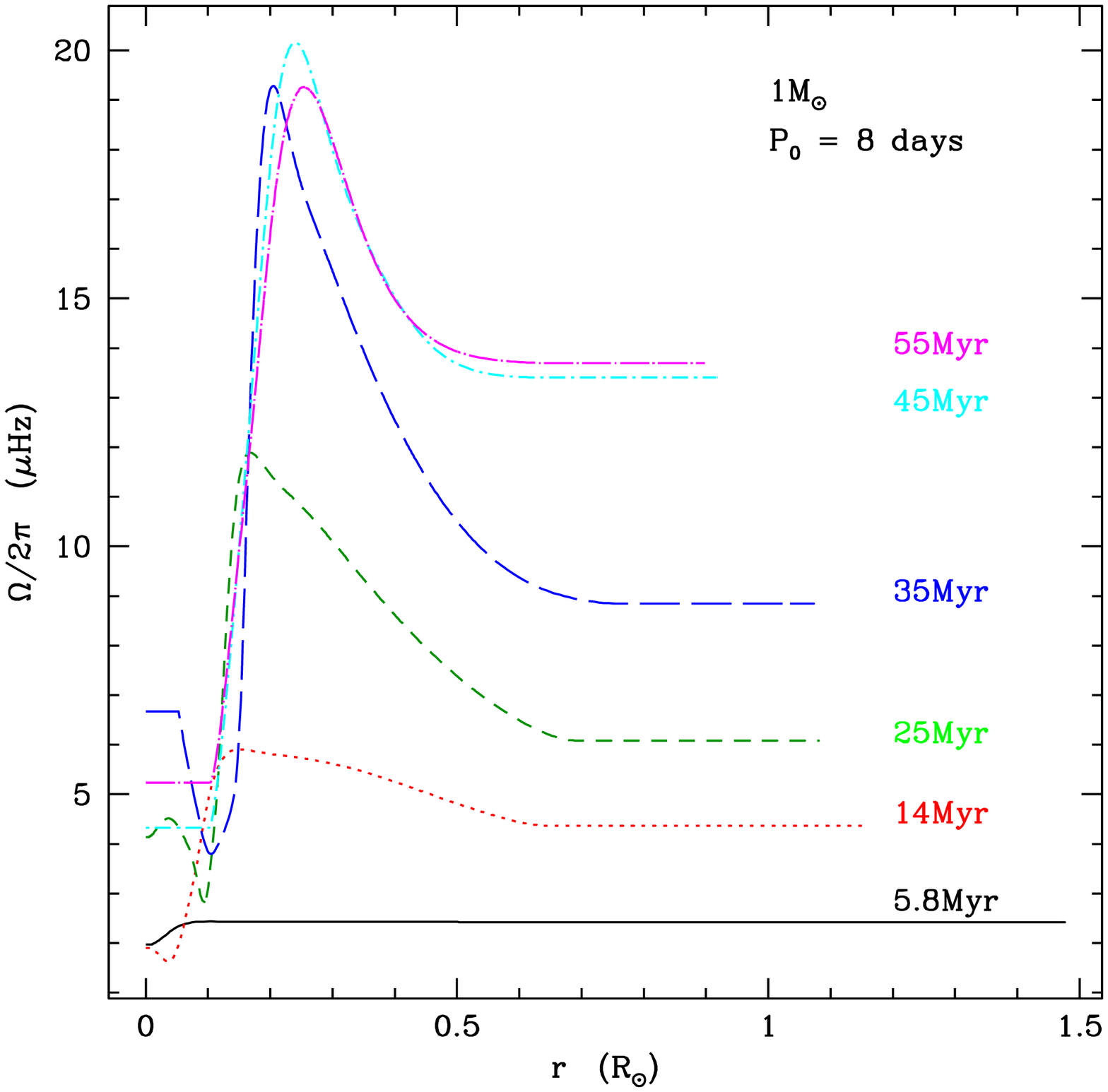}
\caption{Evolution of the internal rotation profile of a 1 M$_\odot$ at solar metallicity with a rotation period of 8 days during the disc-locking phase, and an initial velocity of about 40\% of the critical velocity, that is to say $\upsilon \approx 12$ km.s$^{-1}$. {\sf Left panel} Case where IGW are neglected; {\sf Right panel} Case with rotation-driven transport and IGWs. {\em From \cite{CC13}}.}
\label{f-rotprof1Msun}
\end{figure}

The surface angular velocity evolution during the PMS of solar-type stars is first controlled by the so-called {\em disc-locking}, and the rotation period is imposed in stellar evolution codes so as to best fit the data from young open clusters. The duration of this phase is not completely established and could vary between 3 and 6 to 9 Myrs for solar-type stars according to the initial velocity (see \cite{IB09,GB13}). During that time, the {\em surface angular velocity $\Omega_{\rm eq,surf}$} is maintained constant while the star rapidly contracts. Models indicate that for longer disc-locking phases, the angular velocity at the ZAMS is slower. Once the disc-locking phase is over, the constant surface angular velocity constrain is released, the surface of the star is spun up due to the global reduction of the stellar radius and to the now quasi-conservation of angular momentum. The later the disc-decoupling occurs, the less contraction is expected and so the less spin up. On the other hand, the magnetized winds (see \S~\ref{s-wind}) will also act during this phase and prior to the arrival on the ZAMS, and limit the spin-up of the surface.\\ The evolution of the surface angular velocity from the disc decoupling to the ZAMS is shown in Fig.~\ref{f-rotprof1Msun} where we show the internal rotation profiles at different ages for models only including rotation hydrostatic effects and rotation-driven transport and for models additionally including the transport by IGW (\cite{CC13}).\\

\noindent {\sf Internal rotation profile}\\

When the transport by IGW in the radiative interior is neglected (which should not be done as we shall show below), the meridional circulation and the hydrostatic terms control the transport of angular momentum and build up of differential rotation with a negative gradient of angular velocity (the core rotates faster than the surface). When the surface constrain is released after disc decoupling, the surface is spun up but the transport and the core contraction are efficient enough to maintain and even exacerbate the differential rotation as can be seen on the left panel of Fig.~\ref{f-rotprof1Msun}.\\
According now to what has been explained in \S~\ref{s-IGW}, IGW are likely to be efficient to transport angular momentum during the PMS evolution, and it is actually the case as soon as the radiative core appears in the star and the waves have a radiative region to propagate and dissipate into. The low-degree, low-frequency waves are those experiencing the largest Doppler shift due to differential rotation and they penetrate deep in the core regions where they deposit negative angular momentum, thus efficiently spinning down these regions as can be seen on the right panel of Fig.~\ref{f-rotprof1Msun}. On the other hand, closer to the surface the angular velocity gradient is not modified so that the effect of IGW cannot be detected by confronting rotation periods.\\
Such rotation profiles however will imply a radically different evolution for the further evolution beyond the ZAMS, in particular when the effect of the magnetized winds will be at its maximum.\\
\begin{quote}
{\em IGW dominate the transport of angular momentum in solar-type PMS stars and strongly spin down the central regions as the stars evolve to the ZAMS. The slope inversion of the angular velocity profile and the peculiar configuration with a core rotating slower than the surface needs to be taken into account to study and understand the rotation and chemical evolution of these stars on the early main sequence.} 
\end{quote}

\acknowledgements{I sincerely thank the organizers of this school for
  trusting my pedagogical abilities and giving me the opportunity to
  present this lecture. I also thank the STAREVOL team members without
  whom such a document could not have been published, with a special
  thought to M. Forestini, the TOUPIES team and J. Bouvier, and F. Martins for fruitful discussions on massive stars.}


\end{document}